\documentclass[final,3p,times,authoryear]{elsarticle}

\usepackage[usenames,dvipsnames]{xcolor}
\usepackage[colorlinks=true]{hyperref}
\usepackage{mathtools}
\usepackage{soul}
\usepackage{subcaption}
\usepackage{pgfplots}
\usepackage{cancel}
\usepackage{tabularray}
\usepackage{algorithm, algorithmicx, algpseudocode}
\usepackage{setspace}
\usepackage{relsize}
\usepackage{nomencl}
\usepackage[bottom]{footmisc}
\makenomenclature

\usepackage{etoolbox}
\renewcommand\nomgroup[1]{%
  \item[\bfseries
  \ifstrequal{#1}{A}{Abbreviations}{%
  \ifstrequal{#1}{S}{List of Symbols}}%
]}

\usetikzlibrary{matrix}
\usepgfplotslibrary{groupplots}
\pgfplotsset{compat=newest}

\usepackage{tikz}
\usetikzlibrary{tikzmark}
\usepackage{tzplot}

\usepackage{amssymb}
\usepackage{amsthm}

\usepackage{layout/personal_shortcuts}

\newlength\figH
\newlength\figW

\newcommand{\cs}{\hl{\texttt{Content Suggestion}}}

\newcommand*{\algrule}[1][\algorithmicindent]{\hspace*{.5em}\vrule\vrule
width 0pt height \baselineskip depth .25\baselineskip\hspace*{\dimexpr#1-.5em}}

\makeatletter
\def\ps@pprintTitle{%
  \let\@oddhead\@empty
  \let\@evenhead\@empty
  \let\@oddfoot\@empty
  \let\@evenfoot\@oddfoot
}
\makeatother

\makeatletter
\newcount\ALG@printindent@tempcnta
\def\ALG@printindent{%
    \ifnum \theALG@nested>0
    \ifx\ALG@text\ALG@x@notext
    \else
    \unskip
    \ALG@printindent@tempcnta=1
    \loop
    \algrule[\csname ALG@ind@\the\ALG@printindent@tempcnta\endcsname]%
    \advance \ALG@printindent@tempcnta 1
    \ifnum \ALG@printindent@tempcnta<\numexpr\theALG@nested+1\relax
    \repeat
    \fi
    \fi
}%
\usepackage{etoolbox}
\patchcmd{\ALG@doentity}{\noindent\hskip\ALG@tlm}{\ALG@printindent}{}{\errmessage{failed to patch}}
\makeatother

\AtBeginEnvironment{algorithmic}{\lineskip0pt}

\usepackage[scaled]{beramono}
\usepackage[T1]{fontenc}

\algtext*{EndWhile}
\algtext*{EndFor}
\algtext*{EndIf}
\algdef{SE}[DOWHILE]{Do}{doWhile}{\algorithmicdo}[1]{\algorithmicwhile\ #1}%

\algnewcommand\algorithmicto{\textbf{to}}
\algrenewtext{While}{\textbf{while}\ }
\algrenewtext{Procedure}[2]{\textbf{function}\ \textproc{#1}\ifthenelse{\equal{#2}{}}{}{(#2)}}

\algrenewcommand\algorithmicrequire{\textbf{Input:}}
\algrenewcommand\algorithmicensure{\textbf{Output:}}

\usepackage{eqparbox}
\algrenewcommand{\algorithmiccomment}[1]{\hfill\eqparbox{COMMENT}{\color{gray} \it-- #1}}

\journal{Journal of the Mechanics and Physics of Solids}

\begin{document}

\begin{frontmatter}



\title{Consistent machine learning for topology optimization with microstructure-dependent neural network material models}

\author[label1]{Harikrishnan Vijayakumaran}
\affiliation[label1]{organization={Department of Materials Science Engineering, Delft University of Technology}, 
					city={Delft},
					postcode={2628 CD},
					country={The Netherlands}}

\author[label3]{Jonathan B. Russ}

\author[label3,label4]{Glaucio H. Paulino}
\affiliation[label3]{organization={Department of Civil and Environmental Engineering, Princeton University}, 
	city={Princeton},
	postcode={NJ 08544},
	country={United States}}
\affiliation[label4]{organization={Princeton Materials Institute (PMI), Princeton University},
	city={Princeton},
	postcode={NJ 08544},
	country={United States}}

\author[label5]{Miguel A. Bessa}
\ead{miguel_bessa@brown.edu}
\cortext[cor1]{Corresponding author}
\affiliation[label5]{organization={School of Engineering, Brown University}, 
	city={Providence},
	postcode={RI 02912},
	country={United States}}

\begin{abstract}

  Additive manufacturing methods together with topology optimization have enabled the creation of multiscale structures with controlled spatially-varying material microstructure. 
  However, topology optimization or inverse design of such structures in the presence of nonlinearities remains a challenge due to the expense of computational homogenization methods and the complexity of differentiably parameterizing the microstructural response.
  A solution to this challenge lies in machine learning techniques that offer efficient, differentiable mappings between the material response and its microstructural descriptors.
  This work presents a framework for designing multiscale heterogeneous structures with spatially varying microstructures by merging a homogenization-based topology optimization strategy with a consistent machine learning approach grounded in hyperelasticity theory.
  We leverage neural architectures that adhere to critical physical principles such as polyconvexity, objectivity, material symmetry, and thermodynamic consistency to supply the framework with a reliable constitutive model that is dependent on material microstructural descriptors.
  Our findings highlight the potential of integrating consistent machine learning models with density-based topology optimization for enhancing design optimization of heterogeneous hyperelastic structures under finite deformations.

\end{abstract}




\end{frontmatter}


\graphicspath{{figures/}}

\section{Introduction}
\label{sec:introduction}

The growth of computational resources, together with advances in additive manufacturing (AM), have brought topology optimization (TO) to the forefront of engineering design.
The foundational work by \citet{bendsoee1988gene}, which addressed the optimization of material distribution through a homogenized treatment of the microstructure, provided the basis for the approaches we have today.
Despite this, subsequently developed methods such as the solid isotropic material with penalization (SIMP) method \citep{bendsoee1989opti,zhou1991the} and the level set method \citep{allaire2004stru} were favored over homogenization-based TO because of early challenges related to manufacturability, small length scale effects, and microstructural connectivity.
Today, with advances in additive manufacturing technology, it is now possible to 3D print various graded microstructures (see e.g. \citealp{schumacher2015micr}), which has prompted a resurgence of homogenization-based methods \citep{pantz2008apo, groen2017homo,groen2020deho} in multiscale TO.
Structures of unprecedented complexity have been both designed and physically realized at multiple scales \citep{sanders2021opti}.
Furthermore, AM methods have enabled the realization of multi-material structures \citep{gaynor2014mult}, complementing the development of novel techniques in multimaterial TO \citep{sanders2018mult,sanders2018poly} and its extension to hyperelastic materials under large deformations \citep{zhang2020adap}.

\nomenclature[A]{AM}{Additive manufacturing}
\nomenclature[A]{TO}{Topology optimization}
\nomenclature[A]{3D}{Three-dimensional}

Extending the existing single scale (e.g. \citealp{chi2021univ,senhora2022mach}) to two-scale concurrent design optimization methods for use with nonlinear physics is a particularly challenging task as the relevant physics can no longer be approximated via linear, first-order homogenization theory.
Moreover, analytical homogenization methods are not available for arbitrarily complex microstructures, limiting what can be readily used in a homogenization based TO framework.
The alternative of employing a typical numerical homogenization approach (e.g., $\text{FE}^2$) concurrently is generally computationally prohibitive \citep{xia2016rece}.
Data-driven homogenization techniques involving machine learning (ML) methods \citep{ghaboussi1991know,yvonnet2013comp}, specifically neural networks and Gaussian processes for nonlinear material behavior \citep{bessa2017fram}, provide an elegant alternative to the expensive $\text{FE}^2$ method, and offer a means to account for the dependence of microstructural descriptors in the homogenized response. Notably, neural networks with recurrent units are even capable of modeling history-dependent material responses \citep{mozaffar2019deep}. Furthermore, a natural advantage of employing NNs to capture the homogenized constitutive behavior is the ease of obtaining derivatives for both the tangent stiffness of the material as well as the sensitivities in TO, leveraging automatic differentiation.
Although NNs trained to capture microstructural descriptor-based homogenized responses were previously employed in the context of TO for elastic multiscale designs \citep{white2019mult, chandrasekhar2023grad}, the application of ML driven TO techniques to geometrically and materially nonlinear responses is still in its infancy.

\nomenclature{$\text{FE}^2$}{Finite element squared}
\nomenclature[A]{ML}{Machine learning}
\nomenclature[A]{NN}{Neural network}

The fundamental limitation that has hindered the use of typical \textit{black-box} neural network (NN) constitutive models is their failure to adhere to established physical principles, particularly in data-scarce situations, leading to concerns about accuracy and numerical stability.
In the case of hyperelasticity, this includes thermodynamic consistency, objectivity, existence of a \textit{natural state} \citep{coleman1959ont}, quasiconvexity, and volumetric growth conditions \citep{ball1976conv}, in addition to material dependent symmetry conditions.
{\it Polyconvexity} is a convenient and tractable condition to ensure quasiconvexity (and thereby ellipticity), which is needed to prove the existence of minimizers for the variational problem being solved in the forward analysis.
However, in the case of hyperelastic composites, like the ones we consider here, the homogenization step can lead to a loss of ellipticity (and therefore, quasiconvexity) despite each phase individually being polyconvex \citep{braides1994loss}.
This implies that even under a data-rich scenario, a black-box model trained using homogenized response data can exhibit microstructural instabilities leading to numerical difficulties in the forward analyses performed in the optimization loop.
Thus, it becomes all the more important to wrap the homogenized response with a polyconvex envelope (sometimes refered to as \textit{polyconvexification}, see \citealp{avazmohammadi2016macr}) to ensure that the response is stable during the optimization procedure.
The development of the input convex neural networks (ICNNs) \citep{amos2016inpu,chen2018opti} has enabled significant progress in realizing NN-based models that attempt to incorporate polyconvexity among other relevant physical principles \citep{asad2022ame, klein2022poly}.
In particular, \citet{linden2023neur} have provided a rigorous framework for using ICNN based models in the context of hyperelasticity formulated in terms of strain invariants.

\nomenclature[A]{ICNN}{Input convex neural network}

\begin{figure}[t!]
	\begin{minipage}{\linewidth}
	\centering        
	\def\svgwidth{0.95\linewidth}
	{\footnotesize
	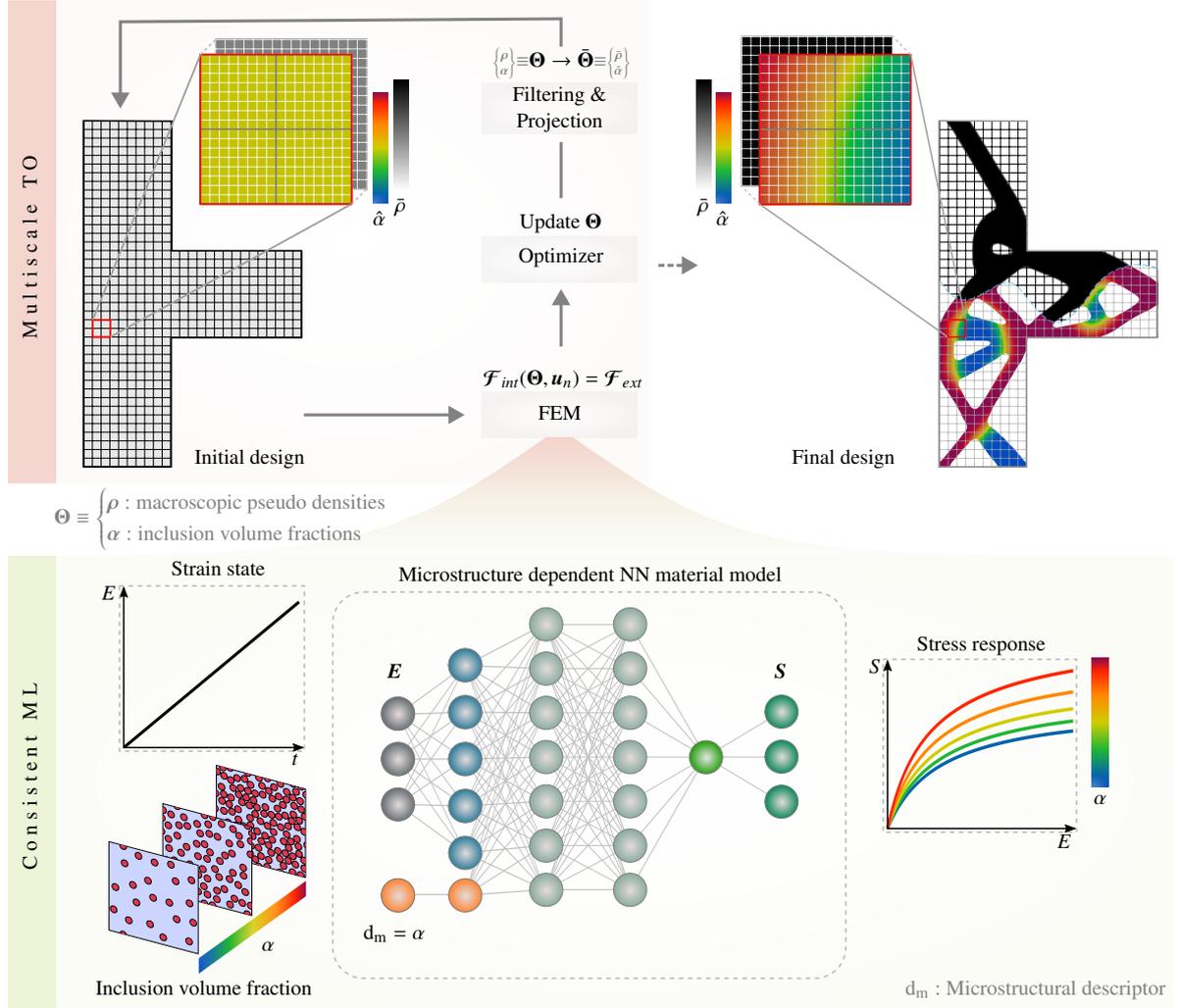
	}
	\caption{An overview of the proposed consistent machine learning-driven topology optimization framework for multiscale hyperelastic structures. The consistent ML block learns the mapping between microstructural descriptors $d_m$ (set as the inclusion volume fraction $\al$) and the homogenized constitutive response in an offline phase adhering to the hyperelastic principles. The multiscale TO block treats $\al$ as an additional design variable using its filtered counterpart to approximate the spatially varying material response. By employing a differentiable ML material model in the forward analyses, the optimizer can efficiently update the design variables using analytically computed sensitivities of the objective/constraint functions obtained through automatic differentiation. The magnified regions in the multiscale TO block illustrate the filtered and projected pseudo-density field $\bar{\rh}$ and the filtered microstructural inclusion volume fraction field $\hat{\al}$ in the initial design state (left) and the final optimized state (right).}
	\label{fig:TO_schematic}
	\end{minipage}
\end{figure}

In the present work, we address multiscale TO in geometrically and materially nonlinear settings by proposing a consistent machine learning-driven framework for hyperelastic composite structure design, as illustrated in Figure~\ref{fig:TO_schematic}.
The framework is composed of the consistent ML block that captures the homogenized constitutive response as a function of microstructural descriptors in
an offline phase adhering to hyperelasticity principles, and a multiscale TO block that treats the microstructural descriptors as additional design variables in the nonlinear TO problem.
Extending the work of \citet{linden2023neur} on ICNN based models, we propose a more general, principal stretch-based hyperelastic model that adheres to the necessary conditions for isotropic hyperelasticity \citep{ball1976conv} for the consistent ML block.
A single microstructural descriptor, the microstructural volume fraction, is considered to describe the microstructural variations; however, the framework can be extended to include multiple microstructural descriptors, such as the orientation of the microstructural phases to capture anisotropy in the homogenized response, among others.
The data-driven design and analysis framework developed by \citet{bessa2017fram,yi2023rves} is employed to streamline the data generation needed for training the consistent ML constitutive models.
We first compare the performance of consistent ML models against classical single-scale phenomenological models through a series of TO examples for maximizing external work, providing confidence in the approach prior to demonstrating the two-scale optimization results.
Subsequently, we demonstrate the two-scale concurrent design optimization examples that result in heterogeneous structures with spatially varying microstructural volume fractions and contrast it with homogeneous structures with fixed microstructural volume fractions.

\section{Constitutive Modeling for Hyperelasticity Materials}

The constitutive model employed in the multiscale TO framework presented here relies on a first order computational homogenization through finite element method (FEM) to compute the effective macroscopic response dataset corresponding to the microscopic representative volume element (RVE). 
As usual, we assume a separation of scales, with the characteristic dimension of the RVE much smaller than that of the macroscopic problem while also being larger than the characteristic dimension of the microstructural features of each material phase.
Throughout this manuscript we refer to the kinematic and kinetic quantities at the macroscopic scale to avoid confusion between the single and multiple scale settings.
We consider a total Lagrangian formulation and choose the Green-Lagrange strain tensor $\bE$ as the primary kinematic variable, computed using the deformation gradient $\bF $ and the right Cauchy-Green deformation tensor $\bC$ via
\begin{equation}
	\label{eq:def_grad_and_r_cauchy_green_tensor}
	\quad \bC \coloneqq \bF^{\mathsf{T}}\bF \quad;\quad \bE \coloneqq \frac{1}{2}(\bC - \Bone) \ ,
\end{equation}
and the second Piola-Kirchhoff stress $\bS$ as the corresponding energetically conjugate variable.
To avoid interpenetration of matter, the condition $J \coloneqq \det\bF~{\scriptstyle \overset{!}{>}}~0$ must hold.
In the multiscale setting, these quantities correspond to their homogenized counterparts in accordance with the Hill--Mandel lemma.
From the solution to the (periodic) boundary value problem at the microscale subject to the macroscopic deformation gradient $\bF$, the macroscopic Piola-Kirchhoff stress is obtained through a volume average of its microscopic counterpart, i.e.  $\bP \equiv \langle \bP_{\mu} \rangle$.
The second Piola-Kirchhoff stress is then obtained from the relation $\bS \coloneqq \bF^{-1} \bP$.
With the definition of consistent macroscopic quantities, the conditions set for the constitutive relation between the second Piola-Kirchhoff stress $\bS$ and the Green-Lagrange strain $\bE$ are identical for the single and multiple scale settings, and hence the following discussion is valid for both.
\nomenclature[A]{FEM}{Finite element method}
\nomenclature[A]{RVE}{Representative volume element}
\nomenclature[S]{$\bE$}{Green-Lagrange strain}
\nomenclature[S]{$\bF$}{Deformation gradient}
\nomenclature[S]{$J$}{Determinant of the deformation gradient}
\nomenclature[S]{$\bC$}{Right Cauchy-Green deformation}
\nomenclature[S]{$\bP$}{First Piola-Kirchhoff stress}
\nomenclature[S]{$\langle \bP_{\mu} \rangle$}{Volume averaged microscopic first Piola-Kirchhoff stress}
\nomenclature[S]{$\bS$}{Second Piola-Kirchhoff stress}

\subsection{Hyperelasticity conditions}
\label{subsec:hyperelasticity_conditions}

The constitutive definition of hyperelastic materials is provided through a strain energy density function $\ps$, whose derivative with respect to a strain measure directly relates to the definition of the \textit{thermodynamically consistent} conjugate stress measure \citep{coleman1959ont}.
If we define $\ps = \hat{\ps}(\bE)$, then the energetically conjugate stress measure -- the second Piola-Kirchhoff stress, is obtained as $\bS = \partial_{\bE}\hat{\ps}$.

\nomenclature[S]{$\ps, \hat{\ps}, \tilde{\ps}$}{Strain energy density}

The reference configuration of the body must correspond to a \textit{natural state} \citep{coleman1959ont}, i.e. a zero-valued minimum stored energy density state that is stress free:
\begin{equation}
	\label{eq:natural_state}
	\hat{\ps} (\bE) \overset{!}{\geq}0 \quad \mathrm{s.t.} \quad \hat{\ps}(\bE)|_{\bE= \Bzero} \overset{!}{=}0 \quad \text{and} \quad \bS(\bE)|_{\bE = \Bzero} \overset{!}{=}\Bzero
\end{equation}

Following the \textit{principle of objectivity}  \citep{coleman1959ont}, the material behavior should be independent of the observer, implying that the strain energy density remains invariant to orthogonal transformations $\bQ$ to the displacement gradient $\bF$.
Since the right Cauchy-Green deformation $\bC$ is objective by definition, i.e. $\bC = \bF^{{\mathsf{T}}}\bF = \bF^{{\mathsf{T}}}\bQ^{{\mathsf{T}}}\bQ\bF = (\bQ\bF)^{{\mathsf{T}}}(\bQ\bF)$, the Green-Lagrange strain $\bE$ is also objective, c.f. \eqref{eq:def_grad_and_r_cauchy_green_tensor}. Thus formulating the strain energy density as $\hat{\ps}(\bE)$ ensures objectivity: 
\begin{equation}
	\label{eq:objectivity}
	\tilde{\ps}(\bQ \bF) = \hat{\ps}(\bE)=  \tilde{\ps}(\bF)
\end{equation}

\nomenclature[S]{$\bQ$}{Orthogonal transformation tensor}

Furthermore, the material body cannot be compressed to a volume of zero, or expanded to infinite volume.
Thus, the strain energy density of hyperelastic materials has to satisfy the volumetric \textit{growth condition}:
\begin{equation}
	\label{eq:volumetric_growth_condition}
	\ps \to \infty \quad \text{as} \quad (J \to 0^{+} \quad \vee \quad J \to \infty)
 \end{equation}

In order to satisfy the requirements for the existence of total potential energy minimizers associated with the boundary value problem, the strain energy density $\ps$ should also be \textit{quasiconvex}.\footnote{
Solving a typical boundary value involving hyperelastic materials corresponds to finding the minimizer for the total elastic energy potential
$ \Pi^{\mathrm{total}} (\Bvph, \bF) = \int_{\CB_{0}} \tilde{\ps} (\bF) \dif V- \Pi^{\mathrm{external}} (\Bvph)$.
If one ignores the body force potential $\Pi^{\mathrm{external}} (\Bvph)$, the total energy potential $\Pi^{\mathrm{total}} (\Bvph, \bF)$ can be viewed as a \textit{functional} of the stored energy density \textit{function} $\tilde{\ps}$.
The existence of a minimizer for the total elastic energy potential relies on the constitutive restriction of the hyperelastic strain energy density to be quasiconvex \citep{ball1976conv}.
This follows from \cite{morrey1952quasi, morrey1966mult} that quasiconvexity of a \textit{function}, together with certain continuity and growth hypothesis is the necessary and sufficient condition for its \textit{functional} to be weakly lower semicontinuous, which in turn establishes the existence of minimizers for the \textit{functional}.
}
For a twice differentiable function, quasiconvexity implies the satisfaction of the \textit{Legendre-Hadamard} condition or \textit{ellipticity} and thereby rank-$1$ convexity.

However, imposing quasiconvexity on a strain energy density $\ps$ directly is challenging, as the non-pointwise nature of the condition makes it intractable to verify whether a given function is quasiconvex.
Furthermore, the stringent growth conditions assumed for quasiconvexity \citep{morrey1952quasi} preclude any singular behavior, such as the strain energy density $\ps \to \infty$ as $J \to 0^{+}$.

Polyconvexity \citep{ball1976conv} is a more tractable condition that arises from generalizing the notions of duality for convex functions to the vectorial context, bridging the above limitations. A strain energy density $\tilde{\ps}(\bF)$ is said to be polyconvex if:
\begin{equation}
	\label{eq:polyconvexity}
	\tilde{\ps}(\bF) = \CG(\bF, \mathrm{cof}\,\bF, \det \bF) \quad \text{where} \quad \CG(\cdot) \ \text{is a convex function}
\end{equation}

\nomenclature[S]{$\CG, \tilde{\CG}$}{Convex function}
\nomenclature[S]{$\mathrm{cof}\,\bF$}{Cofactor of the deformation gradient}

By restricting the strain energy density to be polyconvex, one may arrive at far more physically useful functions within the context of hyperelasticity i.e. not only does polyconvexity imply quasiconvexity, but it also establishes existence of minimizers that are valid under weaker growth conditions necessary to accommodate singular behavior such as $\ps \to \infty$ as $J \to 0^{+}$.

For isotropic hyperelasticity, Theorem 5.2 from \citet{ball1976conv} provides the sufficient conditions for polyconvexity of the strain energy density function.
Let $\la_{1}, \la_{2}, \la_{3}$ be the principal stretches associated with the deformation gradient $\bF$ and $\tilde{\CG}(\cdot)$ a convex function in its arguments. Then a strain energy density function $\ps$ of the form:

\begin{equation}
    \label{eq:polyconvexity_condition_isotropic}
    \tilde{\ps}(\bF) = \tilde{\CG}(\la_{1}, \la_{2}, \la_{3}, \la_{1}\la_{2}, \la_{2}\la_{3}, \la_{3}\la_{1}, J) \quad J = \la_{1}\la_{2}\la_{3} = \det \bF
\end{equation}
is isotropic and polyconvex as per \eqref{eq:polyconvexity} if $\tilde{\CG}$ is symmetric and non-decreasing in the principal stretches $(\la_{1}, \la_{2}, \la_{3})$ as well as their  pair-wise products $(\la_{1}\la_{2}, \la_{2}\la_{3}, \la_{3}\la_{1})$.

Note that such a strain energy density function $\ps$ has to additionally satisfy the natural state conditions \eqref{eq:natural_state} and the volumetric growth conditions \eqref{eq:volumetric_growth_condition} to be a valid and useful hyperelastic model respectively.
We remark that many of the existing isotropic hyperelastic models such as Neo-Hookean, Mooney-Rivlin, Arruda-Boyce, Blatz-Ko etc.
\citep[see][]{holzapfel2002nonl} satisfy these conditions through constructing the strain energy density in terms of the invariants of some objective strain measure such as the right Cauchy-Green deformation tensor $\bC$ or the left Cauchy-Green deformation tensor $\bB$.
However the sufficient conditions still correspond to \eqref{eq:polyconvexity_condition_isotropic}.

\nomenclature[S]{$\la_{1}, \la_{2}, \la_{3}$}{Principal stretches of the deformation gradient}
\nomenclature[S]{$\bB$}{Left Cauchy-Green deformation tensor}

\subsection{Consistent machine learning for material modeling}
\label{subsec:neural_network_for_material_modeling}

Feed-forward neural networks (FFN) have universal approximation properties \citep{hornik1989mult, cybenko1989appr} and can represent any continuous function to arbitrary accuracy provided enough data.
As opposed to the classical approach of formulating a strain energy density function $\ps$ based on physical intuition and then fitting the parameters to experimental data, neural networks, with their superior representation capabilities, can directly learn arbitrarily complex material models from experimental or numerically generated data for nonlinear elastic and plastic properties \citep{bessa2017fram}.
However, as discussed in Section \ref{subsec:hyperelasticity_conditions}, the constitutive model for hyperelasticity should correspond to a {\it thermodynamically consistent  strain energy density function} $\ps$ that satisfies conditions \eqref{eq:natural_state}, \eqref{eq:objectivity}, \eqref{eq:volumetric_growth_condition} and \eqref{eq:polyconvexity} along with the necessary material symmetry conditions (e.g. isotropy).
Standard FFNs do not have inductive biases that guarantee that the strain energy density function they learn satisfies the conditions set forth for hyperelasticity.
Thus, a feasible alternative consists of employing the so-called input convex neural networks (ICNNs) \citep{amos2016inpu} and constructing a polyconvex strain energy density function that satisfies the conditions for hyperelasticity \citep{klein2022poly, asad2022ame, linden2023neur,kalina2023fete, kalina2024neur}.

\nomenclature[A]{FFN}{Feed-forward neural network}
\begin{figure}[t!]
	\begin{minipage}{\linewidth}
	\centering        
	\def\svgwidth{0.8\linewidth}
	{\footnotesize
	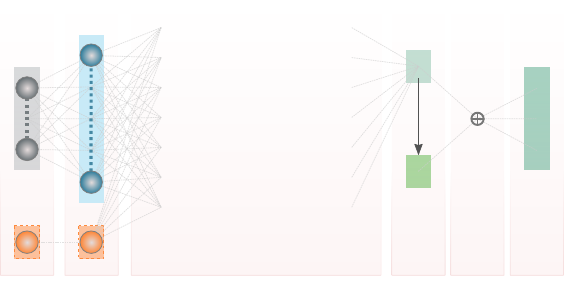
	}
	\vspace*{10pt}
	\caption{Schematic illustration of the consistent machine learning model for isotropic hyperelasticity.
	The input to the neural network is $\bE$, which passes through a fixed transformation layer $\CT$ to arrive at $\tilde{\BLA}$ that subsequently goes through input convex neural layers as in \citet{chen2018opti} to arrive at polyconvex output $\CN$. This output is regularized for the natural state condition \eqref{eq:natural_state} and passed through a gradient layer which obtains the second Piola-Kirchhoff stress $\bS$ as in \eqref{eq:overall_stress_strain_relationship}.
	Extension of the model to include microstructural descriptors $\mathrm{d_m}$ is achieved by appending the input space of the ICNN. The treatment of the microstructural descriptor $\mathrm{d_m}$ is such that the ICNN output (strain energy density) is convex with respect to the microstructural descriptor, which is a reasonable assumption for the volume fraction $\al$ of a stiff inclusion in a soft matrix.}
	\label{fig:01icnn_model}
	\end{minipage}
\end{figure}

In our work, we employ the alternative ICNN formulation from \citet{chen2018opti} to define the strain energy density function $\ps$ for hyperelastic materials (see \ref{section:appendix_A}).
This architecture enforces the sufficient condition for polyconvexity for isotropic materials \eqref{eq:polyconvexity_condition_isotropic} in terms of the principal stretches $\la_{1}, \la_{2}, \la_{3}$.
As presented in \ref{section:appendix_B}, such a polyconvex isotropic ICNN takes in input $\BLA  = \{\la_{1}, \la_{2}, \la_{3}, \la_{1}\la_{2}, \la_{2}\la_{3}, \la_{3}\la_{1}, J, -J \}$ and achieves isotropy through a symmetric weight sharing strategy.
The growth condition corresponding to $\ps \to \infty$ as $J \to \infty$ is handled naturally due to the convex nature of the ICNN.
However, the singularity associated with $J = 0$ requires special treatment due to limited data availability in the region $J \to 0^{+}$.
Yet we argue that the behavior of the strain energy density as $J \to 0^{+}$ has to be learned from the experimental data.
For this reason, rather than adding a fixed coercive function to the strain energy density definition, as suggested in \citet{linden2023neur}  and other conventional hyperelastic models, we add a convex non-decreasing term to the input to our isotropic ICNN.
We therefore choose the term $f(J) = -\ln J$ that is a convex, decreasing function in $J$ ($J~{\scriptstyle \overset{!}{>}}~0$). 
By constructing the strain energy density as an ICNN with inputs $ \tilde{\BLA } = \{\la_{1}, \la_{2}, \la_{3}, \la_{1}\la_{2}, \la_{2}\la_{3}, \la_{3}\la_{1}, J, -J, -\ln J \}$, the singularity point becomes embedded in the model and the volumetric growth condition \eqref{eq:volumetric_growth_condition} becomes satisfied and learnable from the experimental data.

\nomenclature[S]{$\BLA, \tilde{\BLA}$}{ Principal stretch-based transformed input space for ICNN}

Next we address the natural state condition \eqref{eq:natural_state}. This is achieved by adding a correction function $\CN_{0}$ to the ICNN output that ensures that the strain energy density function $\ps$ and the second Piola-Kirchhoff stress $\bS$ vanishes at the natural state $\bE = \Bzero$. The overall strain energy density function $\ps$ can be formulated as:
\begin{equation}
	\label{eq:overall_strain_energy_density}
	\begin{aligned}
	\hat{\ps}(\bE) &= \tilde{\CG}(\tilde{\BLA}(\bE)) = \CN(\tilde{\BLA}(\bE)) + \underbrace{\CN_{0}^{\ \text{stress}}(\bE) + \CN_{0}^{\ \text{energy}}}_{\CN_{0}(\bE)} \quad \text{where} \quad
	\CN_{0}^{\ \text{energy}} &=- \CN(\tilde{\BLA}(\bE)) \Big|_{\bE = \Bzero}
	\end{aligned}
\end{equation}

\nomenclature[S]{$\CN, \CN_{0}$}{Output of the ICNN and correction term}

A stress correction term $\CN_{0}^{\ \text{stress}}(\bE)$ formulated in the form:
\begin{equation}
	\label{eq:stress_correction_term_non_polyconvex}
	\CN_{0}^{\ \text{stress}}(\bE) = - \partial_{\bE}\CN(\tilde{\BLA}(\bE))\Big|_{\bE = \Bzero}\colon\bE
\end{equation}
satisfies the natural state condition similarly to the correction terms proposed in \cite{asad2022ame}, but violates polyconvexity in addition to the material symmetry condition as remarked in \cite{linden2023neur}. To address this, we propose a stress correction term following the projection approach inspired by the weighted sum of derivatives method proposed in \cite{linden2023neur} to obtain a polyconvex stress correction term:
\begin{equation}
	\label{eq:stress_correction_term}
	\begin{aligned}
		\CN_{0}^{\ \text{stress}}(\tilde{\BLA}(\bE)) & = \CN_{0}^{\ \text{stress}}(J) = - c_{0} (J-1) \quad \text{with} \quad 
	c_{0}&=  -\sum_{i=1}^{9}\xi_{i}\partial_{\tilde{\LA}_{i}}\CN(\tilde{\BLA}(\bE))\Big|_{\bE = \Bzero}
	\end{aligned}
\end{equation}
where 
$\Bxi = \{{1}/{3}, {1}/{3}, {1}/{3}, {2}/{3}, {2}/{3}, {2}/{3}, 1, -1, 1\}$ 
correspond to weights for the derivatives with respect to the inputs $\tilde{\BLA}=\{\la_{1}, \la_{2}, \la_{3},\allowbreak \la_{1}\la_{2}, \la_{2}\la_{3}, \la_{3}\la_{1}, J, -J, -\ln J \}$. A detailed derivation of the polyconvex stress correction term is provided \ref{section:appendix_D}.
The second Piola-Kirchhoff stress $\bS$ is then obtained via differentiation of the strain energy density function $\hat{\ps}$ with respect to the Green Lagrange strain measure:
\begin{equation}
	\label{eq:overall_stress_strain_relationship}
	\begin{aligned}[b]
	\bS = \partial_{\bE}\hat{\ps}(\bE) &= \partial_{\bE}\CN(\tilde{\BLA}(\bE)) + \partial_{\bE}\CN_{0}(\tilde{\BLA}(\bE))\\
	&= \partial_{\bE}\CN(\tilde{\BLA}(\bE)) - \sum_{i=1}^{9} \xi \partial_{\tilde{\LA}_{i}}\CN(\tilde{\BLA}(\bE))\Big|_{\bE= \Bzero} J \bC^{-1}
	\end{aligned}
\end{equation}

The overall architecture of the resulting ICNN-based consistent ML model for hyperelasticity is illustrated in Figure \ref{fig:01icnn_model}.
The input to the neural network is the Green-Lagrange strain tensor $\bE$ that is transformed to the input $\tilde{\BLA}$ for the ICNN through a fixed transformation layer $\CT$.
{\it The ICNN then computes the strain energy density function $\hat{\ps}(\bE)$ that satisfies objectivity, material symmetry, polyconvexity as well as the natural state condition \eqref{eq:natural_state} and the thermodynamically consistent second Piola-Kirchhoff stress $\bS$ is obtained as in \eqref{eq:overall_stress_strain_relationship}.}
We remark that it is possible to use the ICNN architecture to define an isotropic hyperelastic model that satisfies the conditions for hyperelasticity (c.f. \citet{linden2023neur}).
However, such a model can have limitations due to the hard enforcement of convexity on the pre-set invariants. A detailed discussion of this point is presented in \ref{section:appendix_B}.

\nomenclature[S]{$\CT$}{Transformation layer in the ICNN-based model}

Extension of the consistent ML hyperelastic model to include microstructural descriptors is achieved by extending the input space of the ICNN, so that the constitutive mapping now corresponds to $\bS (\bE, \mathrm{d_m}) = \partial_{\bE}\hat{\ps}(\bE, \mathrm{d_m})$. 
In the most generalized setting, the microstructural descriptor $\mathrm{d_m}$ could be included in the ML model through the partial input convex neural network (PICNN) \citep{amos2016inpu} wherein no convexity related constraints are imposed on output strain energy density function $\ps$ with respect to the microstructural descriptor.
In the present study involving two phase hyperelastic composite with a soft matrix and stiff inclusion, we consider the microstructural descriptor $\mathrm{d_m}$ to be the volume fraction $\al$ of the stiff inclusion phase, which could be considered to hold a convex relationship with the strain energy density function $\ps$.
As a result the microstructural descriptor $\mathrm{d_m} = \al$ is introduced in the model thorough a simple extension of the input space of the ICNN as illustrated in Figure \ref{fig:01icnn_model}.

\nomenclature[S]{$\mathrm{d_m}$}{Microstructural descriptor}
\nomenclature[S]{$\al$}{Inclusion volume fraction}
\nomenclature[A]{PICNN}{Partial input convex neural network}

\section{Topology Optimization Formulation}
\label{sec:top_opt_formulation}

We employ a standard density-based topology optimization scheme in which each finite element, $e$, is assigned a scalar pseudo density parameter, $\rh_{e} \in [0,1]$, reflecting the presence $\rh_{e} = 1$ or absence $\rh_{e} =0$ of the material within the element.
The design problem is regularized by means of a standard linear filtering technique \citep{bourdin2001filt} with a user-specified length scale, $r_\rh$, in order to mitigate the checkerboard effect and mesh dependency issues, resulting in the filtered design variables $\hat{\rh}_{e}$. This may be mathematically expressed as:

\nomenclature[S]{$e$}{Element index}
\nomenclature[S]{$\rh, \rh_{e}$}{Pseudo-density design variable and its element-wise value}
\nomenclature[S]{$\hat{\rh}, \hat{\rh}_e$}{Filtered pseudo-density design variable and its element-wise value}
\nomenclature[S]{$r_\rh$}{Filter radius for the pseudo-density design variable}

\begin{equation}
    \label{eq:filter_operation}
    \hat{\rh}_{e} = \ph^\rh_{ei} \rh_{i}, \quad \ph^\rh_{ei} = \frac{h_{ei} V_i ~\text{(no sum)}}{\sum_{j=1}^{N_\mathrm{elem}} h_{ej} V_j}, \quad h_{ei} = \max{\left(0, ~r_\rh - \big|\big| \bx_e - \bx_i  \big|\big| \right)}
\end{equation}
where $\bx_e$ represents the centroid of element $e$ and $\left \| \cdot \right \|$ denotes the Euclidean norm.

\nomenclature[S]{$\Bph^\rh, \ph^\rh_{ei}$}{Filter matrix for the pseudo-density design variable}
\nomenclature[S]{$\bh_{ei}$}{Linear filter function in filtering operation}
\nomenclature[S]{$\bx_e, \bx_i$}{Centroid of element $e$ and $i$ respectively}

Subsequently, the filtered design is projected by means of a smooth Heaviside function \citep{wang2010onp}, in order to significantly reduce intermediate densities that may develop as a result of the filtering technique.
\begin{equation}
    \label{eq:projected_density}
    \bar{\rh}_{e} = \frac{\tanh(\be_{\rh}\et_{\rh}) +  \tanh(\be_{\rh}(\hat{\rh}_{e} - \et_{\rh}))}
    {\tanh(\be_{\rh}\et_{\rh}) + \tanh(\be_{\rh}(1 - \et_{\rh}))}
\end{equation}

\nomenclature[S]{$\bar{\rh}, \bar{\rh}_e$}{Projected pseudo-density design variable and its element-wise value}
\nomenclature[S]{$\be_{\rh}$}{Projection strength parameter for the pseudo-density design variable}
\nomenclature[S]{$\et_{\rh}$}{Projection threshold parameter for the pseudo-density design variable}

The filtered and projected pseudo-density field $\bar{\rh}_{e}$ is then used in a SIMP interpolation scheme \citep{zhou1991the,bendsoee1989opti} during the forward analysis.
To ensure numerical stability in the large deformation finite element simulations, we adopt the energy interpolation scheme proposed in \cite{wang2014inte} which takes the form,
\begin{equation}
    \label{eq:simp_iterpolation}
    \ps_{e} = \ch_{e}(\bar{\rh}_{e})\ps_{e,\ga_{e}} \quad \mathrm{with} \quad  \ch_{e}(\bar{\rh}_{e}) = \ep + (1 - \ep)  \bar{\rh}_e^{p}
\end{equation}
where $p$ is SIMP penalization parameter to penalize intermediate densities.
The function $\ps_{e,\ga_{e}}$ is the interpolated strain energy density and is defined as
\begin{equation}
    \label{eq:strain_energy_interpolation}
    \begin{aligned}[b]
     \ps_{e,\ga_{e}} &= \ps (\bE (\bF_{e\ga_{e}})) - \ps_{L} (\ga_{e}\nabla_{\sym} \bu_{e}) + \ps_{L} (\nabla_{\sym} \bu_{e}) \\
     & = \ps (\bE (\bF_{e\ga_{e}})) - (1 - \ga_{e}^{2})\ps_{L} (\nabla_{\sym} \bu_{e})
    \end{aligned}
\end{equation}
which consists of the solid phase strain energy density $\ps(\cdot)$ and small deformation linear elastic stored energy density $\ps_{L}(\cdot)$.
The effective deformation gradient and symmetric gradient operator used above are defined as
\begin{equation}
    \label{eq:strain_energy_interpolation2}
    \bF_{e\ga_{e}} = \nabla \ga_{e}\Bvph_{e} = \Bone + \ga_{e}\nabla\bu_{e} \quad \mathrm{and} \quad \nabla_{\sym} (\cdot) = 1/2 (\nabla (\cdot) + \nabla (\cdot)^\mathsf{T})  
\end{equation}
where the interpolation factor $\ga_{e}$ takes the value $1$ for solid elements and $0$ for void elements.
This results in a geometrically nonlinear treatment for solid elements and small strain kinematics for the void elements.
A smooth Heaviside projection function is used to compute the strain energy density interpolation parameter $\ga_{e}$, providing a differentiable transition between the two kinematic formulations:
\begin{equation}
    \label{eq:strain_energy_interpolation_factor}
    \ga_{e} = \frac{\tanh(\be_{\ps}\et_{\ps}) +  \tanh(\be_{\ps}(\ch_{e} - \et_{\ps}))}
    {\tanh(\be_{\ps}\et_{\ps}) + \tanh(\be_{\ps}(1 - \et_{\ps}))}
\end{equation}

\nomenclature[A]{SIMP}{Solid Isotropic Material with Penalization}
\nomenclature[S]{$p$}{SIMP penalization parameter}
\nomenclature[S]{$\ep$}{SIMP ersatz parameter}
\nomenclature[S]{$\ch_{e}$}{SIMP interpolation factor for element $e$}
\nomenclature[S]{$\ps_{e}$}{Strain energy density in element $e$}
\nomenclature[S]{$\ps_{e,\ga_{e}}$}{Interpolated strain energy density in element $e$}
\nomenclature[S]{$\ps_{L}$}{Small deformation linear elastic strain energy density}
\nomenclature[S]{$\ga_{e}$}{Strain energy density interpolation factor for element $e$}
\nomenclature[S]{$\bF_{e\ga_{e}}$}{Effective deformation gradient in element $e$}
\nomenclature[S]{$\nabla_{\sym}$}{Symmetric gradient operator}
\nomenclature[S]{$\Bvph, \Bvph_{e}$}{Deformation field and its element-wise value}
\nomenclature[S]{$\nabla \bu_{e}$}{Displacement gradient in element $e$}
\nomenclature[S]{$\be_{\ps}$}{Projection strength parameter for the strain energy density interpolation factor}
\nomenclature[S]{$\et_{\ps}$}{Projection threshold parameter for the strain energy density interpolation factor}

The set of design variables controlled by the optimizer, denoted by $\BTH \coloneqq \{ \TH_{e}| \forall e = 1 \ldots N_{elem} \}$ represents the macroscopic density design variable $\rh_{e}$, i.e. $\TH_{e} \equiv \{\rh_{e}\}$ in the single scale setting or additionally include the microstructural design variable, i.e. $\TH_{e} \equiv \{\rh_{e}, \al_{e}\}$ at each element $e$ in the multiscale setting.
In a similar manner to the macroscopic density field, a design length scale on the inclusion volume fraction field is prescribed via a linear filtering operation to produce a filtered inclusion volume fraction, $\hat{\al}_{e}$, in each element to obtain a smooth variation of the field across the topology. Consistent with Equation \eqref{eq:filter_operation}, we represent this operation, $\hat{\al}_{i} = \ph^\al_{ij} \al_{j}$, via the separate filter matrix $\Bph^\al$ constructed with user-specified radius, $r_\al$.
To simplify notation, the set of all \textit{physical design variables}, i.e the filtered (and projected) design variables, is labeled as $\bar{\BTH}$.
Depending upon the problem setting, $\bar{\BTH}$ may be composed of the set of all filtered and projected macroscopic density design variables $\bar{\Brh}$ (single scale) or additionally include the microstructural design variables, $\bar{\BTH} \equiv \{\bar{\Brh}, ~\hat{\Bal} \}$ (multiscale).

\nomenclature[S]{$\BTH, \TH_{e}$}{Set of design variables and element-wise set of design variables}
\nomenclature[S]{$\hat{\al}, \hat{\al}_e$}{Filtered inclusion volume fraction design variable and its element-wise value}
\nomenclature[S]{$\Bph^\al, \ph^\al_{ij}$}{Filter matrix for the inclusion volume fraction design variable}

A total Lagrangian formulation is used to compute the internal force in each element ($e$) of the form:
\begin{equation}
    \label{eq:element_internal_force}
    \boldsymbol{\mathcal{F}}_{int_{e}}(\bar{\TH}_{e},\bu_{e, n}) = \frac{\partial}{\partial \bu_{e, n}} \int_{\Omega_{0_e}} \ps_{e} ~ dV
\end{equation}
at pseudo-time increment, $n$, with integration over the element in the reference configuration, $\Omega_{0_e} \subset \Omega_0$.
Subsequently, the element contributions are then assembled into their global counter part, $\boldsymbol{\mathcal{F}}_{int}$.
Global equilibrium is then achieved via the solution of the residual equations at the $n^{\text{th}}$ pseudo-time increment.
As $\bar{\BTH}$ may be represented as a function of $\BTH$, the global residual equations may be written as:

\begin{equation}
    \label{eq:global_residual}
    \begin{aligned}[b]
    \BR(\BTH,\bu_n, \zeta_n) &=\boldsymbol{\mathcal{F}}_{int}(\BTH,\bu_n) - \zeta_n \bf_0 = \Bzero \\
    \bf_0 \cdot \bu_n &=  c_n &
    \end{aligned}
\end{equation}

The modified generalized displacement control method \citep{leon2014ont} is an effort to mitigate instabilities associated with load control in large strain topology optimization problems, while also providing more flexibility over the load distribution than standard displacement control.
Note that $\bf_0$ is a constant vector containing the load distribution, whereas the load factor, $\zeta_n$, controls the magnitude of the applied load in each pseudo-time increment. The load factor at each step is determined via the additional equation in \eqref{eq:global_residual}, representing a user-specified weighted average of the displacements with the constant nonzero external applied force distribution vector, $\bf_0$. This weighted average displacement is constrained to be equal to the parameter $c_n$ at the $n^{\Rth}$ pseudo-time step. 
The forward analyses begin in an undeformed state corresponding to zero nodal displacements and zero load (i.e., $\bu_{0}=\Bzero$, $\zeta_{0}=0$).
With $\bu_{n-1}, \zeta_{n-1}$ known, the nodal displacements and load factor at the subsequent pseudo-time increment, $n$, are obtained via the procedure 
provided in Algorithm \ref{alg:forward_problem} with evenly spaced generalized applied displacements, $c_n$, between $c_0=0$ and the user-specified value, $c_N$, at the final step, corresponding to $n=N$.

\begin{algorithm}[t]
    \caption{Incremental update of the nodal displacements \& load factor}
    \label{alg:forward_problem}

    \begin{algorithmic}[1]
        \Require Nodal displacements $\bu_{n-1}$ and load factor $\zeta_{n-1}$ at pseudo-time $t_{n-1}$; Generalized applied displacement $c_n$

        \Ensure Nodal displacements $\bu_{n}$ and load factor $\zeta_{n}$ at pseudo-time $t_{n}$

        \State Initialize Newton iteration counter $k = 1$
        \State Initialize $\bu_{n}^{(k-1)} = \bu_{n-1}$, $\zeta_{n}^{(k-1)} = \zeta_{n-1}$
        \State Initialize $\Delta \bu_n^{(k-1)} = \Bzero$, $\Delta \zeta_n^{(k-1)} = 0$, $\Delta c_n = c_n - c_{n-1}$

        \While{ $\big|\big|\BR\left(\BTH,\bu_n^{(k-1)}, \zeta_n^{(k-1)}\right)\big|\big|_2 > \mathrm{tol_{NR}} \cdot \big|\big|\BR\left(\BTH,\bu_n^{(0)}, \zeta_n^{(0)}\right)\big|\big|_2$}
            \State Compute $\BK_T= \frac{\partial \BR}{\partial \bu}(\BTH,\bu_n^{(k-1)}, \zeta_n^{(k-1)})$
            \State Compute $\delta \bu_R = - \BK_T^{-1} \BR(\BTH,\bu_n^{(k-1)}, \zeta_n^{(k-1)})$
            \State Compute $\delta \bu_{f_0} = \BK_T^{-1} \bf_0$
            \State Compute $\delta \zeta = \frac{\Delta c_n - \bf_0 \cdot \Delta \bu_n^{(k-1)} + \bf_0 \cdot \delta \bu_R}{\bf_0 \cdot \delta \bu_{f_0}}$ 
            \State Compute $\delta \bu = \delta \bu_R + \delta \zeta~ \delta \bu_{f_0}$
            \State Update $\Delta \bu_n^{(k)} = \Delta \bu_n^{(k-1)} + \delta \bu$ 
            \State Update $\bu_{n}^{(k)} = \bu_{n-1} + \Delta \bu_n^{(k)}$
        \EndWhile

        \State \Return $\bu_{n} = \bu_{n}^{(k)}$, $\zeta_{n} = \zeta_{n}^{(k)}$

    \end{algorithmic}
    
\end{algorithm}

The optimization problem is formulated as a work maximization problem subject to constraints imposed on the material volume fractions. This is defined mathematically via the following:
\begin{equation} \boxed{
    \label{eq:optimization_statement}
    \begin{aligned}
    \max\limits_{\BTH}~~& W_{ext} = \sum_{n=1}^N \frac{1}{2} \left(\zeta_n + \zeta_{n-1}\right) \left(c_n - c_{n-1} \right) \vspace{6pt} \\
    \text{s.t.}~~& g^{(m)}(\BTH) \le g^{(m)}_{\max}, \quad m=1,\ldots,M \vspace{6pt}\\
    ~~& \BTH_{\min} \preceq \BTH \preceq \BTH_{\max} \vspace{6pt}\\
    \text{with:}~~&\BR(\BTH,\bu_n, \zeta_n) = \boldsymbol{\mathcal{F}}_{int}(\BTH,\bu_n) - \zeta_n \bf_0 = \Bzero \quad \forall n = 1,\ldots,N  \vspace{6pt}\\
    ~~& \bf_0 \cdot \bu_n = c_n
    \end{aligned}}
\end{equation}
where $\BTH_{\min}$ and $\BTH_{\max}$ represent the side constraint bounds on the design variables. For benchmark problems, there is a single constraint (i.e. $M=1$) corresponding to the material volume fraction:

\begin{equation}
    \label{eq:single_volume_fraction_constraint}
    g^{(1)}(\BTH) = g(\BTH) = \frac{1}{\sum_{e=1}^{N_{elem}} V_e} \sum_{e=1}^{N_{elem}} \bar{\rh}_e V_e
\end{equation}
where $V_e$ is the volume of element $e$. The macroscopic design density variables are bounded by $\rh_e \in [0, 1]$ at the element level. 

Subsequently, for the two-scale optimization examples, two separate constraints are included on each material phase (i.e. $M=2$):
\begin{equation}
    \label{eq:constraint_equations}
    \begin{aligned}[b]
    g^{(1)}(\BTH) = g^{\mathrm{inc}}(\BTH) &= \frac{1}{\sum_{e=1}^{N_{elem}} V_e} \sum_{e=1}^{N_{elem}} \hat{\al}_e \bar{\rh}_e V_e \\
    g^{(2)}(\BTH) = g^{\mathrm{mat}}(\BTH) &= \frac{1}{\sum_{e=1}^{N_{elem}} V_e} \sum_{e=1}^{N_{elem}} (1-\hat{\al}_e) \bar{\rh}_e V_e
    \end{aligned}
\end{equation}
where the first constraint equation corresponds to the inclusion volume fraction and the second restricts the quantity of matrix material.
Here, in addition to the macroscopic density design variables side constraint $\rh_e \in [0, 1]$, the inclusion volume fraction design variables are bounded by $\al_e \in [\al_{\ell}, \al_{u}]$ at the element level, where $\al_{\ell}$ and $\al_{u}$ are the lower and upper bounds on the inclusion volume fraction design variables.

\section{Sensitivity analysis}
\label{appendix:sensitivity_analysis}

Here we provide the derivation for the sensitivity of the external work objective function with respect to the design variables, $\BTH$. First, the sequence of algebraic operations reducing to our expression for the external work integrated using the trapezoidal rule are shown below.
\begin{equation}
    \begin{aligned}[b]
        W_{ext} &= \sum_{n=1}^N \frac{1}{2} \left(\bf_{ext}^{(n)} + \bf_{ext}^{(n-1)}\right) \cdot \left(\bu_n - \bu_{n-1} \right) 
        = \sum_{n=1}^N \frac{1}{2} \left(\zeta_n \bf_0 + \zeta_{n-1}\bf_0\right) \cdot \left(\bu_n - \bu_{n-1} \right)  \\
        &= \sum_{n=1}^N \frac{1}{2} \left(\zeta_n + \zeta_{n-1}\right)  \left(\bf_0  \cdot\bu_n - \bf_0  \cdot\bu_{n-1} \right)  
        = \sum_{n=1}^N \frac{1}{2} \left(\zeta_n + \zeta_{n-1}\right) \left(c_n - c_{n-1} \right) 
    \end{aligned}
    \label{eq:external_work}
\end{equation}
To obtain the sensitivity of this function, we directly differentiate the residual equation at time step, $n$.
\begin{equation}
    \begin{aligned}[b]
    \frac{d}{d\BTH}\BR(\BTH,\bu_n, \zeta_n) &= \frac{d}{d\BTH}\boldsymbol{\mathcal{F}}_{int}(\BTH,\bu_n) - \bf_0 \frac{d\zeta_n}{d\BTH} = \Bzero & \\
    &= \frac{\partial \boldsymbol{\mathcal{F}}_{int}}{\partial\BTH} + \frac{\partial \boldsymbol{\mathcal{F}}_{int}}{\partial \bu_n}\frac{d \bu_n}{d \BTH} - \bf_0 \frac{d\zeta_n}{d\BTH} = \Bzero & \\
    &= \frac{\partial \boldsymbol{\mathcal{F}}_{int}}{\partial\BTH} + \BK_{T}\frac{d \bu_n}{d \BTH} - \bf_0 \frac{d\zeta_n}{d\BTH} = \Bzero & \\
    \implies  \frac{d \bu_n}{d \BTH} &= \BK_{T}^{-1} \left( \bf_0 \frac{d\zeta_n}{d\BTH} -  \frac{\partial \boldsymbol{\mathcal{F}}_{int}}{\partial\BTH} \right) &
    \end{aligned}
\end{equation}
Recalling the constraint, $\bf_0 \cdot \bu_n = c_n$, we also have the following useful relationship:
\begin{equation}
\begin{aligned}[b]
    \frac{d}{d\BTH}\left( \bf_0 \cdot \bu_n \right) &= \bf_0 \cdot  \frac{d \bu_n}{d \BTH} = 0& \\
    \implies \bf_0 \cdot \frac{d \bu_n}{d \BTH} &= \bf_0 \cdot \BK_{T}^{-1} \left( \bf_0 \frac{d\zeta_n}{d\BTH} -  \frac{\partial \boldsymbol{\mathcal{F}}_{int}}{\partial\BTH} \right) = 0 &
\end{aligned}
\end{equation}
Rearranging this final equation we have the sensitivity of the load factor at time step, $n$,
\begin{equation}
\label{eq:lf_sensitivity}
    \frac{d\zeta_n}{d\BTH} = \frac{\bf_0 \cdot \BK_{T}^{-1} \cdot  \frac{\partial \boldsymbol{\mathcal{F}}_{int}}{\partial\BTH} }{\bf_0 \cdot \BK_{T}^{-1} \cdot \bf_0} = \frac{\delta \bu_{f_0} \cdot  \frac{\partial \boldsymbol{\mathcal{F}}_{int}}{\partial\BTH} }{\bf_0 \cdot \delta \bu_{f_0}}
\end{equation}
which does not require the solution of any additional system of equations (assuming that the $\delta \bu_{f_0}$ vectors were saved during the forward analysis). All required partial derivatives can be easily obtained using automatic differentiation. Equation \eqref{eq:lf_sensitivity} can then be used to obtain the final sensitivity of the external work as provided in Equation \eqref{eq:external_work} since each $c_n$ is a constant.
\begin{equation}
    \frac{dW_{ext}}{d\BTH} = \sum_{n=1}^N \frac{1}{2} \left(\frac{d\zeta_n}{d\BTH}  + \frac{d\zeta_{n-1}}{d\BTH} \right) \left(c_n - c_{n-1} \right)
\end{equation}
Finally, the chain rule is required to obtain the derivative with respect to the design variables, $\{\Brh, \Bal\}$, through the projection function and the corresponding filters in the standard manner.

\section{Numerical Experiments and Results}
\label{sec:numerical_experiments_and_results}

In the present study, we consider two-phase composite material microstructures with stiff inclusions embedded in a soft matrix. An Arruda-Boyce (AB) hyperelastic model \citep{arruda1993ath} is used to represent the soft rubbery matrix, while the stiff inclusion phase is represented by a Neo-Hookean model \citep[see][]{bergstrom2015mech}.
The Arruda-Boyce model is parameterized by the initial bulk modulus $\ka_{0}$, the initial shear modulus $\mu_{0}$, and  $\la_{m}$, which is associated with the chain locking stretch. The strain energy density function is given by:

\begin{equation}
    \label{eq:arruda_boyce}
    \begin{gathered}\boxed{
        \ps_{AB}  = \mu \sum_{i=1}^{5} a_{i} \ \be^{\ i-1} (\bar{I}^{\ i}_{C1} - 3 ^{i}) + \frac{\ka_{0}}{2} \left ( \frac{J^{2}-1}{2} - \ln J \right )  \quad 
        } \hspace*{0.3cm}
        \text{where} \\
        \mu = \mu_{0} \left ( 1 + \frac{3}{5\la_{m}^{2}} + \frac{99}{175 \la_{m}^{4}} + \frac{513}{875 \la_{m}^{6}} + \frac{42039}{67375 \la_{m}^{8}} \right )^{-1} \quad \text{and} \\
        \be = \frac{1}{\la_{m}^{2}}; \quad a_{1} = \frac{1}{2}; \quad a_{2} = \frac{1}{20}; \quad a_{3} = \frac{11}{1050}; \quad a_{4} = \frac{19}{7000}; \quad a_{5} = \frac{519}{673750}
    \end{gathered}
\end{equation}

Here, $\bar{I}_{C1} =  J^{-2/3} I_{C1}$ is the modified first invariant of the right Cauchy-Green deformation tensor $\bC$ and $J$ is the determinant of the deformation gradient tensor $\bF$. 
The Neo-Hookean model is parameterized by the initial shear modulus $\mu_{0}$ and the initial bulk modulus $\ka_{0}$:

\begin{equation}
    \ps_{NH} = \frac{\mu_{0}}{2} \left ( \bar{I}_{C1} - 3 \right ) + \frac{\ka_{0}}{2} \left ( J - 1 \right )^{2}
\end{equation}

Note that the initial bulk modulus is related to the initial Poisson's ratio and shear modulus $\mu_{0}$ as $\ka_{0} = \frac{2\mu_{0}(1+\nu_{0})}{3(1-2\nu_{0})}$.
The material properties considered for the matrix phase are $\mu^\mathrm{mat}_{0} = 3.098 \times 10^{-1} \U\mathrm{MPa}$, $\la^\mathrm{mat}_{m} = 5.083$ to closely resemble natural rubber \citep{treloar1944stre, bergstrom2015mech} with $\nu^\mathrm{mat}_{0} = 0.45$. The material properties for the inclusion phase chosen such that its Young's modulus is 100 times stiffer than that of the matrix, with a Poisson's ratio $\nu^\mathrm{inc}_{0} = 0.3$, resulting in $\mu^\mathrm{inc}_{0} = 3.455 \times 10^{1} \U\mathrm{MPa}$.
Prior to demonstrating the multiscale optimization results, we first report the performance of the consistent ML models against classical single scale phenomenological models. 
In order to make meaningful comparison, we choose the soft rubbery material for the single scale assessment, using the soft matrix material parameters presented above.
For the sake of brevity, we present the results under 2D plane strain considerations in this section, while a 3D proof of concept example is presented in \ref{appsubsec:fixed_fixed_beam_3D_example_benchmark}

\subsection{Design of experiments for generating dataset for model calibration}
\label{subsec:doe_data_generation_for_ml}

Two datasets are generated for calibrating two distinct models: the dataset $\CD^{s}$ to train the model $\CM^{s}$ for single scale TO evaluations, and the dataset $\CD^{m}$ to train the the microstructure-dependent model $\CM^{m}$ for two-scale TO evaluations.
The dataset $\CD^{s}$ consists of input-output tuples of the Green-Lagrange strain $\bE$ and the corresponding second Piola-Kirchhoff stress $\bS$, with the $i^\Rth$ data sample corresponding to $(\bE, \bS)^{i}$.
The dataset $\CD^{m}$ for the model $\CM^{m}$ for the two-scale TO has an additional microstructural descriptor $\mathrm{d_{m}}$ which is the volume fraction of the inclusion phase, $\al$, in the RVE, making the $i^\Rth$ data sample $((\al, \bE), \bS)^{i}$.
The strain states for data generation are obtained through Sobol sampling of the principal stretch space spanned by ${\la_{d}} \in [0.75, 1.75]$ and random sampling of orthonormal principal directions $\BN_{d}$ for $d \in (1, \cdots, N_D)$ where $N_D$ is the spatial dimensionality of the problem.
The right stretch tensor is then obtained as $\bU = \sum \la_{d} \BN_{d} \otimes \BN_{d}$, from which the Green-Lagrange strain tensor is computed as $\bE = \frac{1}{2}(\bU^T\bU - \bI)$.
This ensures that the strain sampling is space filling in the principal stretch space and non-informative sample points are limited.
Under plane strain conditions, $N_{D} =2$ and $\la_{3}$ is set to unity.
The target second Piola-Kirchhoff stresses for the datasets $\CD^{s}$  comprised of $2^{12}=4096$ sample strain states are obtained by a direct evaluation of the second Piola-Kirchhoff stress tensor $\bS = \partial_{\bE}{\ps_{AB}}$ at the sampled Green-Lagrange strain states.
In order to generate the second Piola-Kirchhoff stresses in the dataset $\CD^{m}$ for the microstructure-dependent model, the data-driven design and analysis framework developed by \citet{bessa2017fram,yi2023rves,van2024f3dasm} is employed to perform RVE simulations with periodic boundary conditions in the commercial FE software ABAQUS \citep{abaqus2021}, including the subsequent volumetric averaging of the stress measure.
The microstructural geometry is fully described by the volume fraction $\al$ of the inclusion phase $(0.1 \leq \al \leq 0.5)$, which is generated by randomly distributing circular inclusions of radius $r_{\mu}$ in a square RVE domain of side length $L_{\mathrm{RVE}}$.
Following the approach discussed by \citet{bessa2017fram}, the macroscopic deformation gradient tensor $\bF$ necessary for applying the RVE boundary conditions is obtained in its rotational invariant form:

\begin{equation}
    \bF = \bU = \bC^{1/2} = (2\bE + \Bone)^{1/2}
\end{equation}

For the considered materials and inclusion radius ($r_{\mu} = 0.15 \Umm$), a $4\Umm$ RVE side length ($L_{\mathrm{RVE}} = 4.0 \Umm$) was found to be sufficiently large to achieve a statistically representative response for volume fraction, $\al$.
The macroscopic second Piola Kirchhoff stresses are obtained corresponding to $2^{10}$ macroscopic Green-Lagrange strain states for representative realizations of the microstructure for each volume fraction $\al \in \{0.1, 0.2, 0.3, 0.4, 0.5 \}$, totalling to $5 \times 2^{10}$ data samples. Upon eliminating data samples corresponding to the non-converged RVE simulations, the final dataset $\CD^{m}$ consists of $4787$ data samples.

\begin{table}[t]
    \caption{Model and training hyperparameters}
    \label{tab:hyperparameters}
    \begin{minipage}{0.45\linewidth}
    \raggedleft
    \footnotesize
    \begin{tblr}{l r r}
        \hline
        Hyperparameter & $\CM^{s}$ & $\CM^{m}$ \\
        \hline[0.15em]
        No. hidden layers & $2$ & $2$ \\
        No. neurons per layer & $8$ & $16$ \\
        Batch size & 128 & 128 \\
        Initial learning rate & $1 \times 10^{-3}$ & $5 \times 10^{-3}$ \\
        \hline[0.1em]
    \end{tblr}
    \end{minipage}
    \hspace{0.05\linewidth}
    \begin{minipage}{0.5\linewidth}
    \raggedright
    \footnotesize
    \begin{tblr}{l r r}
        \hline
        Hyperparameter & $\CM^{s}$ & $\CM^{m}$ \\
        \hline[0.15em]
        Exponential decay rate & $0.5$ & $0.5$ \\
        Decay transition epoch interval & $3000$ & $3000$ \\
        Max epochs & 15000 & 15000 \\
        Early stopping patience & 5000 & 5000 \\
        \hline[0.1em]
    \end{tblr}
    \end{minipage}
\end{table}

\begin{table}[t]
    \caption{Performance metrics of the trained models}
    \label{tab:model_performance}
    \begin{minipage}{0.5\linewidth}
    \centering
    \footnotesize
    \begin{tblr}{c c r r r}
        \hline
        Model & Metric & $\CD^{s}_\mathrm{train}$  & $\CD^{s}_\mathrm{val}$  & $\CD^{s}_\mathrm{test}$  \\
        \hline[0.15em]
        \SetCell[r=3]{c} $\CM^{s}$ & $1 - \mathrm{R}^{2}$ & $3.41 \times 10^{-8}$ & $3.66 \times 10^{-8}$ & $3.22 \times 10^{-8}$ \\
        & RMSE & $1.14 \times 10^{-4}$  & $1.14 \times 10^{-4}$ & $1.14 \times 10^{-4}$ \\
        & MAE & $8.47 \times 10^{-5}$ & $8.51 \times 10^{-5}$ & $8.52 \times 10^{-5}$ \\
        \hline[0.1em]
    \end{tblr}
    \end{minipage}
    \begin{minipage}{0.5\linewidth}
    \centering
    \footnotesize
    \begin{tblr}{c c r r r}
        \hline
        Model & Metric & $\CD^{m}_\mathrm{train}$  & $\CD^{m}_\mathrm{val}$  & $\CD^{m}_\mathrm{test}$  \\
        \hline[0.15em]
        \SetCell[r=3]{c} $\CM^{m}$ & $1 - \mathrm{R}^{2}$ & $1.13 \times 10^{-4}$ & $1.20 \times 10^{-4}$ & $1.53 \times 10^{-4}$ \\
        & RMSE & $1.04 \times 10^{-2}$ & $1.11 \times 10^{-2}$ & $1.19 \times 10^{-2}$ \\
        & MAE & $4.23 \times 10^{-3}$ & $4.32 \times 10^{-3}$ & $4.41 \times 10^{-3}$ \\
        \hline[0.1em]
    \end{tblr}
    \end{minipage}
\end{table}

\nomenclature[A]{$\mathrm{R}^{2}$}{Coefficient of determination}
\nomenclature[A]{RMSE}{Root mean squared error}
\nomenclature[A]{MAE}{Mean absolute error}

\subsection{Model training}

The strategy for training of the ML models is standardized and described in this section. The dataset $\CD$ associated with a model $\CM$ is split into training, validation and test datasets in the ratio of $60:20:20$.
The mean squared error loss function between the predicted second Piola-Kirchhoff stress $\bS$ and the target second Piola-Kirchhoff stress $\bS^{*}$ , given by:
\begin{equation}
    \label{eq:mse_loss}
    \CL_\square = \frac{1}{|\CD_\square|} \sum_{|\CD_\square|} \left \| \bS - \bS^{*} \right \|^{2},
\end{equation}

\noindent is used as the objective function in the training process, where $\square$ is a placeholder for the dataset split under consideration, and $|\CD_\square|$ is the number of samples in the dataset split and $\left \| \cdot \right \|$ denotes the Eucledian norm.
The model training is performed using Adam optimizer \citep{kingma2014adam} with decoupled weight decay regularization \citep{loshchilov2019deco}, together with a exponential decaying learning rate scheduler by minimizing the loss function $\CL_\mathrm{train}$ on the training dataset $\CD_\mathrm{train}$.
The validation loss $\CL_\mathrm{val}$ on the validation dataset $\CD_\mathrm{val}$ is monitored during the training process to prevent overfitting through early stopping, as well as to tune the hyperparameters of the model.
The hyperparameters of the models and the training process, tuned through a simple grid search method, are summarized in Table \ref{tab:hyperparameters}.
The predictive capability of the tuned model is then evaluated on the test dataset $\CD_\mathrm{test}$.
Table \ref{tab:model_performance} summarizes the performance of the trained models on the training, validation and test datasets in terms of root mean squared error (RMSE), mean absolute error (MAE), and coefficient of determination ($\mathrm{R}^{2}$) metrics.\footnote{The machine learning workflow is implemented in JAX \citep{bradbury2018jax}, realized through the Equinox library \citep{kidger2021equi} and leverages the gradient processing and optimization library Optax \citep{deepmind2020the}. 
All the ML trainings are performed on a standard desktop with NVIDIA RTX 3050 Ti GPU.
}

\subsection{Evaluation of model effectiveness}
\label{subsec:effectiveness_of_ml_model}

\begin{figure}[b!]
    \begin{minipage}{\linewidth}
    \centering
    \def\svgwidth{\linewidth}
    \setlength{\figW}{8cm}
    \setlength{\figH}{6cm}
    {\scriptsize
    \input{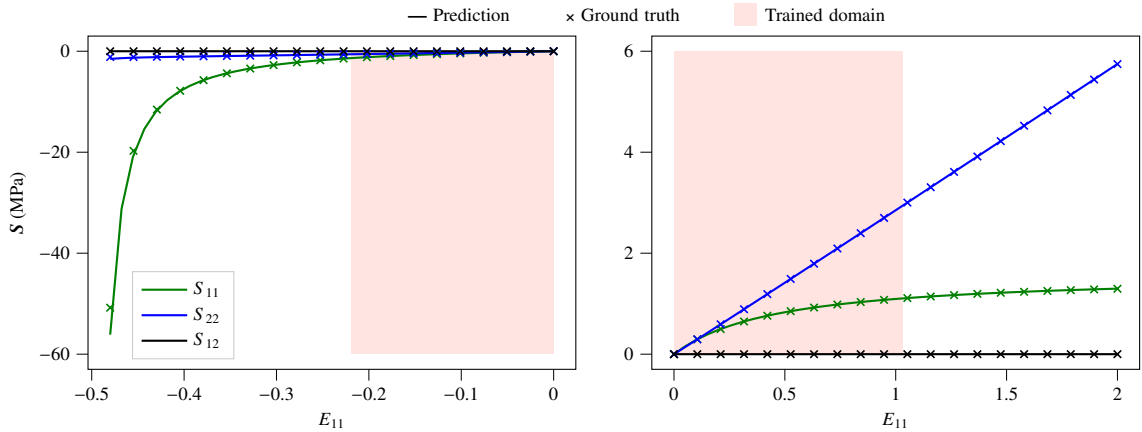}
    }
    \end{minipage}
    \caption{Stress response of the single scale consistent ML model compared to the ground truth phenomenological model for a uniaxial strain loading case. The plot is split into two subplots to better visualize the compressive and tensile loading ranges.}
    \label{fig:ml_model_benchmark_stress_strain}
\end{figure}

\begin{figure}[t!]
    \begin{minipage}{\linewidth}
    \centering
    \def\svgwidth{\linewidth}
    {
    \scriptsize
    \relscale{2.5}
    \scalebox{0.4}{\input{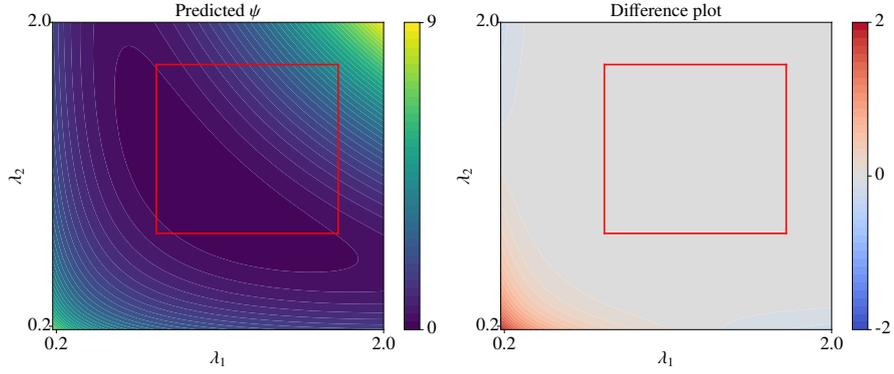}}
    }
    \caption{Strain energy density contour plot of the single scale consistent ML model and the corresponding difference plot to the ground truth phenomenological model. The region within the red box corresponds to the trained domain.}
    \label{fig:ml_model_benchmark_strain_energy_density}
    \end{minipage}
\end{figure}

\begin{figure}[t!]
    \begin{minipage}{\linewidth}
    \centering
    \def\svgwidth{\linewidth}
    \setlength{\figW}{8cm}
    \setlength{\figH}{6cm}
    {\scriptsize
    \input{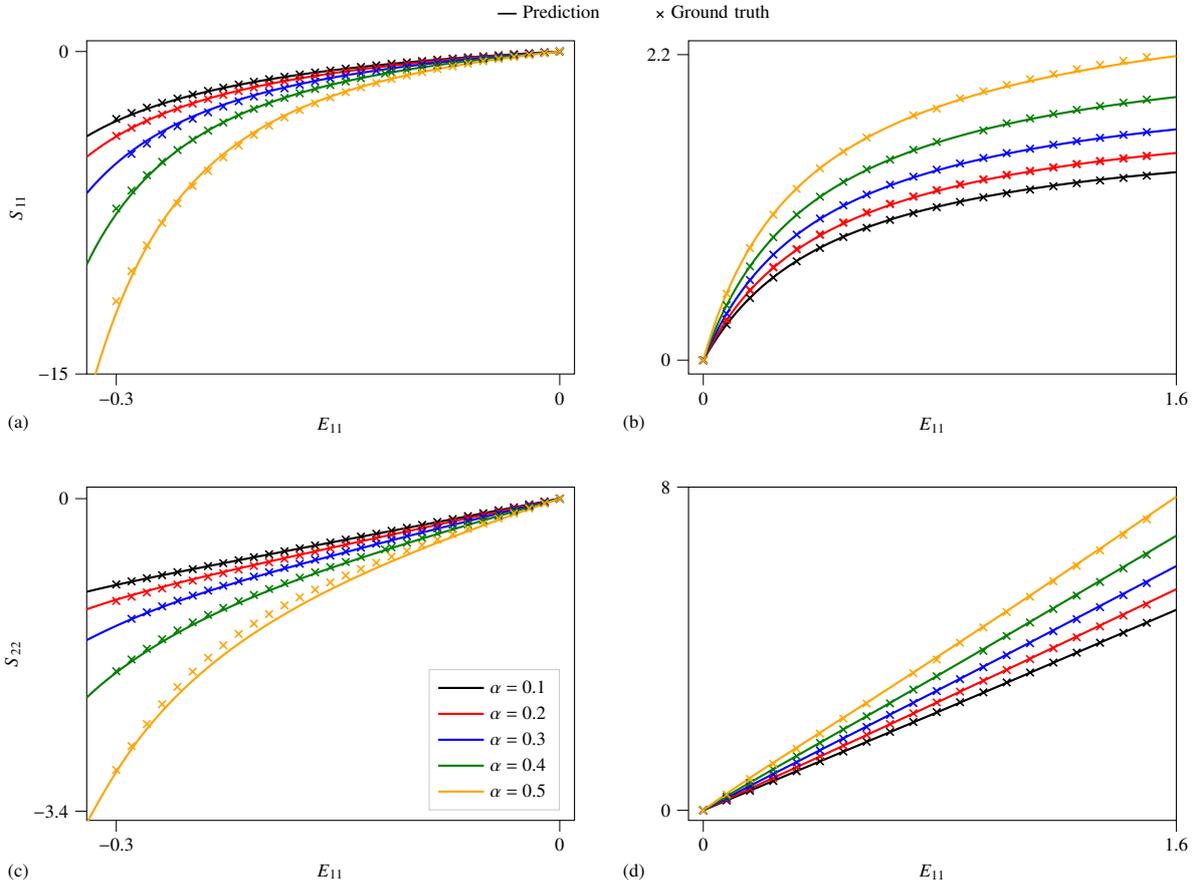}
    
    }
    \caption{Stress response of the microstructure-dependent consistent ML model compared to the ground truth obtained through RVE simulations for a uniaxial strain loading scenario for (left) compressive loading, and (right)  tension loading range. The top and bottom plots show the stress components $S_{11}$ and $S_{22}$ separately.}
    \label{fig:micro_ml_model_benchmark_stress_strain_compression}
    \end{minipage}
\end{figure} 

We begin by evaluating the effectiveness of the consistent ML model in representing classical single scale phenomenological models.
As mentioned earlier, an Arruda Boyce hyperelastic model \eqref{eq:arruda_boyce} corresponding to the soft rubbery matrix phase is considered the ground truth used for evaluating the approximation capability of the consistent ML material models.
Figure \ref{fig:ml_model_benchmark_stress_strain} illustrates the constitutive response of the the trained model $\CM^{s}$ compared to the ground truth phenomenological model for uniaxial loading (i.e., variation of $E_{11}$, with all other strain components set to zero).
The figure highlights how closely the the trained model $\CM^{s}$ approximates the ground truth phenomenological model, which is in agreement with the performance metrics provided in Table \ref{tab:model_performance}.
Importantly, the model shows impressive predictive performance beyond the training range.
In order to visualize the predictive performance of the model in a more comprehensive manner, the strain energy density is compared to the ground truth phenomenological model in Figure \ref{fig:ml_model_benchmark_strain_energy_density}.
From the plot, visualized in the principal stretch space, we see that the trained model $\CM^{s}$ exhibits high interpolation accuracy, with acceptable extrapolation performance attributed largely to the relevant restrictions in the consistent ML model architecture.
The extrapolation quality is observed to deteriorate as the principal strain state $\{\la_{1}, \la_{2}\} \to \Bzero$ which corresponds to the asymptote as $J \to 0$, yet the model is able to predict the strain energy density response in the vicinity of the singularity with reasonable accuracy.

Next we assess the ability of the consistent ML model to capture the microstructure-dependent response variations encapsulated within the two-phase composite material dataset $\CD^{m}$. 
We are now specifically interested in evaluating how well the model $\CM^{m}$ learns the mapping between the inclusion volume fraction $\al$ of the microstructure and the homogenized material response.
The stress response corresponding to uniaxial loading for inclusion volume fractions $\al \in \{0.1, 0.2, 0.3, 0.4, 0.5\}$ in both compression and tension is shown in Figure \ref{fig:micro_ml_model_benchmark_stress_strain_compression}.
Note that this data is not present in the training dataset.
Naturally, similar to the the single scale case, the microstructure-dependent model $\CM^{m}$ preserves the hyperelasticity conditions. More importantly, the model captures the dependence of $\al$ on the stress response very well, suggesting that the assumption of convexity with respect to $\al$ in the model architecture was appropriate.

\subsection{Single-scale topology optimization}
\label{sec:singleTO_main_text}

We now verify the performance of the consistent ML model in the context of topology optimization via direct comparison against results obtained using the ``classical'' or phenomenological model. The intent of this exercise is to provide confidence in the trained ML model for use in topology optimization by demonstrating nearly identical designs in a setting with directly comparable conventional results.
Note that the update of the design variables is performed using the method of moving asymptotes (MMA) \citep{svanberg1987the}.

In the main text we present the results for two examples -- the T-bracket and a cantilever beam.
In \ref{appsubsec:portal_frame_example_benchmark} we also provide the results for a portal frame example, and a three-dimensional fixed-fixed beam example is presented in \ref{appsubsec:fixed_fixed_beam_3D_example_benchmark}.  
The results in these appendices are consistent with the examples discussed in the main text.

We define each design domain with reference to a base dimension, $L = 100 \Umm$, and the parameters provided in Table \ref{tab:TO_hyperparameters} are used unless otherwise specified.
A continuation scheme is employed on both the SIMP penalty exponent and the projection strength parameter.
The SIMP penalty exponent is initialized to $p = 1$ and increased in increments of $\DE p = 1$ each continuation step up to a maximum value of $p = 4$.
In subsequent continuation steps the projection strength parameter is doubled up to a maximum value computed according to \citet{silva2019stre}, from an initial value of $\be_{p} = 1$.
The next continuation step begins when either the percent difference in the objective function over the previous 5 iterations has fallen below a tolerance of $0.1$ or a maximum number of $50$ iterations is reached. A minimum of $20$ optimization iterations is also enforced in each step.

\begin{table}[b!]
    \caption{Topology optimization problem hyperparameters}
    \label{tab:TO_hyperparameters}
    \begin{minipage}{0.46\linewidth}
    \centering
    \footnotesize
    \begin{tblr}{l c r }
        \hline
        Hyperparameter & Symbol & Value \\
        \hline[0.15em]
        Filter radius$^\ast$ & $r_{\rh}$ & 4.0 \\
        SIMP penalty parameter (initial value) & $p$ & 1.0 \\
        SIMP ersatz parameter & $\ep$ & 1e-5 \\
        Projection strength  (initial value) & $\be_{\rh}$ & 1.0 \\
        Projection threshold & $\et_{\rh}$ & 0.5 \\
        Energy interpolation transition strength & $\ga_{\rh}$ & 32.0 \\
        \hline[0.1em]
    \end{tblr}

    $\ast$ Filter radius for \textit{cantilever beam} problem is 6.0
    \end{minipage}
    \begin{minipage}{0.54\linewidth}
        \centering
        \footnotesize
        \begin{tblr}{l c r }
            \hline
            Hyperparameter & Symbol & Value \\
            \hline[0.15em]
            Energy interpolation transition threshold & $\et_{\ga}$ & 0.01 \\
            Linearized elastic energy Young's modulus & $E_{\mathrm{lin}}$ & 0.898385 \\
            Linearized elastic energy Poisson's ratio & $\nu_{\mathrm{lin}}$ & 0.45 \\
            Forward problem number of time steps & $N$ & 8 \\
            Relative tolerance for Newton convergence & $\mathrm{tol_{NR}}$ & $10^{-6}$ \\
            MMA move limit & ~ & 0.15 \\
            \hline[0.1em]
        \end{tblr}
        
        \vspace{\baselineskip}
    \end{minipage}
\end{table}

\begin{figure}[t!]
    \begin{minipage}{0.3\linewidth}
    \centering        
    \def\svgwidth{\linewidth}
    {\footnotesize
    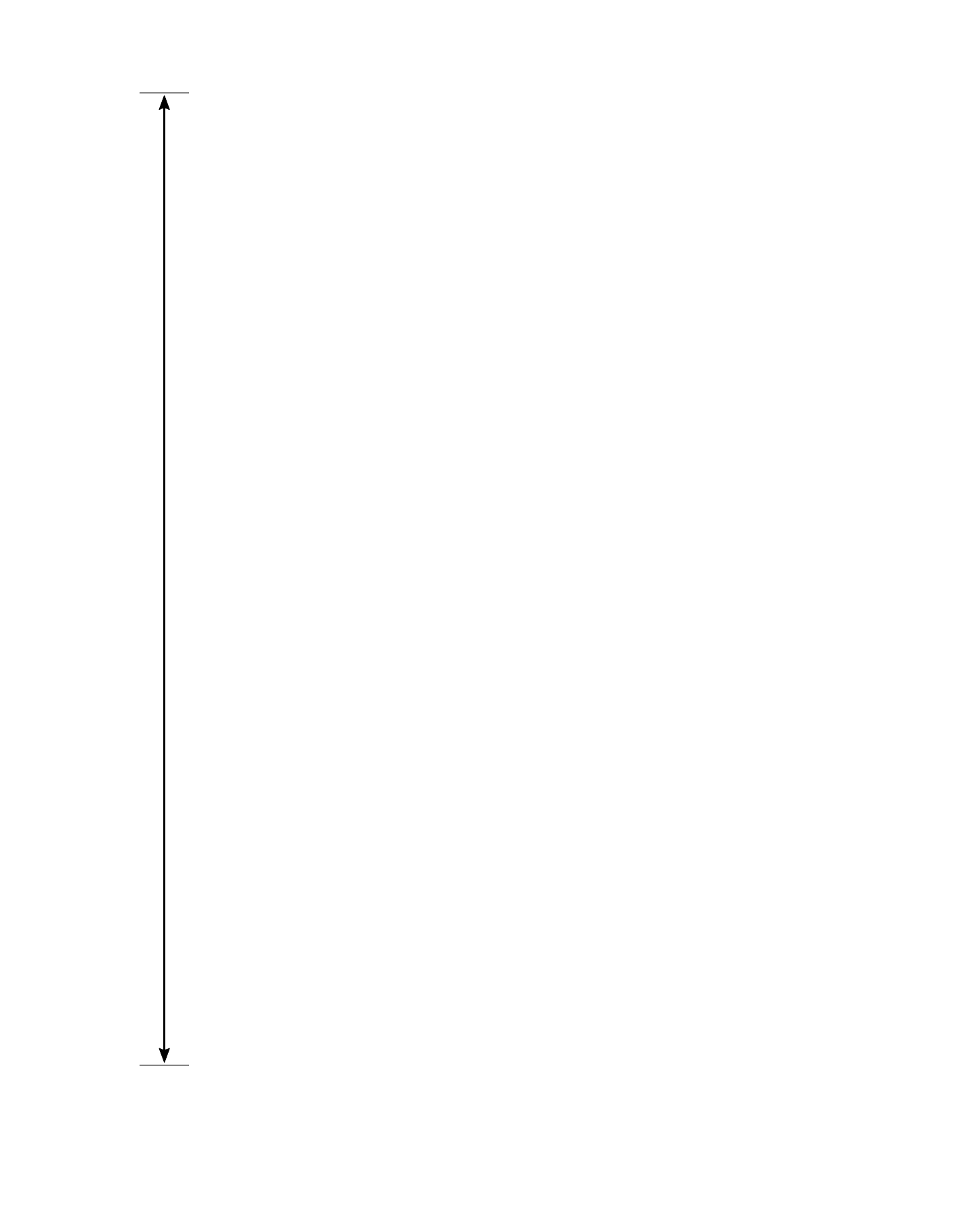
    \subcaption*{(a)}
    }
    \end{minipage}
    \begin{minipage}{0.3\linewidth}
    \centering
    \def\svgwidth{\linewidth}
    {\scriptsize
\begingroup%
  \makeatletter%
  \providecommand\color[2][]{%
    \errmessage{(Inkscape) Color is used for the text in Inkscape, but the package 'color.sty' is not loaded}%
    \renewcommand\color[2][]{}%
  }%
  \providecommand\transparent[1]{%
    \errmessage{(Inkscape) Transparency is used (non-zero) for the text in Inkscape, but the package 'transparent.sty' is not loaded}%
    \renewcommand\transparent[1]{}%
  }%
  \providecommand\rotatebox[2]{#2}%
  \newcommand*\fsize{\dimexpr\f@size pt\relax}%
  \newcommand*\lineheight[1]{\fontsize{\fsize}{#1\fsize}\selectfont}%
  \ifx\svgwidth\undefined%
    \setlength{\unitlength}{709.80999756bp}%
    \ifx\svgscale\undefined%
      \relax%
    \else%
      \setlength{\unitlength}{\unitlength * \real{\svgscale}}%
    \fi%
  \else%
    \setlength{\unitlength}{\svgwidth}%
  \fi%
  \global\let\svgwidth\undefined%
  \global\let\svgscale\undefined%
  \makeatother%
  \begin{picture}(1,1.22932618)%
    \lineheight{1}%
    \setlength\tabcolsep{0pt}%
    \put(0,0){\includegraphics[width=\unitlength,page=1]{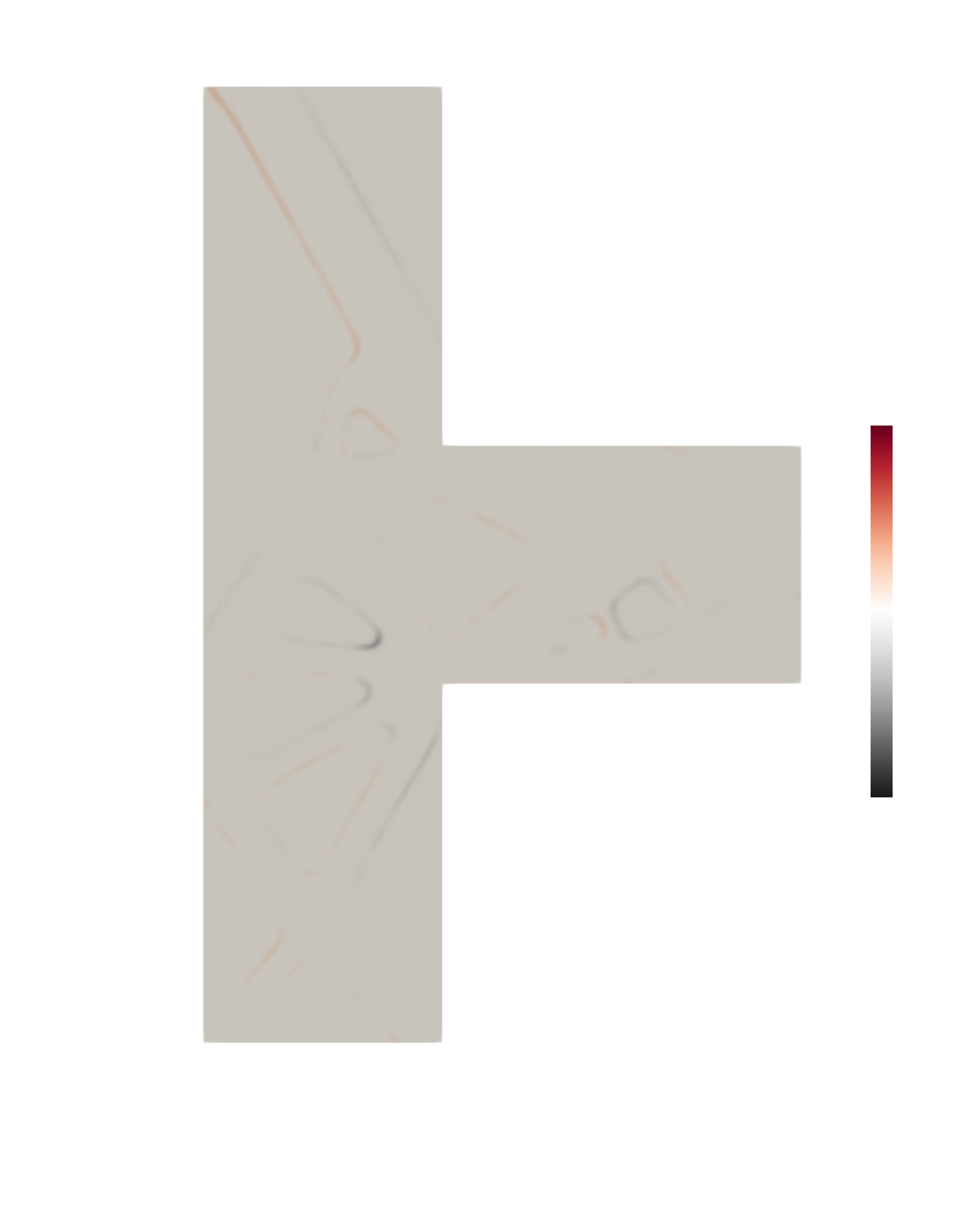}}%
    \put(0.93198063,0.42011559){\color[rgb]{0,0,0}\makebox(0,0)[lt]{\lineheight{1.25}\smash{\begin{tabular}[t]{l}-1.\end{tabular}}}}%
    \put(0.93198063,0.79909025){\color[rgb]{0,0,0}\makebox(0,0)[lt]{\lineheight{1.25}\smash{\begin{tabular}[t]{l}1.\end{tabular}}}}%
    \put(0.91,0.3){\color[rgb]{0,0,0}\makebox(0,0)[t]{\lineheight{1.25}\smash{\begin{tabular}[t]{c}$\de_{\rh}$\end{tabular}}}}%
    \put(0,0){\includegraphics[width=\unitlength,page=2]{t_bracket_final_design_ml_analytical_mse.pdf}}%
    \put(0.93198062,0.60909433){\color[rgb]{0,0,0}\makebox(0,0)[lt]{\lineheight{1.25}\smash{\begin{tabular}[t]{l}0.\end{tabular}}}}%
    \put(0,0){\includegraphics[width=\unitlength,page=3]{t_bracket_final_design_ml_analytical_mse.pdf}}%
  \end{picture}%
\endgroup%

    \subcaption*{(b)}
    }
    \end{minipage}
    \begin{minipage}{0.3\linewidth}
    \centering        
    \def\svgwidth{\linewidth}
    {\scriptsize
\begingroup%
  \makeatletter%
  \providecommand\color[2][]{%
    \errmessage{(Inkscape) Color is used for the text in Inkscape, but the package 'color.sty' is not loaded}%
    \renewcommand\color[2][]{}%
  }%
  \providecommand\transparent[1]{%
    \errmessage{(Inkscape) Transparency is used (non-zero) for the text in Inkscape, but the package 'transparent.sty' is not loaded}%
    \renewcommand\transparent[1]{}%
  }%
  \providecommand\rotatebox[2]{#2}%
  \newcommand*\fsize{\dimexpr\f@size pt\relax}%
  \newcommand*\lineheight[1]{\fontsize{\fsize}{#1\fsize}\selectfont}%
  \ifx\svgwidth\undefined%
    \setlength{\unitlength}{709.80993992bp}%
    \ifx\svgscale\undefined%
      \relax%
    \else%
      \setlength{\unitlength}{\unitlength * \real{\svgscale}}%
    \fi%
  \else%
    \setlength{\unitlength}{\svgwidth}%
  \fi%
  \global\let\svgwidth\undefined%
  \global\let\svgscale\undefined%
  \makeatother%
  \begin{picture}(1,1.22932576)%
    \lineheight{1}%
    \setlength\tabcolsep{0pt}%
    \put(0,0){\includegraphics[width=\unitlength,page=1]{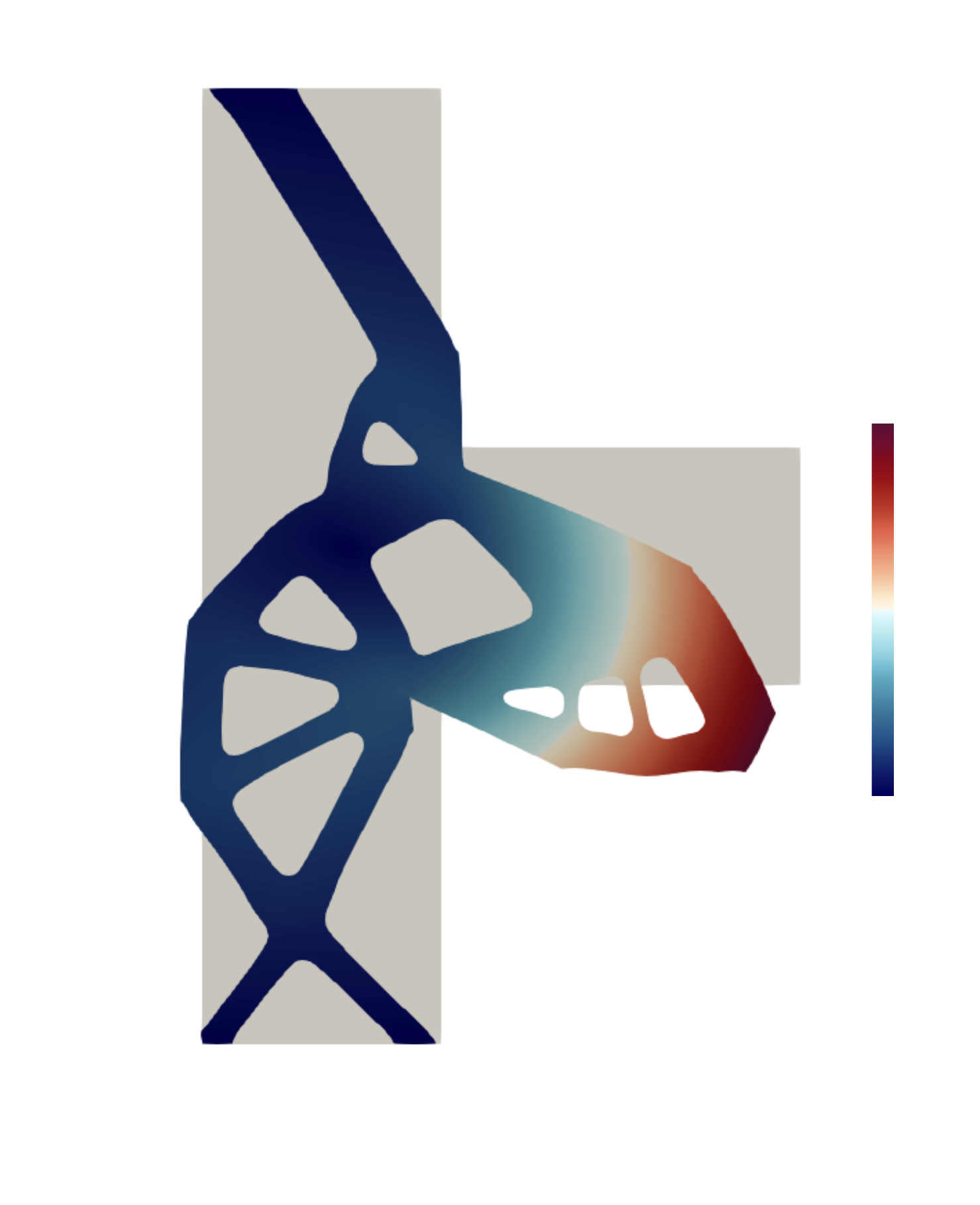}}%
    \put(0.93313848,0.42186413){\color[rgb]{0,0,0}\makebox(0,0)[lt]{\lineheight{1.25}\smash{\begin{tabular}[t]{l}0.\end{tabular}}}}%
    \put(0.93313848,0.80083882){\color[rgb]{0,0,0}\makebox(0,0)[lt]{\lineheight{1.25}\smash{\begin{tabular}[t]{l}31.\end{tabular}}}}%
    \put(0.91,0.3){\color[rgb]{0,0,0}\makebox(0,0)[t]{\lineheight{1.25}\smash{\begin{tabular}[t]{c}$||\bu||$\end{tabular}}}}%
    \put(0,0){\includegraphics[width=\unitlength,page=2]{t_bracket_final_design_ML.pdf}}%
  \end{picture}%
\endgroup%

    \subcaption*{(c)}
    }
    \end{minipage}
    \caption{(a) T-bracket boundary value problem. (b) Difference plot between the design obtained using the ML model and the design obtained using the ground truth phenomenological model, where $\de_{\rh} = 1$ indicates material addition, and $\de_{\rh} = -1$ indicates material removal. (c) Deformed configuration of the design obtained using the ML model.}
    \label{fig:t_bracket_bvp}
\end{figure}

\begin{figure}[t!]
    \begin{minipage}{0.3\linewidth}
        \centering
        \vspace{-172pt}
        \def\svgwidth{1.15\linewidth}
        {\footnotesize
        \hspace{-40pt}
        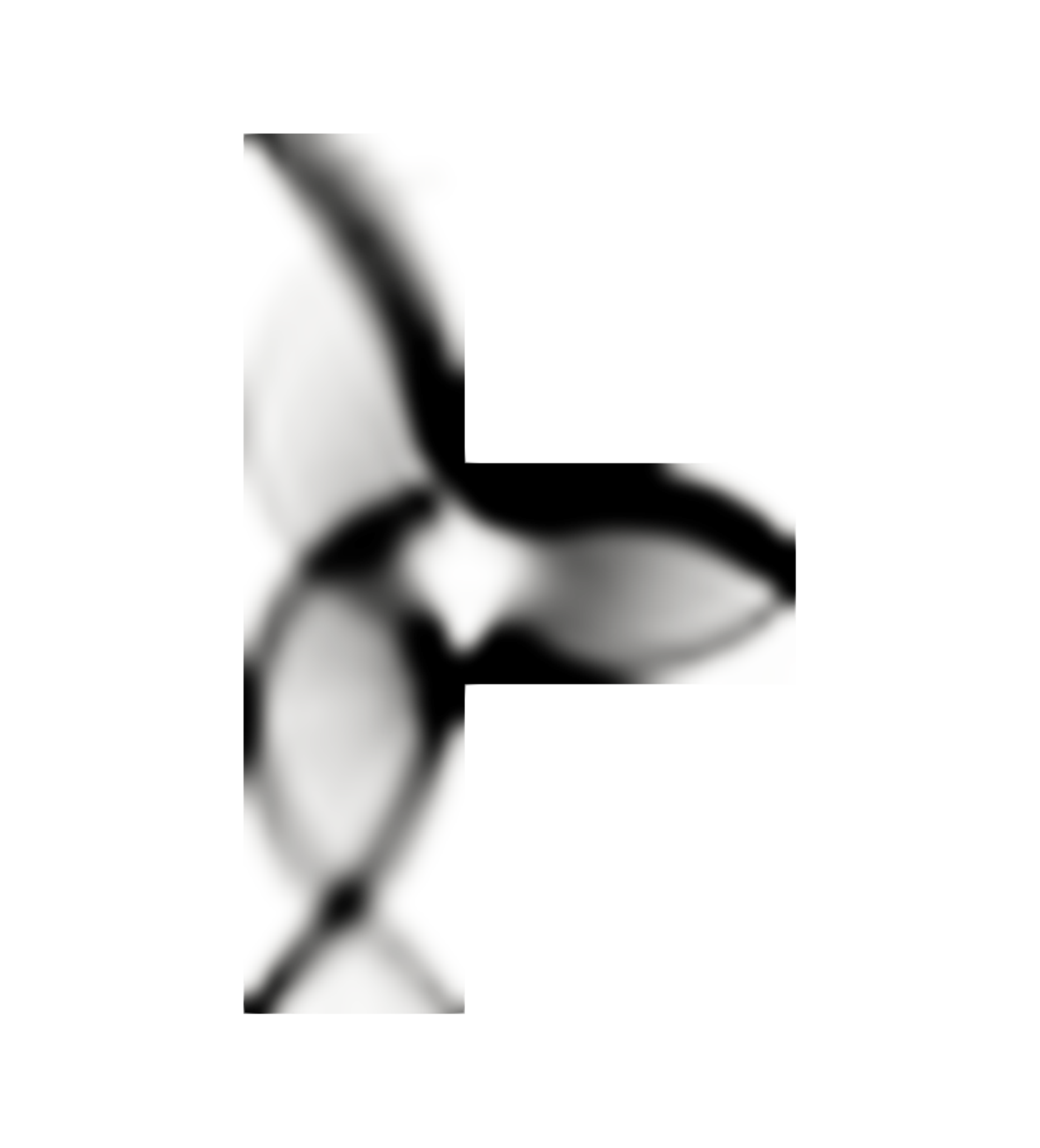
        }
        \vspace{-35pt}
        \subcaption*{\hspace{10pt}$i = 19$}
    \end{minipage}
    \begin{minipage}{0.3\linewidth}
        \centering
        \vspace{-172pt}
        \def\svgwidth{1.15\linewidth}
        {\footnotesize
        \hspace{-60pt}
        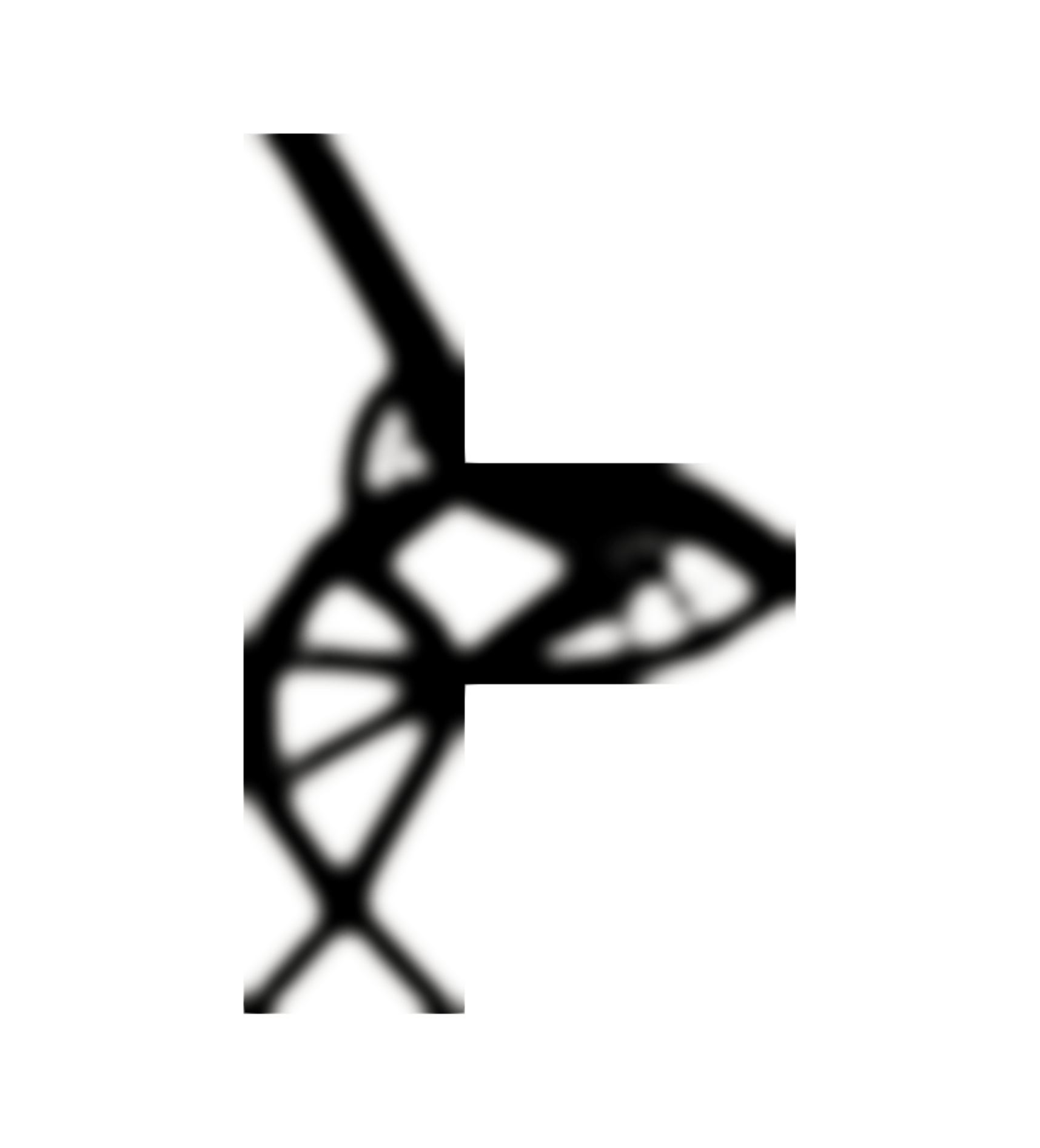
        }
        \vspace{-35pt}
        \subcaption*{\hspace{-5pt}$i = 43$}
 
    \end{minipage}
    \hspace{-50pt}
    \begin{minipage}{0.30\linewidth}
        \centering
        \def\svgwidth{\linewidth}
        \setlength{\figW}{8cm}
        \setlength{\figH}{6cm}
        {\scriptsize
        \input{figures/tbracket_refined_history.tex}
        }
    \end{minipage}
    
    \begin{minipage}{0.3\linewidth}
        \centering
        \vspace{-160pt}
        \def\svgwidth{1.15\linewidth}
        {\footnotesize
        \hspace{-35pt}
        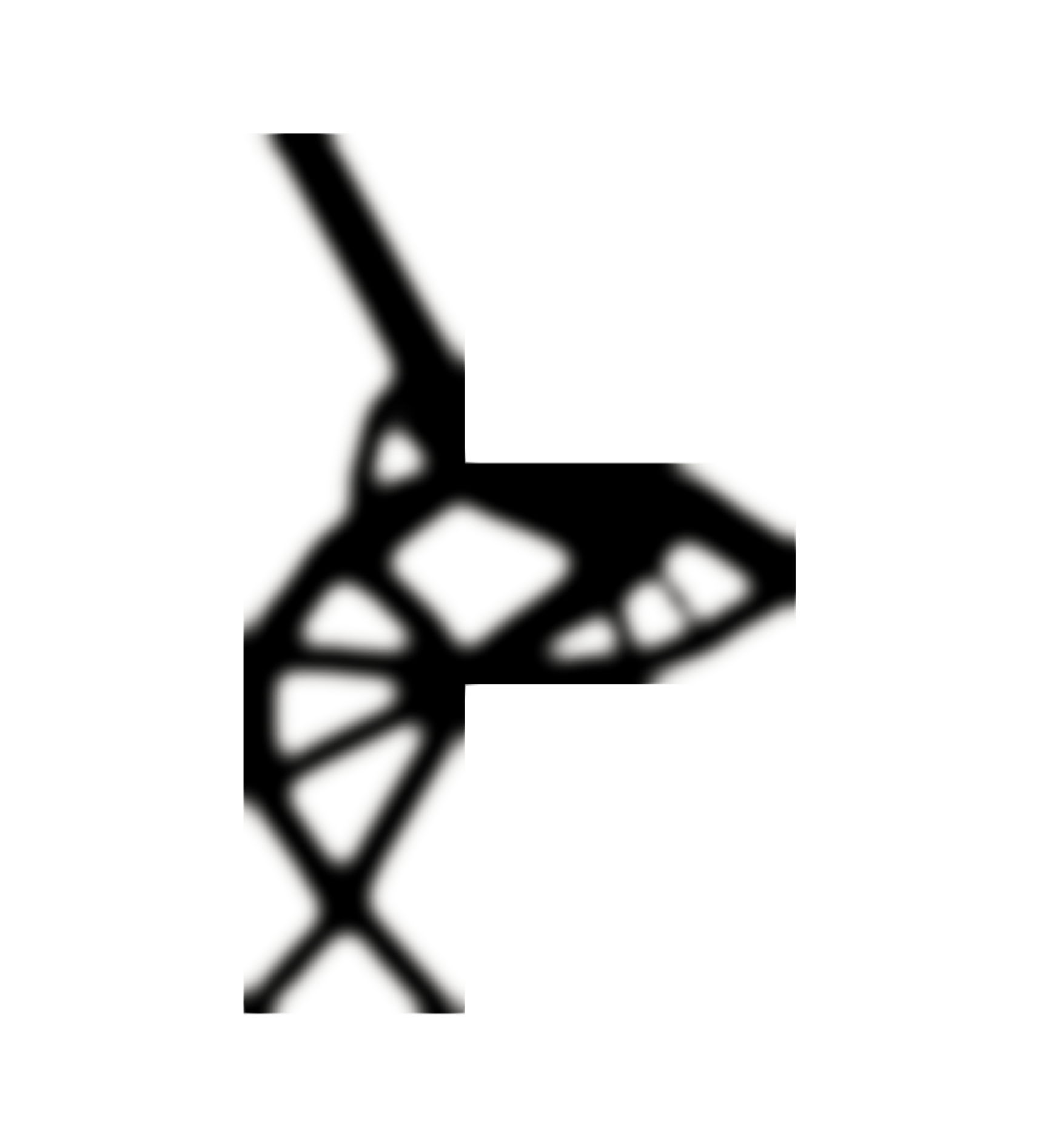
        }
        \vspace{-35pt}
        \subcaption*{\hspace{10pt}$i = 83$}
    \end{minipage}
    \begin{minipage}{0.3\linewidth}
        \centering
        \vspace{-160pt}
        \def\svgwidth{1.15\linewidth}
        {\footnotesize
        \hspace{-60pt}
        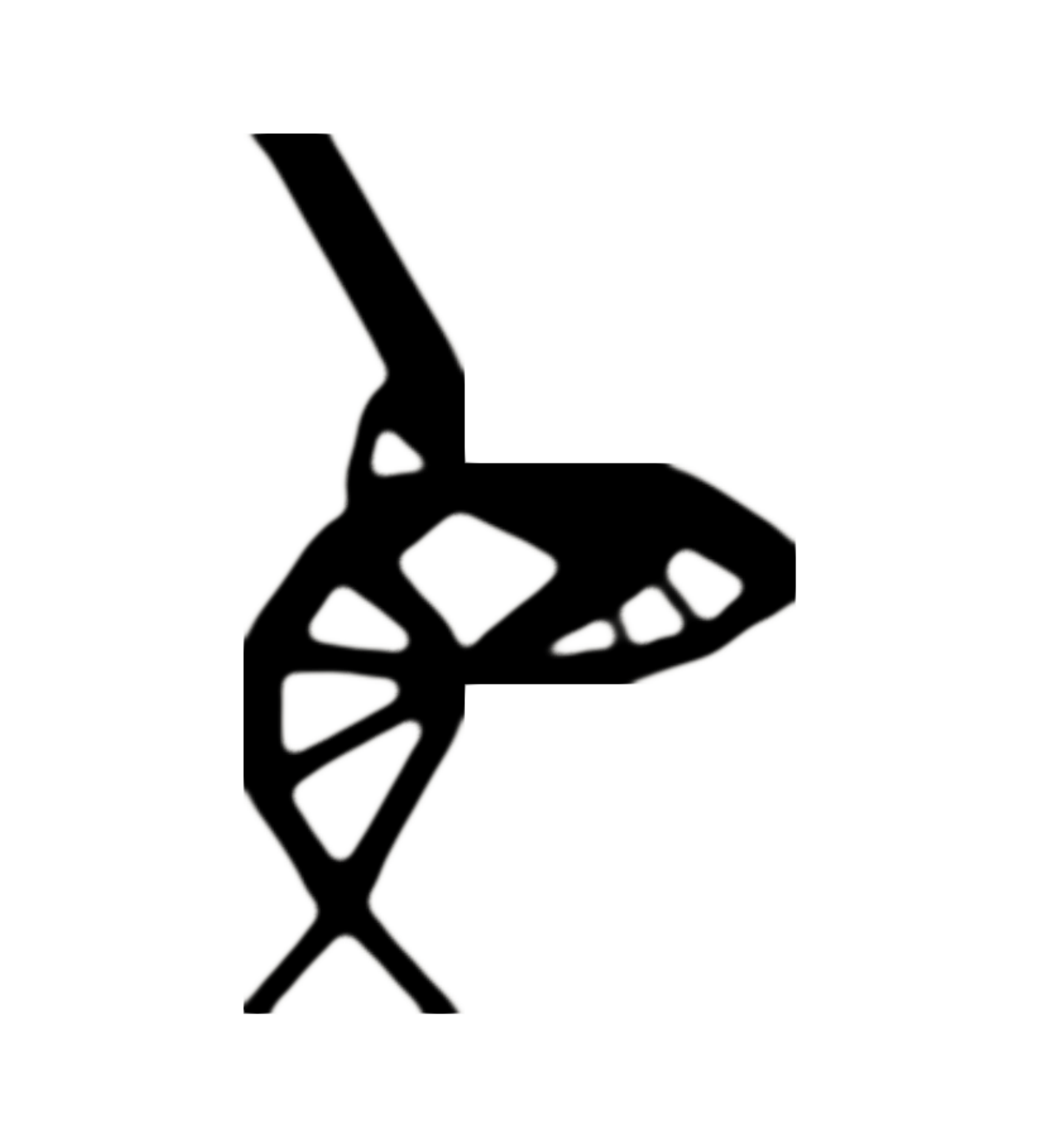
        }
        \vspace{-35pt}
        \subcaption*{\hspace{-5pt}$i = 164$}
 
    \end{minipage}
    
    \caption{Convergence history for the T-bracket example. Illustrated designs correspond to the optimization formulation using the  ML model at iterations with selected continuation steps.}
    \label{fig:tbracket_topopt_benchmark}
\end{figure}

\subsubsection{Single scale topology optimization: T-bracket example} The T-bracket design domain and boundary conditions are illustrated in Figure \ref{fig:t_bracket_bvp}a, corresponding to an average downward displacement of $c_{N} = 0.3 L$ applied at the center of the right-most edge.
The domain is discretized into 35,400 four-node quadrilateral elements within which the displacement field is interpolated using standard bilinear shape functions. In this example, we employ a volume fraction upper bound of $g_{\max} = 0.5$.
The convergence history is provided in Figure \ref{fig:tbracket_topopt_benchmark},
where the continuation updates on the SIMP penalty exponent $p$ and projection strength parameter $\be_{\rh}$ are apparent through the sharp changes in the objective function values.
Apart from the expected jumps in the objective function values at optimization iterations corresponding to the continuation steps, the convergence history is smooth with few oscillations.
The visualized designs in Figure \ref{fig:tbracket_topopt_benchmark} show the progression of the topology as the optimization process proceeds using the ML material model.
The final design  closely resembles the design obtained using the ground truth phenomenological model as evident from the difference plot provided in Figure \ref{fig:t_bracket_bvp}b and the very similar convergence history shown in Figure \ref{fig:tbracket_topopt_benchmark}.
A value of $\de_{\rh} = 1$ indicates material addition and $\de_{\rh} = -1$ indicates material removal from the design obtained using the ground truth phenomenological model.
The deformed configuration of the final design is shown in Figure \ref{fig:t_bracket_bvp}c.

\subsubsection{Single scale topology optimization: Cantilever beam example}
\label{subsubsec:cantilever_beam_example_benchmark}

\begin{figure}[t!]
    \begin{minipage}{0.5\linewidth}
        \centering        
        \def\svgwidth{0.9\linewidth}
        {\footnotesize
\begingroup%
  \makeatletter%
  \providecommand\color[2][]{%
    \errmessage{(Inkscape) Color is used for the text in Inkscape, but the package 'color.sty' is not loaded}%
    \renewcommand\color[2][]{}%
  }%
  \providecommand\transparent[1]{%
    \errmessage{(Inkscape) Transparency is used (non-zero) for the text in Inkscape, but the package 'transparent.sty' is not loaded}%
    \renewcommand\transparent[1]{}%
  }%
  \providecommand\rotatebox[2]{#2}%
  \newcommand*\fsize{\dimexpr\f@size pt\relax}%
  \newcommand*\lineheight[1]{\fontsize{\fsize}{#1\fsize}\selectfont}%
  \ifx\svgwidth\undefined%
    \setlength{\unitlength}{1023.50103423bp}%
    \ifx\svgscale\undefined%
      \relax%
    \else%
      \setlength{\unitlength}{\unitlength * \real{\svgscale}}%
    \fi%
  \else%
    \setlength{\unitlength}{\svgwidth}%
  \fi%
  \global\let\svgwidth\undefined%
  \global\let\svgscale\undefined%
  \makeatother%
  \begin{picture}(1,0.46825151)%
    \lineheight{1}%
    \setlength\tabcolsep{0pt}%
    \put(0,0){\includegraphics[width=\unitlength,page=1]{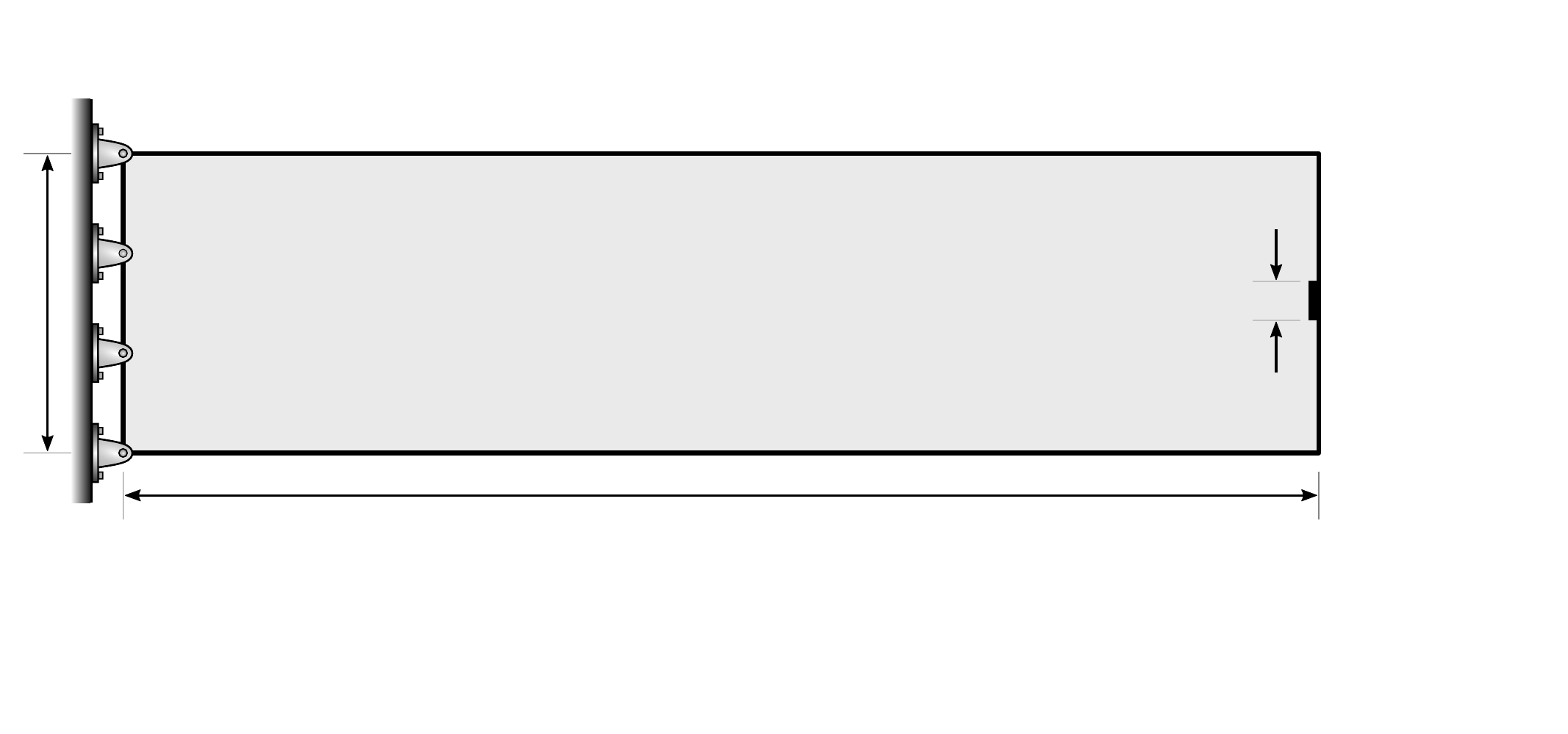}}%
    \put(0.44,0.115){\color[rgb]{0,0,0}\makebox(0,0)[lt]{\lineheight{1.25}\smash{\begin{tabular}[t]{l}$2L$\end{tabular}}}}%
    \put(-0.003,0.2648898){\color[rgb]{0,0,0}\makebox(0,0)[lt]{\lineheight{1.25}\smash{\begin{tabular}[t]{l}$\frac{L}{2}$\end{tabular}}}}%
    \put(0.77,0.22){\color[rgb]{0,0,0}\makebox(0,0)[lt]{\lineheight{1.25}\smash{\begin{tabular}[t]{l}$\frac{L}{10}$\end{tabular}}}}%
    \put(0,0){\includegraphics[width=\unitlength,page=2]{cantilever_bvp.pdf}}%
  \end{picture}%
\endgroup%

        }
        \vspace{-20pt}
        \subcaption*{(a)}
    \end{minipage}
    \begin{minipage}{0.5\linewidth}
        \centering        
        \def\svgwidth{.9\linewidth}
        {\scriptsize
\begingroup%
  \makeatletter%
  \providecommand\color[2][]{%
    \errmessage{(Inkscape) Color is used for the text in Inkscape, but the package 'color.sty' is not loaded}%
    \renewcommand\color[2][]{}%
  }%
  \providecommand\transparent[1]{%
    \errmessage{(Inkscape) Transparency is used (non-zero) for the text in Inkscape, but the package 'transparent.sty' is not loaded}%
    \renewcommand\transparent[1]{}%
  }%
  \providecommand\rotatebox[2]{#2}%
  \newcommand*\fsize{\dimexpr\f@size pt\relax}%
  \newcommand*\lineheight[1]{\fontsize{\fsize}{#1\fsize}\selectfont}%
  \ifx\svgwidth\undefined%
    \setlength{\unitlength}{1023.50097656bp}%
    \ifx\svgscale\undefined%
      \relax%
    \else%
      \setlength{\unitlength}{\unitlength * \real{\svgscale}}%
    \fi%
  \else%
    \setlength{\unitlength}{\svgwidth}%
  \fi%
  \global\let\svgwidth\undefined%
  \global\let\svgscale\undefined%
  \makeatother%
  \begin{picture}(1,0.46825155)%
    \lineheight{1}%
    \setlength\tabcolsep{0pt}%
    \put(0,0){\includegraphics[width=\unitlength,page=1]{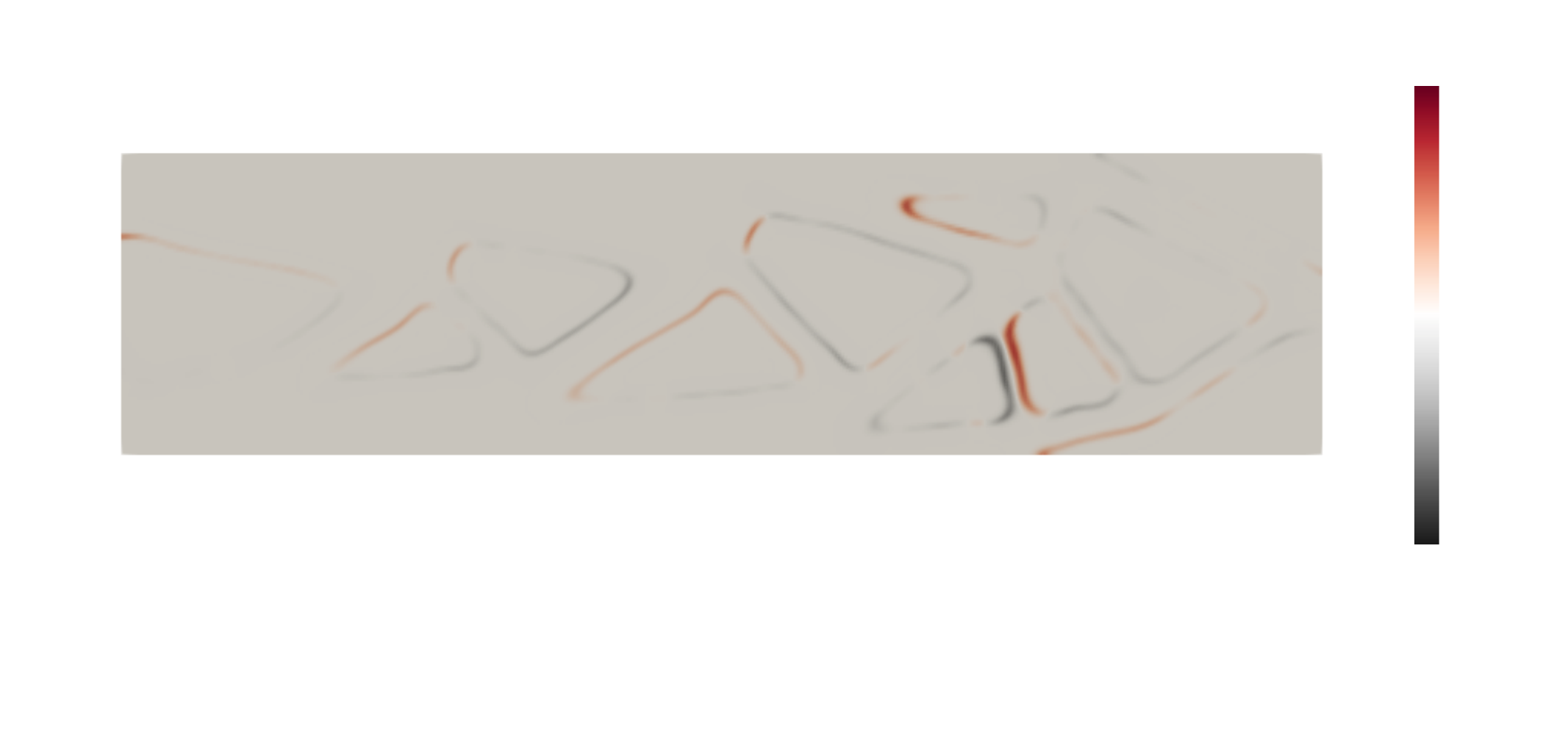}}%
    \put(0.93170396,0.12379161){\color[rgb]{0,0,0}\makebox(0,0)[lt]{\lineheight{1.25}\smash{\begin{tabular}[t]{l}-1.\end{tabular}}}}%
    \put(0.93170396,0.41592617){\color[rgb]{0,0,0}\makebox(0,0)[lt]{\lineheight{1.25}\smash{\begin{tabular}[t]{l}1.\end{tabular}}}}%
    \put(0.93170396,0.26937037){\color[rgb]{0,0,0}\makebox(0,0)[lt]{\lineheight{1.25}\smash{\begin{tabular}[t]{l}0.\end{tabular}}}}%
    \put(0.921,0.05){\color[rgb]{0,0,0}\makebox(0,0)[t]{\lineheight{1.25}\smash{\begin{tabular}[t]{c}$\de_{\rh}$\end{tabular}}}}%
    \put(0,0){\includegraphics[width=\unitlength,page=2]{cantilever_final_design_ml_analytical_mse.pdf}}%
  \end{picture}%
\endgroup%

        }
        \vspace{-20pt}
        \subcaption*{(b)}
    \end{minipage}

    \begin{minipage}{\linewidth}
        \centering        
        \def\svgwidth{0.45\linewidth}
        {\scriptsize
\begingroup%
  \makeatletter%
  \providecommand\color[2][]{%
    \errmessage{(Inkscape) Color is used for the text in Inkscape, but the package 'color.sty' is not loaded}%
    \renewcommand\color[2][]{}%
  }%
  \providecommand\transparent[1]{%
    \errmessage{(Inkscape) Transparency is used (non-zero) for the text in Inkscape, but the package 'transparent.sty' is not loaded}%
    \renewcommand\transparent[1]{}%
  }%
  \providecommand\rotatebox[2]{#2}%
  \newcommand*\fsize{\dimexpr\f@size pt\relax}%
  \newcommand*\lineheight[1]{\fontsize{\fsize}{#1\fsize}\selectfont}%
  \ifx\svgwidth\undefined%
    \setlength{\unitlength}{1023.50097656bp}%
    \ifx\svgscale\undefined%
      \relax%
    \else%
      \setlength{\unitlength}{\unitlength * \real{\svgscale}}%
    \fi%
  \else%
    \setlength{\unitlength}{\svgwidth}%
  \fi%
  \global\let\svgwidth\undefined%
  \global\let\svgscale\undefined%
  \makeatother%
  \begin{picture}(1,0.46825155)%
    \lineheight{1}%
    \setlength\tabcolsep{0pt}%
    \put(0,0){\includegraphics[width=\unitlength,page=1]{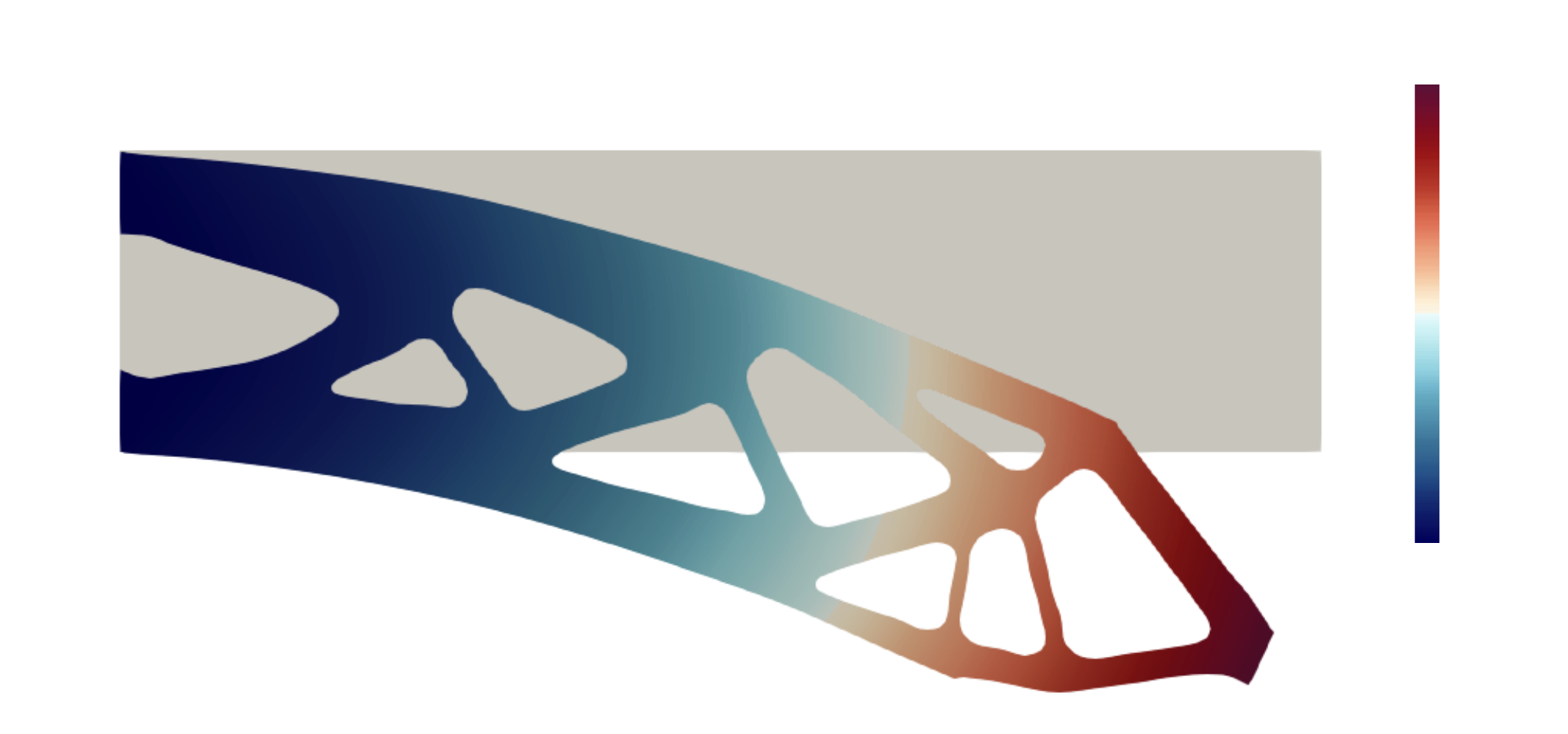}}%
    \put(0.93194593,0.12481716){\color[rgb]{0,0,0}\makebox(0,0)[lt]{\lineheight{1.25}\smash{\begin{tabular}[t]{l}0.\end{tabular}}}}%
    \put(0.93194593,0.41695172){\color[rgb]{0,0,0}\makebox(0,0)[lt]{\lineheight{1.25}\smash{\begin{tabular}[t]{l}62.\end{tabular}}}}%
    \put(0.910,0.04){\color[rgb]{0,0,0}\makebox(0,0)[t]{\lineheight{1.25}\smash{\begin{tabular}[t]{c}$||\bu||$\end{tabular}}}}%
    \put(0,0){\includegraphics[width=\unitlength,page=2]{cantilever_final_design_ML.pdf}}%
  \end{picture}%
\endgroup%

        }
        \subcaption*{(c)}
    \end{minipage}

    \caption{(a) Cantilever boundary value problem. (b) Difference plot of the design obtained using the ML model to the design obtained using the ground truth phenomenological model, where $\de_{\rh} = 1$ indicates material addition, and $\de_{\rh} = -1$ indicates material removal. (c) Deformation of the design obtained using the ML model.}
    \label{fig:cantilever_bvp}
\end{figure}

\begin{figure}[t!]
    \begin{minipage}{0.5\linewidth}
        \centering
        \vspace{-240pt}
        \subcaption*{\hspace{170pt} $i = 19$}
        \vspace{-40pt}
        \def\svgwidth{0.72\linewidth}
        {\footnotesize
        \hspace{-22pt}
        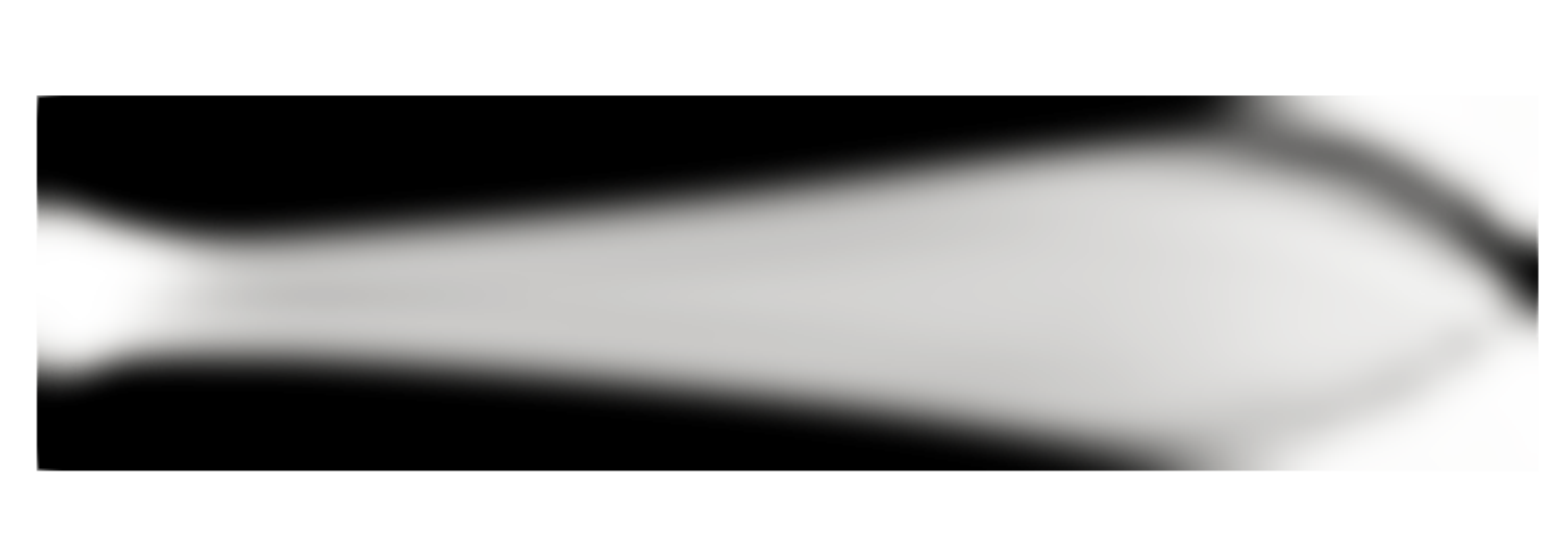
        }
    \end{minipage}
    \begin{minipage}{0.30\linewidth}
        \centering
        \def\svgwidth{\linewidth}
        \setlength{\figW}{8cm}
        \setlength{\figH}{6cm}
        {\scriptsize
        \input{figures/cantilever_refined_history.tex}
        }
    \end{minipage}
    
    \begin{minipage}{0.5\linewidth}
        \centering
        \def\svgwidth{0.72\linewidth}
        \vspace{-420pt}
        \subcaption*{\hspace{170pt} $i = 75$}
        \vspace{-40pt}
        {\footnotesize
        \hspace{-22pt}
        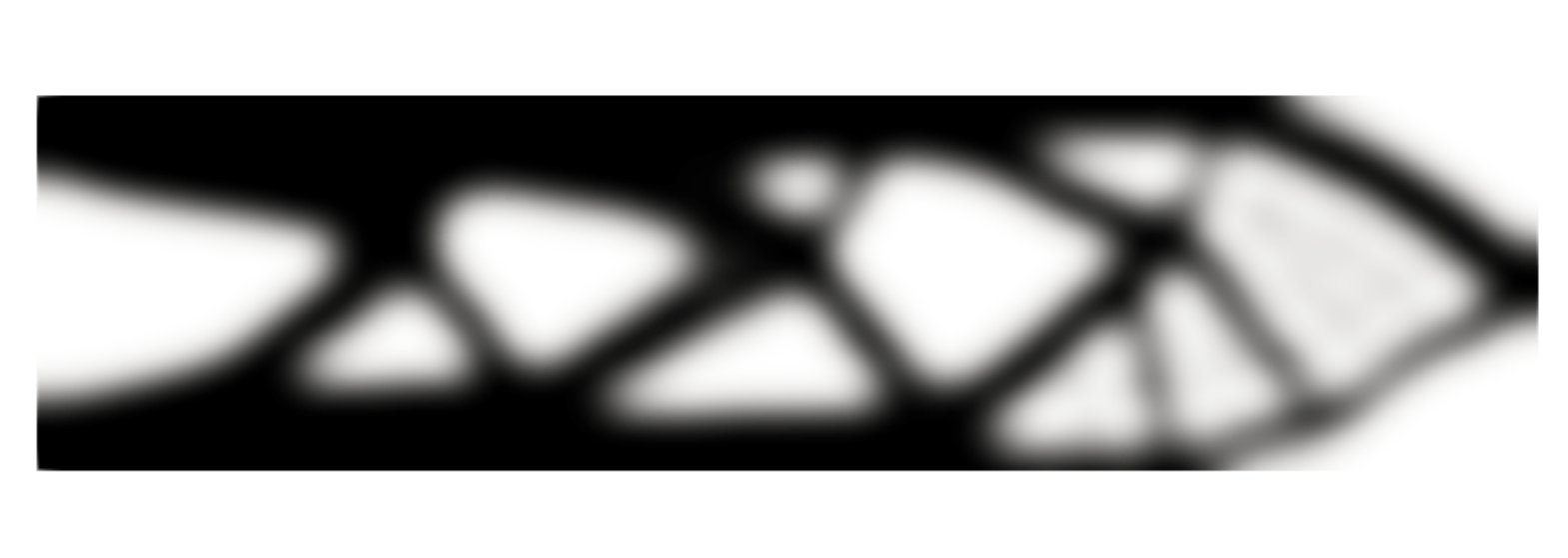
        }
    \end{minipage}

    \begin{minipage}{0.5\linewidth}
        \centering
        \def\svgwidth{0.72\linewidth}
        \vspace{-320pt}
        \subcaption*{\hspace{170pt} $i = 95$}
        \vspace{-40pt}
        {\footnotesize
        \hspace{-22pt}
        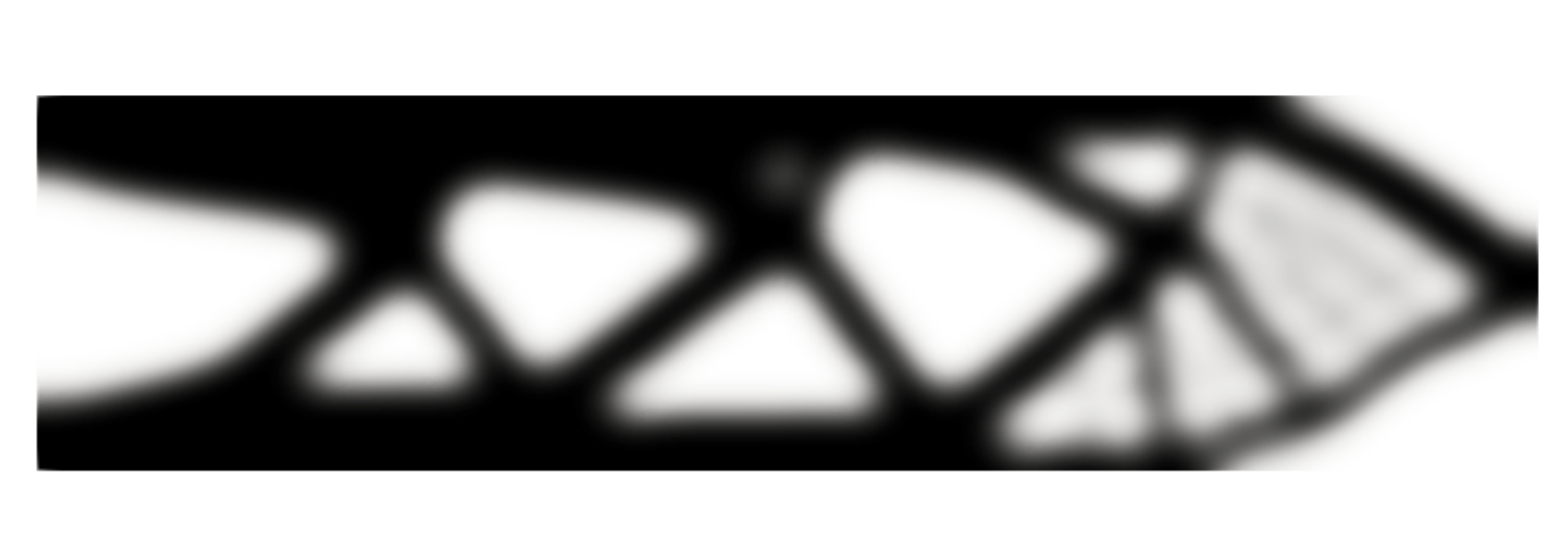
        }
    \end{minipage}
    
    \begin{minipage}{0.5\linewidth}
        \centering
        \def\svgwidth{0.72\linewidth}
        \vspace{-220pt}
        \subcaption*{\hspace{170pt} $i = 155$}
        \vspace{-40pt}
        {\footnotesize
        \hspace{-22pt}
        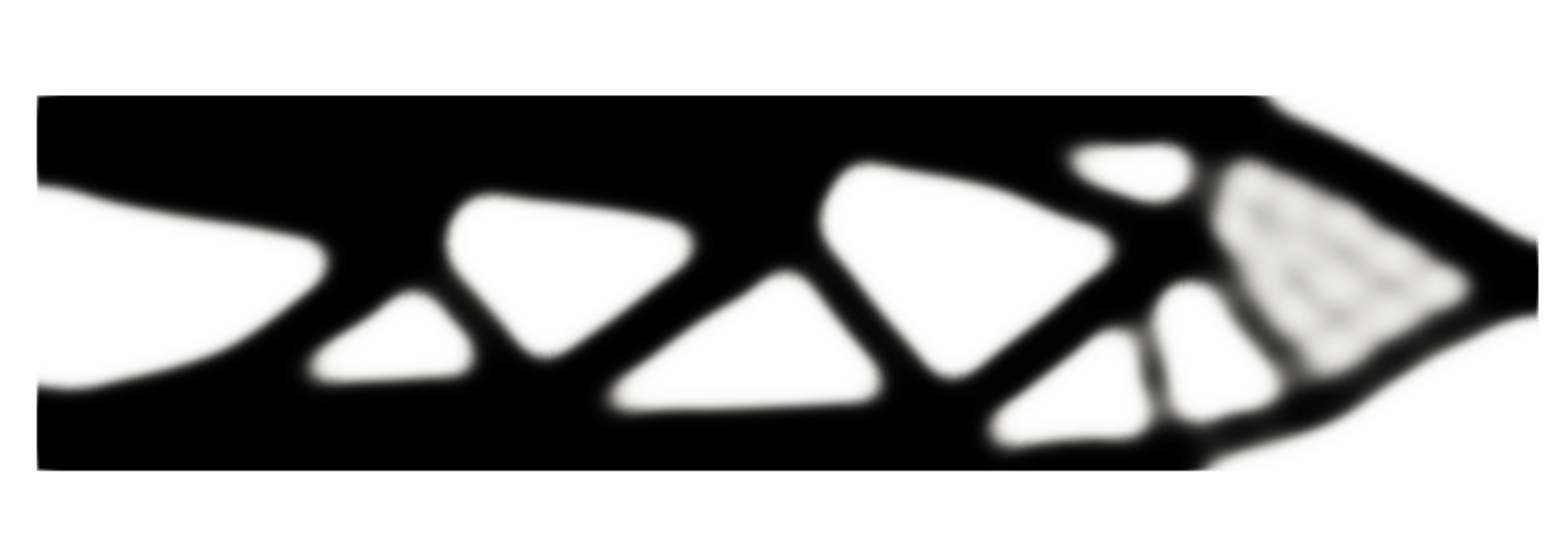
        }
    \end{minipage}

    \begin{minipage}{0.5\linewidth}
        \centering
        \def\svgwidth{0.72\linewidth}
        \vspace{-120pt}
        \subcaption*{\hspace{170pt} $i = 195$}
        \vspace{-40pt}
        {\footnotesize
        \hspace{-22pt}
        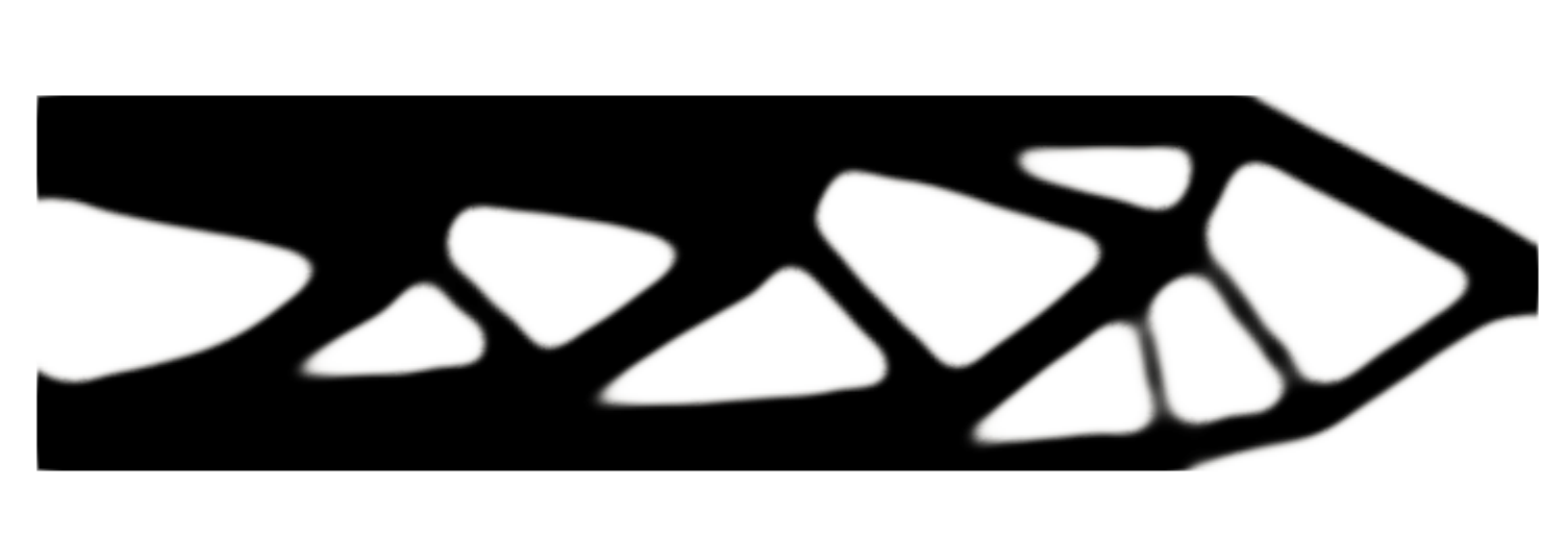
        }
    \end{minipage}

    \vspace{-25pt}
    \caption{Convergence history for cantilever beam example. Illustrated designs correspond to the optimization formulation using the  ML model at iterations with selected continuation steps.}
    \label{fig:cantilever_topopt_benchmark}
\end{figure}

We now consider a cantilever beam with the design domain as shown in Figure \ref{fig:cantilever_bvp}(a).
The cantilever beam is fixed along the left edge and a vertically downward average displacement $c_{N} = 0.6 L$ is applied near the center of the right edge.
The domain is discretized using 4 node quadrilateral elements with a uniform mesh of edge length $0.5 \Umm$ adding up to a total of $40000$ elements.
This example is particularly interesting as the cantilever beam problem subject to large deformation is a challenging example for 
hyperelastic materials due to the presence of buckling instabilities.
We remark that for this problem we considered hyperparameters, viz. material volume fraction constraint $g_{\max} = 0.6$ and the filter radius $r_{\rh} = 6.0$, that lead to final designs not susceptible to large buckling instabilities. Although not the focus of this article, future work could include buckling criteria explicitly as part of the optimization formulation to systematically prevent such behavior.

Nonetheless, minor instabilities occur during the optimization process which has a visible effect on the convergence history as well as the intermediate designs highlighting the challenging nature of the cantilever beam problem, as shown in Figure \ref{fig:cantilever_topopt_benchmark}.
We observe that while the continuation updates on the SIMP penalty exponent $p$ produces sharp changes in the objective function values, the projection strength parameter $\be_{\rh}$ updates, especially in the later stages of the optimization process, do not lead to significantly abrupt changes in the objective.
This is due to the interference of minor instabilities altering the deformation state of the design, resulting in prolonged presence of grey regions in the design, which is only slowly resolved with the $\be_{\rh}$ updates.
See the progression of the topology as the optimization process proceeds using the ML material model in Figure \ref{fig:cantilever_topopt_benchmark} for reference.
Interestingly, this leads the convergence histories to differ very slightly, although the final objective values remain close to one another.
As a result, we see that the final design obtained using the ML model resembles the design obtained using the ground truth phenomenological model, although there are subtle differences around the boundaries of the two design as shown in the difference plot in Figure \ref{fig:cantilever_bvp}(b).
The deformed configuration of the final design obtained using the ML model is depicted in Figure \ref{fig:cantilever_bvp}(c).

\subsection{Multi-scale topology optimization}
\begin{figure}[t!]
    \begin{minipage}{0.3\linewidth}
    \centering        
    \def\svgwidth{0.85\linewidth}
    {\scriptsize
    \hspace{-25pt}
\begingroup%
  \makeatletter%
  \providecommand\color[2][]{%
    \errmessage{(Inkscape) Color is used for the text in Inkscape, but the package 'color.sty' is not loaded}%
    \renewcommand\color[2][]{}%
  }%
  \providecommand\transparent[1]{%
    \errmessage{(Inkscape) Transparency is used (non-zero) for the text in Inkscape, but the package 'transparent.sty' is not loaded}%
    \renewcommand\transparent[1]{}%
  }%
  \providecommand\rotatebox[2]{#2}%
  \newcommand*\fsize{\dimexpr\f@size pt\relax}%
  \newcommand*\lineheight[1]{\fontsize{\fsize}{#1\fsize}\selectfont}%
  \ifx\svgwidth\undefined%
    \setlength{\unitlength}{577.79309082bp}%
    \ifx\svgscale\undefined%
      \relax%
    \else%
      \setlength{\unitlength}{\unitlength * \real{\svgscale}}%
    \fi%
  \else%
    \setlength{\unitlength}{\svgwidth}%
  \fi%
  \global\let\svgwidth\undefined%
  \global\let\svgscale\undefined%
  \makeatother%
  \begin{picture}(1,1.50744384)%
    \lineheight{1}%
    \setlength\tabcolsep{0pt}%
    \put(0,0){\includegraphics[width=\unitlength,page=1]{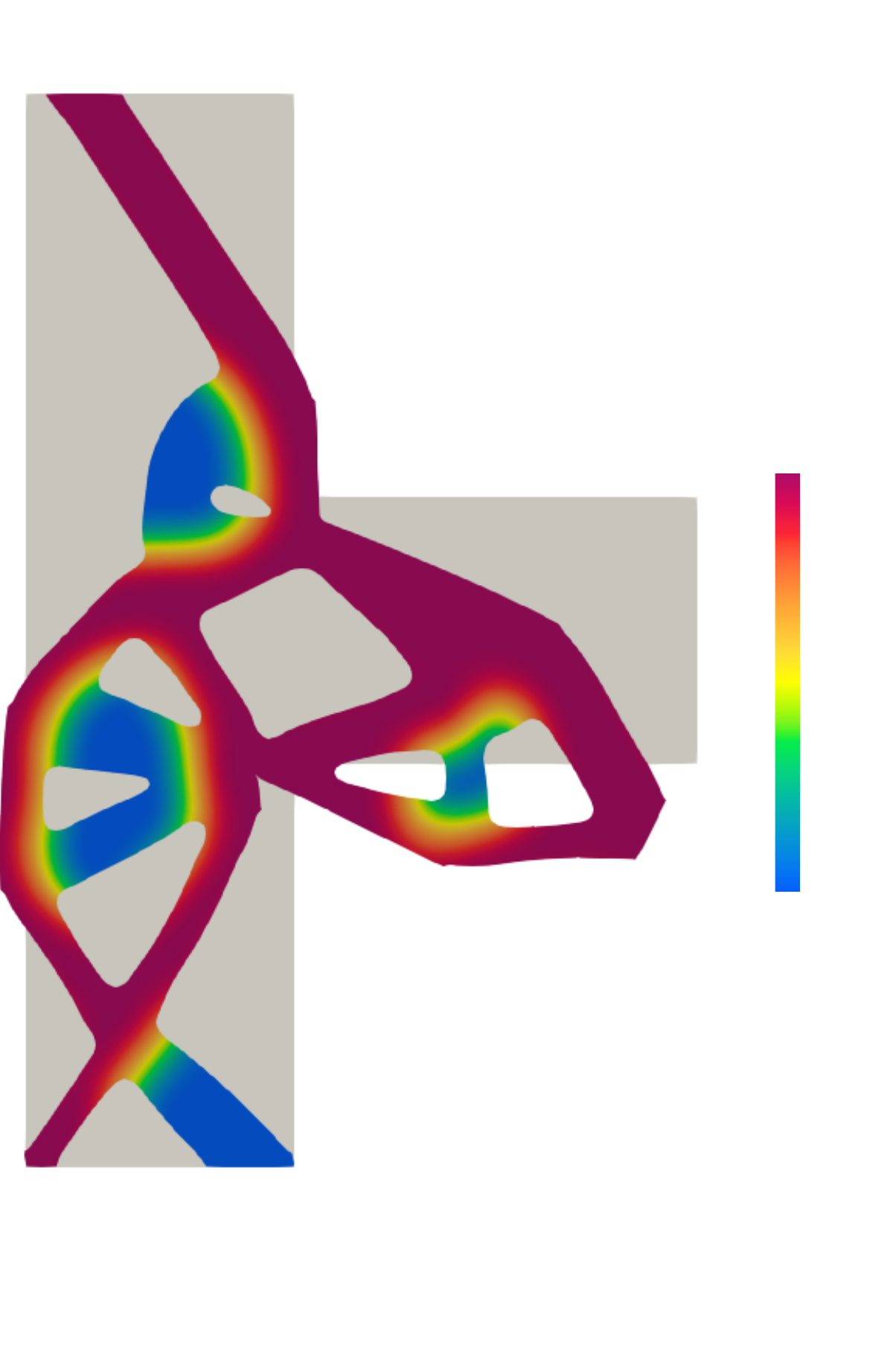}}%
    \put(0.91872691,0.51808543){\color[rgb]{0,0,0}\makebox(0,0)[lt]{\lineheight{1.25}\smash{\begin{tabular}[t]{l}0.1\end{tabular}}}}%
    \put(0.91872691,0.98365002){\color[rgb]{0,0,0}\makebox(0,0)[lt]{\lineheight{1.25}\smash{\begin{tabular}[t]{l}0.5\end{tabular}}}}%
    \put(0.91872691,0.63231317){\color[rgb]{0,0,0}\makebox(0,0)[lt]{\lineheight{1.25}\smash{\begin{tabular}[t]{l}0.2\end{tabular}}}}%
    \put(0.91872691,0.75000236){\color[rgb]{0,0,0}\makebox(0,0)[lt]{\lineheight{1.25}\smash{\begin{tabular}[t]{l}0.3\end{tabular}}}}%
    \put(0.91872691,0.86596083){\color[rgb]{0,0,0}\makebox(0,0)[lt]{\lineheight{1.25}\smash{\begin{tabular}[t]{l}0.4\end{tabular}}}}%
    \put(0.89,0.4){\color[rgb]{0,0,0}{\makebox(0,0)[t]{\lineheight{1.25}\smash{\begin{tabular}[t]{c}$\al$\end{tabular}}}}}%
    \put(0,0){\includegraphics[width=\unitlength,page=2]{t_bracket_final_design_micro.pdf}}%
  \end{picture}%
\endgroup%

    \subcaption*{(a)}
    }
    \end{minipage}
    \begin{minipage}{0.3\linewidth}
        \def\svgwidth{0.85\linewidth}
        {\scriptsize
        \hspace{-20pt}
\begingroup%
  \makeatletter%
  \providecommand\color[2][]{%
    \errmessage{(Inkscape) Color is used for the text in Inkscape, but the package 'color.sty' is not loaded}%
    \renewcommand\color[2][]{}%
  }%
  \providecommand\transparent[1]{%
    \errmessage{(Inkscape) Transparency is used (non-zero) for the text in Inkscape, but the package 'transparent.sty' is not loaded}%
    \renewcommand\transparent[1]{}%
  }%
  \providecommand\rotatebox[2]{#2}%
  \newcommand*\fsize{\dimexpr\f@size pt\relax}%
  \newcommand*\lineheight[1]{\fontsize{\fsize}{#1\fsize}\selectfont}%
  \ifx\svgwidth\undefined%
    \setlength{\unitlength}{577.79309082bp}%
    \ifx\svgscale\undefined%
      \relax%
    \else%
      \setlength{\unitlength}{\unitlength * \real{\svgscale}}%
    \fi%
  \else%
    \setlength{\unitlength}{\svgwidth}%
  \fi%
  \global\let\svgwidth\undefined%
  \global\let\svgscale\undefined%
  \makeatother%
  \begin{picture}(1,1.50761237)%
    \lineheight{1}%
    \setlength\tabcolsep{0pt}%
    \put(0,0){\includegraphics[width=\unitlength,page=1]{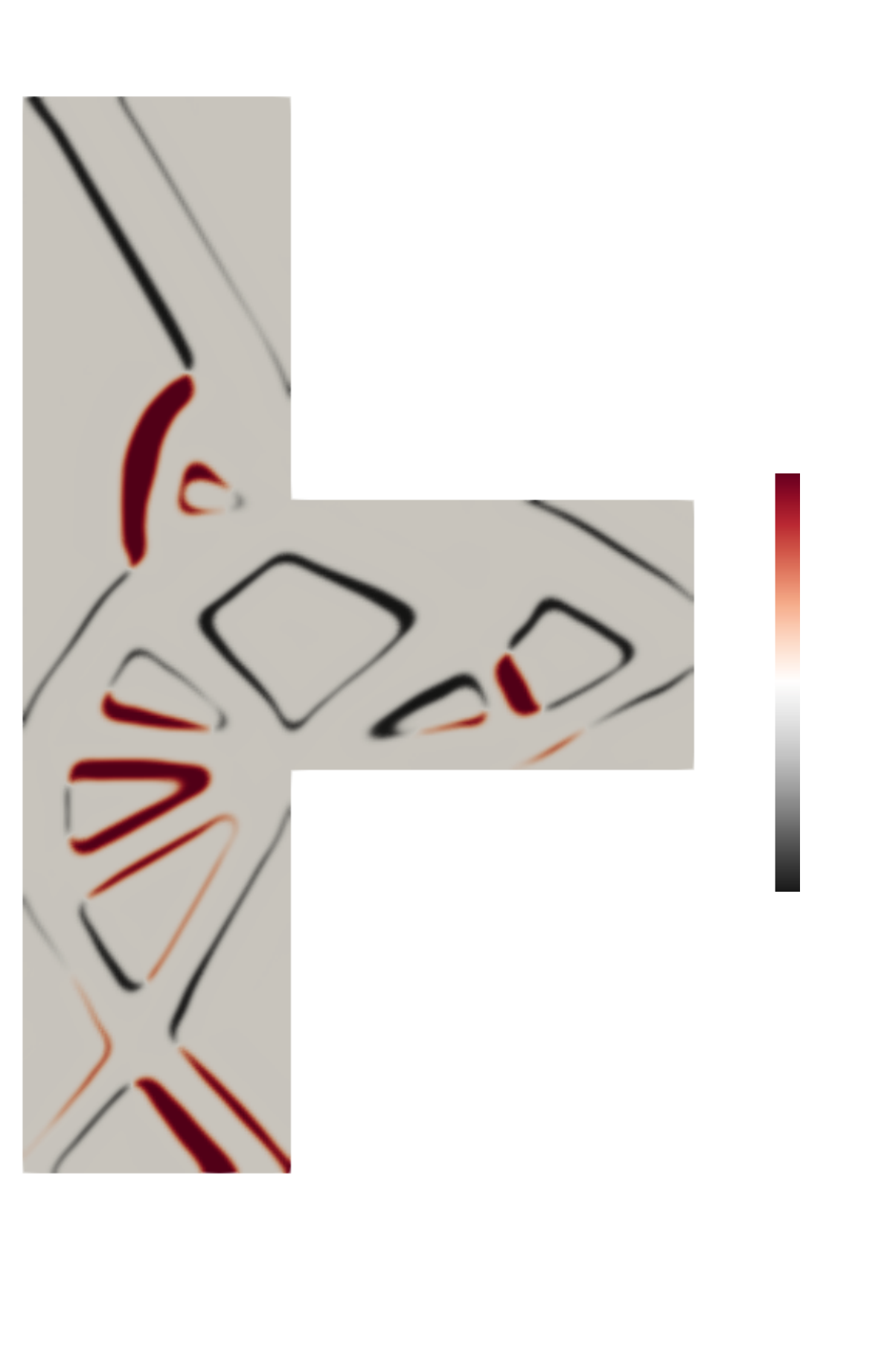}}%
    \put(0.91872686,0.51825401){\color[rgb]{0,0,0}\makebox(0,0)[lt]{\lineheight{1.25}\smash{\begin{tabular}[t]{l}-1.\end{tabular}}}}%
    \put(0.91872686,0.9838186){\color[rgb]{0,0,0}\makebox(0,0)[lt]{\lineheight{1.25}\smash{\begin{tabular}[t]{l}1.\end{tabular}}}}%
    \put(0,0){\includegraphics[width=\unitlength,page=2]{t_bracket_final_design_uniform_micro_mse.pdf}}%
    \put(0.89,0.4){\color[rgb]{0,0,0}\makebox(0,0)[t]{\lineheight{1.25}\smash{\begin{tabular}[t]{c}$\de_{\rh}$\end{tabular}}}}%
    \put(0.91872686,0.75041151){\color[rgb]{0,0,0}\makebox(0,0)[lt]{\lineheight{1.25}\smash{\begin{tabular}[t]{l}0.\end{tabular}}}}%
    \put(0,0){\includegraphics[width=\unitlength,page=3]{t_bracket_final_design_uniform_micro_mse.pdf}}%
  \end{picture}%
\endgroup%

        \subcaption*{(b)}
        }
    \end{minipage}
    \hspace{-40pt}
    \begin{minipage}{0.30\linewidth}
        \centering
        \def\svgwidth{\linewidth}
        \setlength{\figW}{8cm}
        \setlength{\figH}{6cm}
        {\scriptsize
        \vspace{13pt}
\begin{tikzpicture}

\definecolor{gray}{RGB}{128,128,128}
\definecolor{lightgray204}{RGB}{204,204,204}
\definecolor{mediumblue}{RGB}{0,0,205}
\pgfplotsset{grid style={dashed,lightgray204}}

\begin{axis}[
height=\figH,
legend cell align={left},
legend style={fill opacity=0.8, draw opacity=1, text opacity=1, draw=lightgray204},
tick align=outside,
tick pos=left,
width=\figW,
xlabel={Optimization Iteration Number},
xmajorgrids,
xmin=-8.4, xmax=176.4,
xtick = {0, 168},
ytick = {-16.6, -25.8, -14.8, -21.2, -24.2},
xtick style={color=black},
ylabel={Objective},
ymajorgrids,
ymin=-26.38745, ymax=-12.7,
ytick style={color=black}
]
\addplot [very thick, magenta]
table {%
0 -16.634
1 -19.61
2 -22.145
3 -23.427
4 -24.172
5 -24.554
6 -24.815
7 -25.031
8 -25.221
9 -25.378
10 -25.498
11 -25.592
12 -25.672
13 -25.735
14 -25.793
15 -25.825
16 -25.823
17 -25.813
18 -25.818
19 -25.837
20 -14.828
21 -15.669
22 -16.67
23 -17.276
24 -17.786
25 -18.232
26 -18.642
27 -19.047
28 -19.495
29 -20.007
30 -20.451
31 -20.674
32 -20.849
33 -21.009
34 -21.149
35 -21.277
36 -21.396
37 -21.498
38 -21.575
39 -21.659
40 -21.732
41 -21.802
42 -21.856
43 -21.893
44 -21.922
45 -21.945
46 -21.963
47 -21.979
48 -21.992
49 -19.74
50 -19.858
51 -19.962
52 -20.011
53 -20.052
54 -20.079
55 -20.1
56 -20.119
57 -20.133
58 -20.145
59 -20.158
60 -20.169
61 -20.181
62 -20.193
63 -20.204
64 -20.215
65 -20.226
66 -20.237
67 -20.248
68 -20.259
69 -18.963
70 -18.771
71 -18.769
72 -18.784
73 -18.937
74 -19.008
75 -19.055
76 -19.081
77 -19.095
78 -19.105
79 -19.113
80 -19.121
81 -19.128
82 -19.134
83 -19.14
84 -19.145
85 -19.151
86 -19.156
87 -19.161
88 -19.165
89 -19.938
90 -19.795
91 -19.801
92 -19.805
93 -19.808
94 -19.811
95 -19.814
96 -19.817
97 -19.82
98 -19.822
99 -19.825
100 -19.828
101 -19.83
102 -19.833
103 -19.837
104 -19.84
105 -19.843
106 -19.846
107 -19.849
108 -19.852
109 -21.531
110 -21.28
111 -21.297
112 -21.306
113 -21.313
114 -21.32
115 -21.325
116 -21.33
117 -21.334
118 -21.337
119 -21.34
120 -21.342
121 -21.344
122 -21.346
123 -21.348
124 -21.35
125 -21.352
126 -21.353
127 -21.355
128 -21.356
129 -23.16
130 -22.986
131 -23.022
132 -23.04
133 -23.058
134 -23.072
135 -23.086
136 -23.098
137 -23.109
138 -23.117
139 -23.123
140 -23.129
141 -23.135
142 -23.138
143 -23.137
144 -23.138
145 -23.136
146 -23.146
147 -23.152
148 -23.155
149 -24.164
150 -24.121
151 -24.134
152 -24.141
153 -24.148
154 -24.153
155 -24.159
156 -24.164
157 -24.17
158 -24.175
159 -24.179
160 -24.182
161 -24.186
162 -24.189
163 -24.192
164 -24.195
165 -24.197
166 -24.2
167 -24.201
168 -24.193
};
\addlegendentry{ML: varying microstructure}
\addplot [very thick, mediumblue, dash pattern=on 4pt off 2pt]
table {%
0 -16.634
1 -18.586
2 -20.388
3 -21.615
4 -22.308
5 -22.584
6 -22.73
7 -22.82
8 -22.881
9 -22.925
10 -22.959
11 -22.973
12 -23.001
13 -23.019
14 -23.03
15 -23.039
16 -23.041
17 -23.048
18 -23.049
19 -23.053
20 -15.895
21 -16.498
22 -17.115
23 -17.497
24 -17.814
25 -18.071
26 -18.256
27 -18.4
28 -18.523
29 -18.634
30 -18.742
31 -18.847
32 -18.944
33 -19.036
34 -19.124
35 -19.205
36 -19.266
37 -19.312
38 -19.361
39 -19.402
40 -19.432
41 -19.454
42 -19.469
43 -19.479
44 -19.486
45 -17.546
46 -17.596
47 -17.543
48 -17.07
49 -17.583
50 -17.676
51 -17.729
52 -17.758
53 -17.775
54 -17.786
55 -17.795
56 -17.804
57 -17.811
58 -17.818
59 -17.825
60 -17.832
61 -17.838
62 -17.845
63 -17.853
64 -17.861
65 -15.627
66 -16.397
67 -16.074
68 -15.87
69 -16.367
70 -16.481
71 -16.66
72 -16.73
73 -16.772
74 -16.798
75 -16.815
76 -16.828
77 -16.837
78 -16.841
79 -16.759
80 -16.819
81 -16.845
82 -16.855
83 -16.861
84 -16.865
85 -16.868
86 -17.547
87 -17.447
88 -17.453
89 -17.456
90 -17.457
91 -17.459
92 -17.461
93 -17.462
94 -17.464
95 -17.466
96 -17.467
97 -17.469
98 -17.47
99 -17.472
100 -17.473
101 -17.474
102 -17.476
103 -17.477
104 -17.478
105 -17.479
106 -18.958
107 -18.786
108 -18.801
109 -18.808
110 -18.813
111 -18.818
112 -18.822
113 -18.825
114 -18.828
115 -18.83
116 -18.832
117 -18.834
118 -18.836
119 -18.837
120 -18.835
121 -18.806
122 -18.802
123 -18.784
124 -18.823
125 -18.837
126 -18.839
127 -20.416
128 -20.299
129 -20.315
130 -20.322
131 -20.328
132 -20.333
133 -20.337
134 -20.341
135 -20.345
136 -20.348
137 -20.351
138 -20.355
139 -20.358
140 -20.361
141 -20.363
142 -20.366
143 -20.368
144 -20.37
145 -20.372
146 -20.374
147 -21.216
148 -21.185
149 -21.199
150 -21.205
151 -21.21
152 -21.215
153 -21.22
154 -21.224
155 -21.227
156 -21.229
157 -21.231
158 -21.232
159 -21.234
160 -21.236
161 -21.238
162 -21.239
163 -21.24
164 -21.241
165 -21.242
166 -21.243
};
\addlegendentry{ML: fixed microstructure}
\end{axis}

\end{tikzpicture}
        \subcaption*{(c)}
        }
    
    \end{minipage}
    \caption{(a) Deformed configuration of the final design obtained using the ML model with varying microstructure (b) Difference plot of the design obtained using the ML model with varying microstructure to the design obtained using the ML model with fixed microstructure. (c) Convergence history for the T-bracket TO problem with varying microstructure versus fixed microstructure.}
    \label{fig:t_bracket_micro_bvp}
\end{figure}

We now demonstrate the ability of the framework to perform simultaneous design optimization at two spatial scales.
The optimization problem considered here consists of maximizing the external work \eqref{eq:optimization_statement} subject to material volume fraction constraints placed on the matrix and inclusion phases separately, as given in \eqref{eq:constraint_equations}.
Note that in this section we no longer have the benefit of a ground truth phenomenological model with which we can compare our results.
For a fair evaluation of any potential benefits of the two-scale optimization, i.e. optimizing both the pseudo density field $\rho$ and the inclusion phase volume fraction $\al$, we compare the results with a baseline case in which the microstructure is fixed (i.e. $\al = \al_{0}$ is fixed) while also satisfying the same material volume fraction constraints.
In both cases, we employ the microstructure dependent ML model $\CM^{m}$ to represent the hyperelastic composite.
The value of $\al_{0}$ for the baseline case in each example is set based on the side constraint bounds on the total volume fraction of the inclusion phase $(g^{\mathrm{inc}}_{\max})$ and matrix phase $(g^{\mathrm{mat}}_{\max})$, i.e. $\al_{0} = g^{\mathrm{inc}}_{\max}/(g^{\mathrm{inc}}_{\max} + g^{\mathrm{mat}}_{\max})$.
Similarly the total material volume fraction for the baseline case is set to $g_{\max, 0} = g^{\mathrm{inc}}_{\max} + g^{\mathrm{mat}}_{\max}$.

Two examples are considered in this section (the T-bracket and the cantilever beam) which can be compared with the single scale topology optimization results.
The geometry, finite element discretization and the boundary conditions for these examples are discussed in Section \ref{sec:singleTO_main_text}.
We remark that \ref{appsubsec:portal_frame_example_benchmark} contains an additional multiscale TO example for the portal frame boundary value problem.

\subsubsection{Multi-scale topology optimization: T-bracket example}

The T bracket example is subject to total volume fraction side constraint bounds of $g^{\mathrm{mat}}_{\max} = 0.3$ and $g^{\mathrm{inc}}_{\max} = 0.2$.
Thereofre, the baseline case with fixed microstructure has a fixed inclusion volume fraction $\al_{0} = 0.4$ and a total volume fraction constraint of $g_{\max, 0} = 0.5$.
The convergence history of the objective function for this example is shown in Figure \ref{fig:t_bracket_micro_bvp}c.
Consistent with intuition, the objective value for the case in which microstructural variations are allowed is indeed lower.
From the final design with varying microstructure (Figure \ref{fig:t_bracket_micro_bvp}a) and the difference plot with the baseline case with fixed microstructure (Figure \ref{fig:t_bracket_micro_bvp}b), we observe that the number of macroscopic voids remains unchanged, but their shapes change.
We also notice that the material addition is occurring close to regions chosen to have lower inclusion volume fraction $\al$ by the optimizer and the material removal is made in regions dominated by higher inclusion volume fractions.

\begin{figure}[t!]
    \begin{minipage}{0.4\linewidth}
    \centering        
    \def\svgwidth{1.19\linewidth}
    {\scriptsize
    \vspace*{-70pt}
\begingroup%
  \makeatletter%
  \providecommand\color[2][]{%
    \errmessage{(Inkscape) Color is used for the text in Inkscape, but the package 'color.sty' is not loaded}%
    \renewcommand\color[2][]{}%
  }%
  \providecommand\transparent[1]{%
    \errmessage{(Inkscape) Transparency is used (non-zero) for the text in Inkscape, but the package 'transparent.sty' is not loaded}%
    \renewcommand\transparent[1]{}%
  }%
  \providecommand\rotatebox[2]{#2}%
  \newcommand*\fsize{\dimexpr\f@size pt\relax}%
  \newcommand*\lineheight[1]{\fontsize{\fsize}{#1\fsize}\selectfont}%
  \ifx\svgwidth\undefined%
    \setlength{\unitlength}{1023.50097656bp}%
    \ifx\svgscale\undefined%
      \relax%
    \else%
      \setlength{\unitlength}{\unitlength * \real{\svgscale}}%
    \fi%
  \else%
    \setlength{\unitlength}{\svgwidth}%
  \fi%
  \global\let\svgwidth\undefined%
  \global\let\svgscale\undefined%
  \makeatother%
  \begin{picture}(1,0.46825155)%
    \lineheight{1}%
    \setlength\tabcolsep{0pt}%
    \put(0,0){\includegraphics[width=\unitlength,page=1]{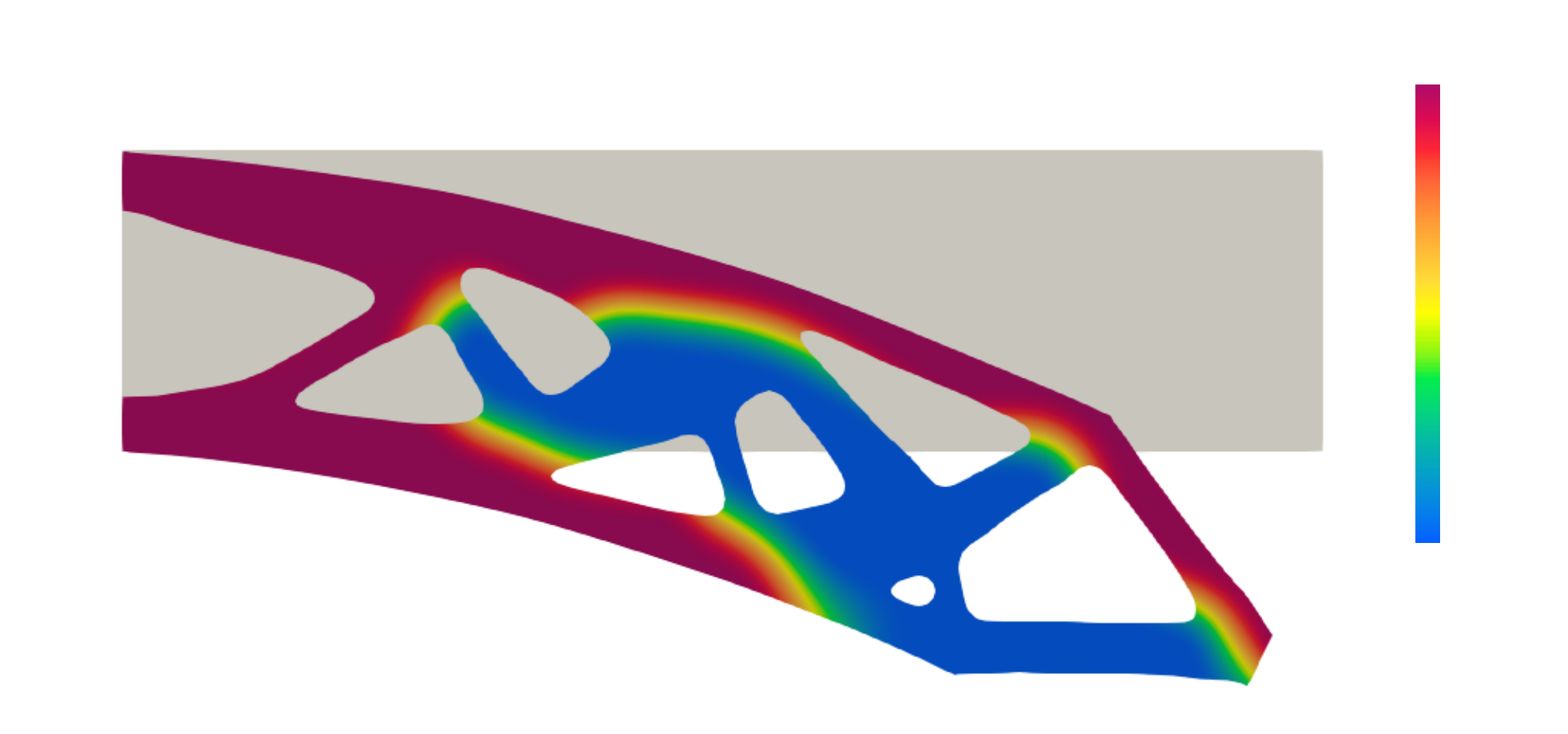}}%
    \put(0.93230184,0.12507591){\color[rgb]{0,0,0}\makebox(0,0)[lt]{\lineheight{1.25}\smash{\begin{tabular}[t]{l}0.1\end{tabular}}}}%
    \put(0.93230184,0.41721047){\color[rgb]{0,0,0}\makebox(0,0)[lt]{\lineheight{1.25}\smash{\begin{tabular}[t]{l}0.5\end{tabular}}}}%
    \put(0.93230184,0.19737677){\color[rgb]{0,0,0}\makebox(0,0)[lt]{\lineheight{1.25}\smash{\begin{tabular}[t]{l}0.2\end{tabular}}}}%
    \put(0.93230184,0.27065467){\color[rgb]{0,0,0}\makebox(0,0)[lt]{\lineheight{1.25}\smash{\begin{tabular}[t]{l}0.3\end{tabular}}}}%
    \put(0.93230184,0.34393257){\color[rgb]{0,0,0}\makebox(0,0)[lt]{\lineheight{1.25}\smash{\begin{tabular}[t]{l}0.4\end{tabular}}}}%
    \put(0.915,0.04){\color[rgb]{0,0,0}\makebox(0,0)[t]{\lineheight{1.25}\smash{\begin{tabular}[t]{c}$\al$\end{tabular}}}}%
    \put(0,0){\includegraphics[width=\unitlength,page=2]{cantilever_final_design_micro.pdf}}%
  \end{picture}%
\endgroup%

    \vspace*{-10pt}
    \subcaption*{(a)}
    }
    \end{minipage}
    \hspace{0.08\linewidth}
    \begin{minipage}{0.30\linewidth}
        \centering
        \def\svgwidth{\linewidth}
        \setlength{\figW}{8cm}
        \setlength{\figH}{6cm}
        {\scriptsize
\begin{tikzpicture}

\definecolor{gray}{RGB}{128,128,128}
\definecolor{lightgray204}{RGB}{204,204,204}
\definecolor{mediumblue}{RGB}{0,0,205}
\pgfplotsset{grid style={dashed,lightgray204}}

\begin{axis}[
height=\figH,
legend cell align={left},
legend style={fill opacity=0.8, draw opacity=1, text opacity=1, draw=lightgray204},
tick align=outside,
tick pos=left,
width=\figW,
xlabel={Optimization Iteration Number},
xmajorgrids,
xmin=-9.7, xmax=203.7,
xtick = {0, 194},
ytick = {-9.6, -18.19, -12.93,-17.26},
xtick style={color=black},
ylabel={Objective},
ymajorgrids,
ymin=-18.62385, ymax=-8.2,
ytick style={color=black}
]
\addplot [very thick, magenta]
table {%
0 -9.597
1 -11.515
2 -13.803
3 -15.453
4 -16.258
5 -16.642
6 -16.866
7 -17.028
8 -17.223
9 -17.449
10 -17.694
11 -17.789
12 -17.858
13 -17.921
14 -17.979
15 -18.036
16 -18.079
17 -18.113
18 -18.138
19 -18.161
20 -18.178
21 -18.189
22 -18.194
23 -14.593
24 -14.692
25 -14.783
26 -14.825
27 -14.9
28 -14.989
29 -15.069
30 -15.123
31 -15.154
32 -15.177
33 -15.191
34 -15.214
35 -15.246
36 -15.272
37 -15.308
38 -15.339
39 -15.366
40 -15.401
41 -15.435
42 -15.465
43 -15.493
44 -15.523
45 -15.554
46 -15.584
47 -15.612
48 -15.638
49 -15.66
50 -15.68
51 -15.697
52 -15.711
53 -15.721
54 -15.729
55 -14.046
56 -14.175
57 -14.269
58 -14.33
59 -14.377
60 -14.416
61 -14.45
62 -14.479
63 -14.503
64 -14.523
65 -14.54
66 -14.556
67 -14.571
68 -14.584
69 -14.596
70 -14.608
71 -14.619
72 -14.629
73 -14.639
74 -14.649
75 -13.74
76 -13.773
77 -13.796
78 -13.814
79 -13.833
80 -13.852
81 -13.871
82 -13.887
83 -13.9
84 -13.911
85 -13.922
86 -13.931
87 -13.941
88 -13.95
89 -13.958
90 -13.966
91 -13.973
92 -13.98
93 -13.986
94 -13.992
95 -14.517
96 -14.39
97 -14.398
98 -14.403
99 -14.407
100 -14.41
101 -14.412
102 -14.415
103 -14.417
104 -14.419
105 -14.42
106 -14.422
107 -14.424
108 -14.425
109 -14.427
110 -14.428
111 -14.429
112 -14.43
113 -14.431
114 -14.432
115 -15.529
116 -15.315
117 -15.334
118 -15.343
119 -15.35
120 -15.356
121 -15.361
122 -15.365
123 -15.369
124 -15.371
125 -15.374
126 -15.375
127 -15.377
128 -15.379
129 -15.38
130 -15.381
131 -15.383
132 -15.384
133 -15.385
134 -15.386
135 -16.507
136 -16.365
137 -16.391
138 -16.403
139 -16.412
140 -16.42
141 -16.426
142 -16.432
143 -16.437
144 -16.441
145 -16.444
146 -16.447
147 -16.45
148 -16.452
149 -16.453
150 -16.455
151 -16.457
152 -16.458
153 -16.459
154 -16.461
155 -17.048
156 -17.01
157 -17.021
158 -17.027
159 -17.032
160 -17.038
161 -17.041
162 -17.045
163 -17.047
164 -17.05
165 -17.052
166 -17.054
167 -17.055
168 -17.056
169 -17.057
170 -17.057
171 -17.056
172 -17.058
173 -17.061
174 -17.063
175 -17.254
176 -17.246
177 -17.249
178 -17.251
179 -17.252
180 -17.253
181 -17.254
182 -17.254
183 -17.255
184 -17.256
185 -17.257
186 -17.257
187 -17.258
188 -17.259
189 -17.259
190 -17.26
191 -17.26
192 -17.26
193 -17.259
194 -17.261
};
\addlegendentry{ML: varying microstructure}
\addplot [very thick, mediumblue, dash pattern=on 4pt off 2pt]
table {%
0 -9.597
1 -10.743
2 -12.119
3 -13.027
4 -13.363
5 -13.508
6 -13.572
7 -13.6
8 -13.607
9 -13.633
10 -13.637
11 -13.645
12 -13.652
13 -13.653
14 -13.663
15 -13.666
16 -13.662
17 -13.668
18 -13.672
19 -13.671
20 -11.101
21 -11.263
22 -11.409
23 -11.456
24 -11.506
25 -11.555
26 -11.608
27 -11.65
28 -11.675
29 -11.695
30 -11.711
31 -11.723
32 -11.733
33 -11.742
34 -11.751
35 -11.76
36 -11.77
37 -11.78
38 -11.79
39 -11.804
40 -9.806
41 -9.8322
42 -9.8987
43 -10.024
44 -10.166
45 -10.277
46 -10.352
47 -10.406
48 -10.451
49 -10.496
50 -10.544
51 -10.595
52 -10.647
53 -10.698
54 -10.745
55 -10.791
56 -10.835
57 -10.878
58 -10.918
59 -10.955
60 -10.993
61 -11.032
62 -11.068
63 -11.098
64 -11.121
65 -11.137
66 -11.147
67 -11.155
68 -11.16
69 -11.166
70 -10.389
71 -10.418
72 -10.448
73 -10.474
74 -10.501
75 -10.52
76 -10.533
77 -10.544
78 -10.556
79 -10.569
80 -10.583
81 -10.596
82 -10.608
83 -10.619
84 -10.627
85 -10.633
86 -10.638
87 -10.643
88 -10.647
89 -10.65
90 -11.003
91 -10.938
92 -10.922
93 -10.933
94 -10.942
95 -10.946
96 -10.947
97 -10.948
98 -10.949
99 -10.95
100 -10.951
101 -10.95
102 -10.952
103 -10.953
104 -10.954
105 -10.954
106 -10.955
107 -10.955
108 -10.956
109 -10.956
110 -11.713
111 -11.585
112 -11.6
113 -11.604
114 -11.607
115 -11.61
116 -11.613
117 -11.615
118 -11.616
119 -11.618
120 -11.619
121 -11.621
122 -11.622
123 -11.623
124 -11.624
125 -11.624
126 -11.619
127 -11.621
128 -11.619
129 -11.625
130 -12.404
131 -12.313
132 -12.328
133 -12.334
134 -12.339
135 -12.343
136 -12.346
137 -12.35
138 -12.353
139 -12.356
140 -12.358
141 -12.361
142 -12.363
143 -12.358
144 -12.346
145 -12.36
146 -12.366
147 -12.369
148 -12.37
149 -12.369
150 -12.76
151 -12.745
152 -12.758
153 -12.764
154 -12.769
155 -12.772
156 -12.765
157 -12.742
158 -12.769
159 -12.777
160 -12.78
161 -12.738
162 -12.746
163 -12.775
164 -12.782
165 -12.785
166 -12.786
167 -12.788
168 -12.758
169 -12.749
170 -12.909
171 -12.909
172 -12.914
173 -12.916
174 -12.919
175 -12.912
176 -12.895
177 -12.915
178 -12.919
179 -12.922
180 -12.923
181 -12.925
182 -12.925
183 -12.915
184 -12.889
185 -12.915
186 -12.923
187 -12.926
188 -12.928
189 -12.929
};
\addlegendentry{ML: fixed microstructure}
\end{axis}

\end{tikzpicture}
        \vspace{-10pt}
        \subcaption*{(c)}
        }
    \end{minipage}

    \begin{minipage}{0.4\linewidth}
        \def\svgwidth{1.19\linewidth}
        {\scriptsize
        \vspace{-60pt}
\begingroup%
  \makeatletter%
  \providecommand\color[2][]{%
    \errmessage{(Inkscape) Color is used for the text in Inkscape, but the package 'color.sty' is not loaded}%
    \renewcommand\color[2][]{}%
  }%
  \providecommand\transparent[1]{%
    \errmessage{(Inkscape) Transparency is used (non-zero) for the text in Inkscape, but the package 'transparent.sty' is not loaded}%
    \renewcommand\transparent[1]{}%
  }%
  \providecommand\rotatebox[2]{#2}%
  \newcommand*\fsize{\dimexpr\f@size pt\relax}%
  \newcommand*\lineheight[1]{\fontsize{\fsize}{#1\fsize}\selectfont}%
  \ifx\svgwidth\undefined%
    \setlength{\unitlength}{1023.50097656bp}%
    \ifx\svgscale\undefined%
      \relax%
    \else%
      \setlength{\unitlength}{\unitlength * \real{\svgscale}}%
    \fi%
  \else%
    \setlength{\unitlength}{\svgwidth}%
  \fi%
  \global\let\svgwidth\undefined%
  \global\let\svgscale\undefined%
  \makeatother%
  \begin{picture}(1,0.46825155)%
    \lineheight{1}%
    \setlength\tabcolsep{0pt}%
    \put(0,0){\includegraphics[width=\unitlength,page=1]{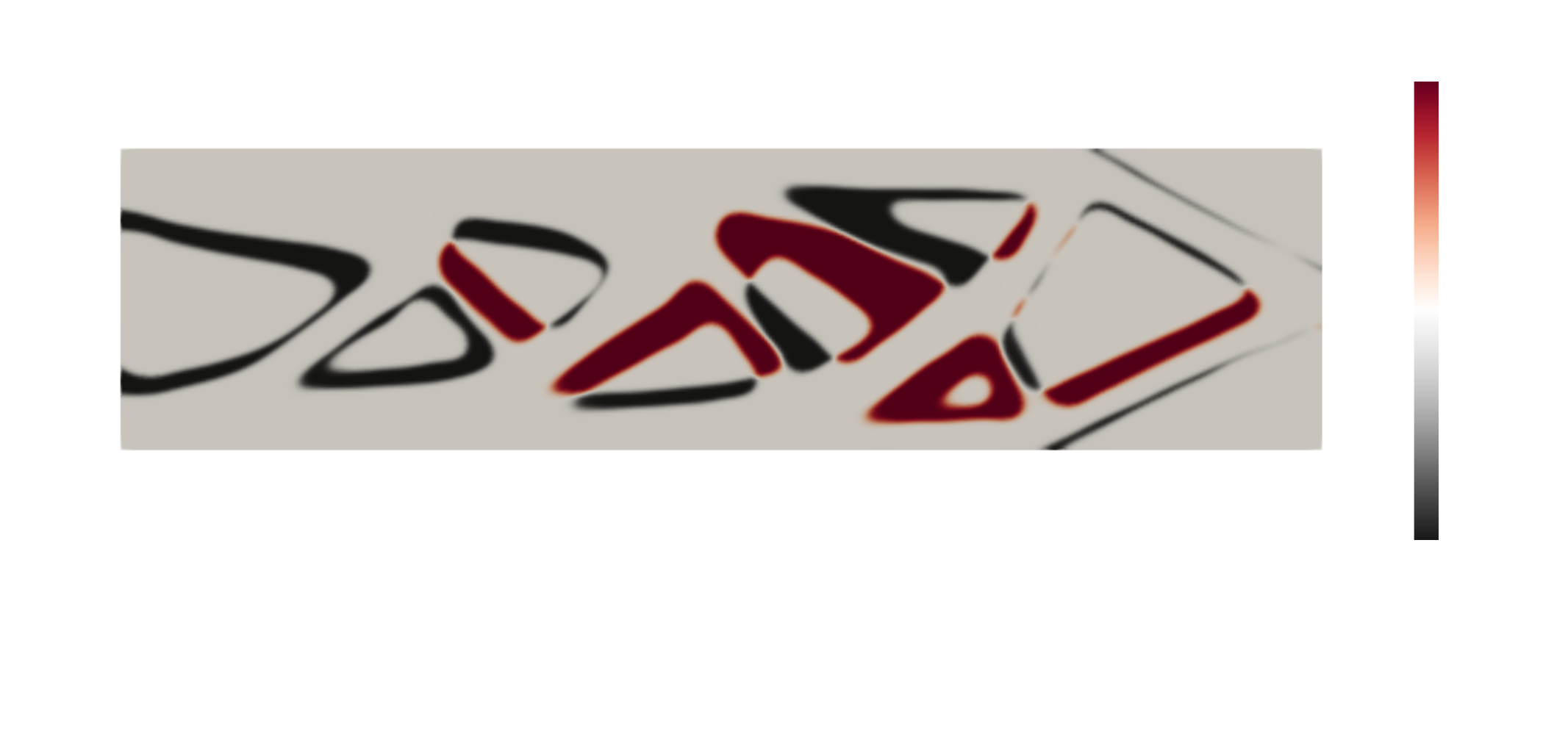}}%
    \put(0.93142461,0.12692587){\color[rgb]{0,0,0}\makebox(0,0)[lt]{\lineheight{1.25}\smash{\begin{tabular}[t]{l}-1.\end{tabular}}}}%
    \put(0.93142461,0.41906042){\color[rgb]{0,0,0}\makebox(0,0)[lt]{\lineheight{1.25}\smash{\begin{tabular}[t]{l}1.\end{tabular}}}}%
    \put(0.93142461,0.27250463){\color[rgb]{0,0,0}\makebox(0,0)[lt]{\lineheight{1.25}\smash{\begin{tabular}[t]{l}0.\end{tabular}}}}%
    \put(0.915,0.04){\color[rgb]{0,0,0}\makebox(0,0)[t]{\lineheight{1.25}\smash{\begin{tabular}[t]{c}$\de_{\rh}$\end{tabular}}}}%
    \put(0,0){\includegraphics[width=\unitlength,page=2]{cantilever_final_design_uniform_micro_mse.pdf}}%
  \end{picture}%
\endgroup%

        \vspace*{-15pt}
        \subcaption*{(b)}
        }
    \end{minipage}
    
    \caption{(a) Deformed configuration of the final design obtained using the ML model with varying microstructure (b) Difference plot of the design obtained with varying microstructure to the design obtained with fixed microstructure. (c) Convergence history for the cantilever TO problem with varying microstructure vs. fixed microstructure}
    \label{fig:cantilever_micro_bvp}
\end{figure}

\subsubsection{Multi-scale topology optimization: Cantilever beam example}
The final example considered here corresponds to the cantilever beam.
Consistent with the example in \ref{subsubsec:cantilever_beam_example_benchmark}, the volume fraction upper bounds for the matrix and inclusion phases are set to $g^{\mathrm{mat}}_{\max} = 0.4$ and $g^{\mathrm{inc}}_{\max} = 0.2$, corresponding to a total volume fraction upper bound of $g_{\max, 0} = 0.6$ for the baseline case with fixed microstructure. 
The fixed microstructure inclusion volume fraction that satisfies the material volume fraction constraint is then set to $\al_{0} = \frac{1}{3}$.
It can be seen from the final design with varying microstructure (Figure \ref{fig:cantilever_micro_bvp}a) and the difference plot with the baseline case (Figure \ref{fig:cantilever_micro_bvp}b) that the former achieves an improved objective value by adjusting the internal material-void distribution in addition to the variation of the inclusion volume fraction $\al$.
Again, the material addition predominantly occurs close to regions of lower inclusion volume fraction $\al$ whereas material removal happens close to regions with higher inclusion volume fraction. Similar to the previous examples, we observe a lower objective value allowing the microstructure to vary, demonstrating the added benefit of the proposed approach.

\section{Conclusions}
\label{sec:conclusions}

A framework is presented for designing multiscale heterogeneous structures with spatially varying hyperelastic microstructures.
Central to this work is the ML material model, which captures the homogenized response of the microstructure as a differentiable function of microstructural descriptors while adhering to important physical principles.

Our results show that ensuring polyconvexity, among other relevant restrictions, not only results in consistent ML models of high accuracy, but also produces reliable final designs in the topology optimization process.
An equally important benefit from adhering to these physical principles is the guarantee of the existence of solutions in the forward problem.
The differentiable mapping of the microstructural descriptors to the homogenized response enables the computation of the tangent stiffness, as well as the sensitivities of the relevant objective/constraint functions with respect to the design variables, facilitating efficient, simultaneous design optimization at two spatial scales.
The presented approach provides an effective method for designing future functionally-graded structures and materials with impactful microstructural details.
Our findings indicate that two-scale optimization can significantly improve performance compared to the baseline case with spatially uniform microstructure.
The examples presented produce optimized designs that align with our intuition and demonstrate that expanding the design space to multiple scales allows the optimizer to achieve higher design performance.

The proposed framework is general and can be extended to include additional microstructural descriptors, such as the orientation of microstructural phases to capture anisotropy in the homogenized response.
Future work will focus on this extension using partial input convex neural networks \citep{chen2018opti}, while also considering manufacturability and variability in loading conditions.
An extension of the consistent ML to handle path and history-dependent material for topology optimization is also envisioned wherein we also anticipate potential acceleration of the optimization process due to the absence of Newton-Raphson iterations \citep{mozaffar2019deep}.

\paragraph{Acknowledgements.}
H.V. and M.A.B. gratefully acknowledge the financial support from the European Union’s Horizon 2020 research and innovation programme under the Marie Sklodowska-Curie grant agreement `LIGHTEN  -- 
Ultralight membrane structures toward a sustainable environment’ H2020-MSCA-ITN-2020-LIGHTEN-956547.

\bibliographystyle{layout/elsarticle-harv} 


\pagebreak

\appendix

\section{Modification to ICNN for modeling hyperelasticity}
\label{section:appendix_A}

Here, we recall three fundamental facts about convexity from \citet{boyd2004conv} that are useful within the context of input convex neural networks.

\begin{enumerate}
\item If $f_{i}: \bR^{n} \to \bR$ are convex functions and $c_{i} \geq 0$, then the function $g: \bR^{n} \to \bR$ defined as:
\begin{equation}
    \label{eq:non_negative_sum}
    g(\by) = \sum_{i=1}^{m} c_{i}f_{i}(\by), \quad \mathbf{dom}\,g = \bigcap_{i=1}^{m} \mathbf{dom}\,f_{i}
\end{equation}
is convex.
\item The composition of a convex function with an affine mapping preserves convexity.
If the function $f: \bR^{n} \to \bR$ is a convex function, then a function $g: \bR^{m} \to \bR$ defined as 
\begin{equation}
    \label{eq:affine_composition}
    g(\by) = f(\bW \by + \bb), \quad \mathbf{dom}\,g = \{\by \, |\, \bW \by + \bb \in \mathbf{dom}\, f\} 
\end{equation}
is also convex, with $\bW \in \bR^{n\times m}$, $\bb \in \bR^{n}$.
\item The composition of a convex function with a convex non-decreasing function is convex.
If $h: \bR^{m} \to \bR^{n}$ is a convex function and $f: \bR^{n} \to \bR$ is a convex non-decreasing function, then the composition $g: \bR^{m} \to \bR$ defined as:
\begin{equation}
    \label{eq:convex_composition}
    g(\by) = f(h(\by)), \quad \mathbf{dom}\,g = \{\by \in \, \mathbf{dom}\,h\, |\, h(\by) \in \mathbf{dom}\, f\}
\end{equation}
is convex.
\end{enumerate}

\begin{figure}[ht]
	\begin{minipage}{\linewidth}
	\centering        
	\def\svgwidth{0.8\linewidth}
	{\footnotesize
	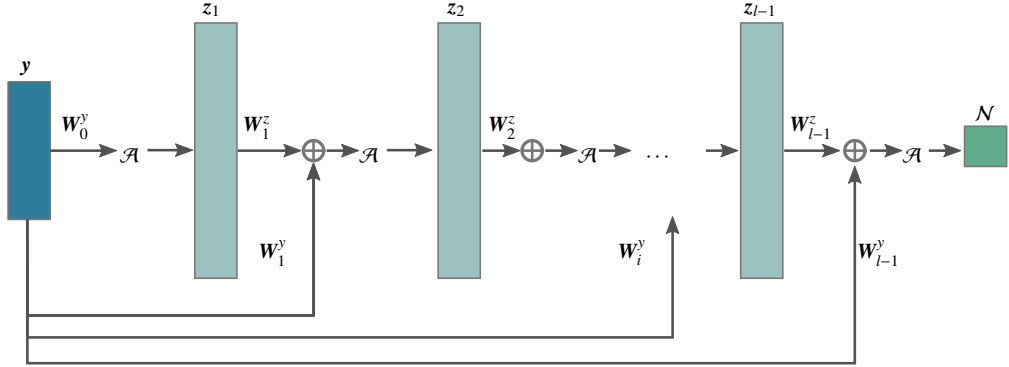
	}
	\caption{Input Neural Network architecture}
	\label{fig:A01icnn_model}
	\end{minipage}
\end{figure}

The ICNN proposed by \citet{amos2016inpu} uses \eqref{eq:affine_composition} together with \eqref{eq:convex_composition} in the first layer to achieve convexity.
In each of the subsequent layers \eqref{eq:convex_composition} is used together with \eqref{eq:non_negative_sum} for each core sub-layer, and \eqref{eq:affine_composition} for each passthrough sub-layer.
Its mathematical definition for general vectorial input is defined as $f(\by; \Bth):$

\begin{equation}
	\label{eq:icnn_amos}
	\begin{split}
	\bz_{i+1} &= \CA_{i}(\bW_{i}^{(z)}\bz_{i} + \bW_{i}^{(y)}\by + \bb_{i}) \quad \text{for} \quad i = 0, \ldots, l-1 \\
	f(\by, \Bth) &= \bz_{l}
	\end{split}
\end{equation}
where $\bz_{i}$ denotes the layers (with $\bz_{0}, \bW_{0}^{z} \equiv \Bzero$) to which convex and non-decreasing element-wise activation functions $\CA_{i}(\cdot)$ are applied. The learnable parameters are $\Bth= \{ \bW_{0:l-1}^{(z)}, \bW_{1:l-1}^{(y)}, \bb_{0:L-1} \}$ where $\bW_{0:l-1}^{(z)}$ are non-negative.
For ensuring C2 continuity, the activation function $\CA_{i}(\cdot)$ is chosen to be the softplus function and the positive weights are enforced by wrapping the weights $\bW_{i}^{(z)}$ in with a squared softplus function, following \citet{asad2022ame}.

As noted in \ref{subsec:hyperelasticity_conditions}, one basic requirement in hyperelastic material modeling is that the strain energy density function $\ps$ is polyconvex. Namely,
\begin{equation}
\tilde{\ps}(\bF) = \CG(\bF, \mathrm{cof}\,\bF, \det \bF) \quad
\end{equation}
where $\CG(\cdot)$ is a convex function in $(\bF, \mathrm{cof}\,\bF, \det \bF)$. In conventional constitutive modeling, the strain energy density function is often expressed as a (convex) function of the invariants of an objective strain measure.
The classic example of this corresponds to the invariants $\Bio = \{I_{C1},I_{C2},I_{C3} \}$  of the the right Cauchy-Green tensor, $\bC = \bF^{\mathsf{T}}\bF$, since they satisfy polyconvexity and have desirable symmetry features.
However, such invariants cannot be used as input in the original ICNN implementation.
The first layer and the passthrough layers involve an \textit{unrestricted}  transformation \eqref{eq:affine_composition} with weights that can be positive or negative, which when applied to the invariants, does not preserve convexity in the original parameters.
Therefore, if it is desired to use the invariants, the treatment should be in accordance with the convex non-decreasing composition \eqref{eq:convex_composition}, which implies that the transformation \eqref{eq:affine_composition} should be restricted to non-negative weights.
However, if the invariant $J = \sqrt{I_{C3}} = \det \bF$ is used instead of $I_{C3}$, there need not be any restrictions on the specific weights mapping $J$.
In effect, when using invariants as input, the ICNN should be modified such that it results in a convex function that is convex non-decreasing in the invariants.

A similar argument holds when using the principal stretches or any of its convex combinations as input, as shown in \ref{section:appendix_B}.
The only exception to this is the invariant $J$, since the strain energy density function need not be non-decreasing in $J$.
The modified version of the ICNN per \citet{chen2018opti} provides a straightforward implementation for handling convexity and ensuring that the output is non-decreasing in particular inputs.
In this definition, an ICNN is expressed as a function, $f(\by; \Bth)$ defined by:
\begin{equation}
	\label{eq:icnn_chen}
	\begin{split}
	\bz_{i+1} &= \CA_{i}(\bW_{i}^{(z)}\bz_{i} + \bW_{i}^{(\hat{y})}\hat{\by} + \bb_{i}) \quad \text{for} \quad i = 0, \ldots, l-1 \\
	f(\hat{\by}, \Bth) &= \bz_{l}
	\end{split}
\end{equation}
where $\hat{\by}$ represents the expanded input $\hat{\by} =[ \by, -\by ]^\mathsf{T}$,   $\bz_{i}$ denotes the layers (with $\bz_{0}, \bW_{0}^{z} \equiv \Bzero$) to which convex and non-decreasing element-wise activation functions $\CA_{i}(\cdot)$ are applied, and $\Bth= \{ \bW_{0:l-1}^{(z)}, \bW_{1:l-1}^{(\hat{y})}, \bb_{0:L-1} \} $ denotes the learnable parameters where $\bW_{0:l-1}^{(z)}$ and $\bW_{0:l-1}^{(\hat{y})}$ are non-negative.
Since all layers, including the passthrough layers, have non-negative weights, if the expanded input is not used (i.e. $\hat{\by} \equiv \by$), the output of the ICNN as defined by \citet{chen2018opti} becomes convex and non-decreasing in the inputs.
By \textit{not} using the expanded inputs, except for $J$, the ICNN \eqref{eq:icnn_chen} can be used to model hyperelastic materials with strain energy density functions that are convex  in $J$ and convex non-decreasing in the remaining inputs.

Amongst the ICNN based neural network models for hyperelasticity present in literature, \citet{linden2023neur} and their subsequent works have addressed this aspect by having non-negative weights overall, however the authors omit the skip connections presented in \citet{chen2018opti}.

\section{Limitations of invariant-based ICNN models}
\label{section:appendix_B}

When choosing a set of invariants as inputs, an ICNN should be convex, non-decreasing in those invariants (except for $J$, as $\ps$ need not be non-decreasing in $J$) c.f. \ref{section:appendix_A}.
We theorize that such an ICNN would be limited in its ability to model an arbitrary polyconvex strain energy density function $\ps$.

Here, we limit the current discussion to isotropic case where the conditions discussed in \eqref{eq:polyconvexity_condition_isotropic} hold.
We aim to provide a simple counter example to show that a strain energy density in accordance with \eqref{eq:polyconvexity_condition_isotropic} cannot always be represented by a composition of convex and non-decreasing functions of pre-chosen invariants (convex in $J$).
Consider the Ogden model \citep{holzapfel2002nonl} which has a strain energy density function given by:
\begin{equation}
    \label{eq:ogden_model}
    \begin{split}
    \ps_{O}(\bF) &= \ps_\mathrm{iso}(\bar{\la}_{1}, \bar{\la}_{2}, \bar{\la}_{3}) + \ps_\mathrm{vol}(J) \\
    &=\Big\{\sum_{i=1}^{N} \frac{\mu_{i}}{a_{i}}(\bar{\la}_{1}^{a_{i}} + \bar{\la}_{2}^{a_{i}} + \bar{\la}_{3}^{a_{i}} - 3)\Big \} + \frac{\ka}{\be^{2}}(\be \ln J + J^{-\be} -1)
    \end{split}
\end{equation}

We adopt $\beta = -2$ as in \citet{simo1992asso}. For the sake of simplicity, we consider the case of $N=1$, $a_{1}= a = 6$, and replace $\mu_{1} = \mu$.
Note that $\bar{\la}_{i} = J^{-1/3} \la_{i}$ and $J = \la_{1}\la_{2}\la_{3}$. Then we have:

\begin{equation}
    \label{eq:ogden_model_simple}
    \begin{split}
        \ps_{O}(\bF) &= \frac{\mu}{6}(\bar{\la}_{1}^{6} + \bar{\la}_{2}^{6} + \bar{\la}_{3}^{6} - 3) + \frac{\ka}{4}(J^{2} - 1 - 2\ln J)\\
    &= \frac{\mu}{6}(J^{-2}(\la_{1}^{6} + \la_{2}^{6} + \la_{3}^{6}) - 3) + \frac{\ka}{4}(J^{2} - 1 - 2\ln J)
    \end{split}
\end{equation}

This model is polyconvex by definition in \eqref{eq:polyconvexity_condition_isotropic}, however this can also be verified with a simple test.
The strain energy density \eqref{eq:ogden_model_simple} should be convex non-decreasing in $\{ \la_{1}, \la_{2}, \la_{3} \}$ and convex in $J$. 
Let us obtain the derivatives of the strain energy density function \eqref{eq:ogden_model_simple} with respect to $\BLA= \{ \la_{1}, \la_{2}, \la_{3}, J \}$:

\begin{equation}
    \label{eq:ogden_model_simple_derivative}
    \nabla_{\BLA} \ps_{O} = \begin{bmatrix}
    \mu J^{-2} \la_{1}^{5} \\
    \mu J^{-2} \la_{2}^{5} \\
    \mu J^{-2} \la_{3}^{5} \\
    -\frac{1\mu}{3}J^{-3}(\la_{1}^{6} + \la_{2}^{6} + \la_{3}^{6}) + \frac{\ka}{2} (J - J^{-1})
    \end{bmatrix}
\end{equation}
For all $\mu >0$, the derivatives of $\{\la_{1}, \la_{2}, \la_{3}\}$ are non-negative.
Hence the strain energy density function \eqref{eq:ogden_model_simple} satisfies the non-decreasing condition in $\{\la_{1}, \la_{2}, \la_{3}\}$.

The hessian of the strain energy density function \eqref{eq:ogden_model_simple} with respect to $\LA$ is:
\begin{equation}
    \label{eq:ogden_model_simple_hessian}
    \nabla^{2}_{\BLA} \ps_{O} = \begin{bmatrix}
    5\mu J^{-2} \la_{1}^{4} & 0 & 0 & -2\mu J^{-3} \la_{1}^{5} \\
    0 & 5\mu J^{-2} \la_{2}^{4} & 0 & -2\mu J^{-3} \la_{2}^{5} \\
    0 & 0 & 5\mu J^{-2} \la_{3}^{4} & -2\mu J^{-3} \la_{3}^{5} \\
    -2\mu J^{-3} \la_{1}^{5} & -2\mu J^{-3} \la_{2}^{5} & -2\mu J^{-3} \la_{3}^{5} &  \mu J^{-4}(\la_{1}^{6} + \la_{2}^{6} + \la_{3}^{6}) + \frac{\ka}{2}(1 + J^{-2})
    \end{bmatrix}
\end{equation}

The principal minors $M_{11}, M_{22}, M_{33}$ are clearly positive for $\mu >0$. It is therefore sufficient that $M_{44}>0$ for the hessian \eqref{eq:ogden_model_simple_hessian} to be positive semi-definite.
Checking this, we have:

\begin{equation}
    \label{eq:ogden_model_simple_hessian_determinant}
    M_{44} = \begin{vmatrix}
        5\mu J^{-2} \la_{1}^{4} & 0 & 0 & -2\mu J^{-3} \la_{1}^{5} \\
        0 & 5\mu J^{-2} \la_{2}^{4} & 0 & -2\mu J^{-3} \la_{2}^{5} \\
        0 & 0 & 5\mu J^{-2} \la_{3}^{4} & -2\mu J^{-3} \la_{3}^{5} \\
        -2\mu J^{-3} \la_{1}^{5} & -2\mu J^{-3} \la_{2}^{5} & -2\mu J^{-3} \la_{3}^{5} &  \mu J^{-4}(\la_{1}^{6} + \la_{2}^{6} + \la_{3}^{6}) + \frac{\ka}{2}(1 + J^{-2})
    \end{vmatrix}\\
\end{equation}

which, after simplification, produces:
\begin{equation}
    \label{eq:ogden_model_simple_hessian_4th minor}
    M_{44} = \frac{25\mu^{3}}{2}J^{-6}(2 \mu(\la_{1}^{6} + \la_{2}^{6} + \la_{3}^{6}) + 5 \ka (J^{4} + J^{2}))
\end{equation}
For $\mu, \ka >0$, the minor $M_{44}$ is positive. Hence the strain energy density function \eqref{eq:ogden_model_simple} is convex in $\BLA$.

Now, let the chosen invariants be $\check{\Bio} = \{I_{C1}, I_{C2}, J\}$:
\begin{equation}
    \begin{split}
    I_{C1} & = \la_{1}^{2} + \la_{2}^{2} + \la_{3}^{2} \\
    I_{C2} & = \la_{1}^{2}\la_{2}^{2} + \la_{2}^{2}\la_{3}^{2} + \la_{3}^{2}\la_{1}^{2} \\
    J & = \la_{1}\la_{2}\la_{3}
    \end{split}
\end{equation}

 The strain energy density function \eqref{eq:ogden_model_simple} can be expressed as a function of these invariants as:

 \begin{equation}
    \label{eq:ogden_model_simple_invariants}
    \begin{split}
    \check{\ps}_{O}(\check{\Bio}) &= \frac{\mu}{6}(J^{-2}(I_{C1}^{3}- 3I_{C1}I_{C2} + 3J^{2}) - 3) + \frac{\ka}{4}(J^{2} - 1 - 2\ln J) \\
    &= \frac{\mu}{6}J^{-2}(I_{C1}^{3}- 3I_{C1}I_{C2}) + \frac{\ka}{4}(J^{2} - 1 - 2\ln J) 
    \end{split}
\end{equation}

Here, we provide the extra steps for clarity:
\begin{equation}
    \begin{split}
    I_{C1}^3&= (\la_{1}^{2} + \la_{2}^{2} + \la_{3}^{2})^{3} \\
    &= \la_{1}^{6} + \la_{2}^{6} + \la_{3}^{6} \\
    &+ 3(\la_{1}^{4}\la_{2}^{2} + \la_{1}^{2}\la_{2}^{4} +
     \la_{2}^{4}\la_{3}^{2} + \la_{2}^{2}\la_{3}^{4} +
      \la_{3}^{4}\la_{1}^{2} + \la_{3}^{2}\la_{1}^{4}) + 6\la_{1}^{2}\la_{2}^{2}\la_{3}^{2}\\
    I_{C1}I_{C2}&= (\la_{1}^{2} + \la_{2}^{2} + \la_{3}^{2})(\la_{1}^{2}\la_{2}^{2} + \la_{2}^{2}\la_{3}^{2} + \la_{3}^{2}\la_{1}^{2})\\
    &= \la_{1}^{4}\la_{2}^{2} + \la_{1}^2\la_{2}^2\la_{3}^2 + \la_{1}^{4}\la_{3}^{2}\\
    &+ \la_{2}^{4}\la_{1}^{2} + \la_{1}^2\la_{2}^{2}\la_{3}^{2} + \la_{2}^{4}\la_{3}^{2}\\
    &+ \la_{1}^2\la_{2}^2\la_{3}^2 + \la_{3}^4\la^{2} + \la_{3}^4\la_{1}^2\\
    &= (\la_{1}^{4}\la_{2}^{2} + \la_{1}^{2}\la_{2}^{4} +
     \la_{2}^{4}\la_{3}^{2} + \la_{2}^{2}\la_{3}^{4} +
      \la_{3}^{4}\la_{1}^{2} + \la_{3}^{2}\la_{1}^{4}) + 3\la_{1}^{2}\la_{2}^{2}\la_{3}^{2}\\
    3I_{C1}I_{C2} &= 3(\la_{1}^{4}\la_{2}^{2} + \la_{1}^{2}\la_{2}^{4} +
     \la_{2}^{4}\la_{3}^{2} + \la_{2}^{2}\la_{3}^{4} +
      \la_{3}^{4}\la_{1}^{2} + \la_{3}^{2}\la_{1}^{4}) + 9\la_{1}^{2}\la_{2}^{2}\la_{3}^{2}\\
    I_{C1}^{3} - 3I_{C1}I_{C2} &= \la_{1}^{6} + \la_{2}^{6} + \la_{3}^{6} -3\la_{1}^{2}\la_{2}^{2}\la_{3}^{2}\\
    3J^2 & = 3\la_{1}^{2}\la_{2}^{2}\la_{3}^{2}\\
    I_{C1}^{3} - 3I_{C1}I_{C2} + 3J^2 &= \la_{1}^{6} + \la_{2}^{6} + \la_{3}^{6}
    \end{split}
\end{equation}

The derivative of the strain energy density function \eqref{eq:ogden_model_simple_invariants} with respect to these invariants is:

\begin{equation}
    \label{eq:ogden_model_simple_invariants_derivative}
    \nabla_{\check{\Bio}} \check{\ps}_{O} = \begin{bmatrix}
    \frac{\mu}{2} J^{-2}(I_{C1}^{2} - I_{C2}) \\
    -\frac{\mu}{2} J^{-2}I_{C1} \\
    -\frac{\mu}{3}J^{-3}(I_{C1}^{3}- 3I_{C1}I_{C2}) + \frac{\ka}{2} (J - J^{-1})
    \end{bmatrix}
\end{equation}

The derivative w.r.t $I_{C2}$ is negative for all $I_{C1}$ for any $\mu>0$. This implies that the strain energy density function \eqref{eq:ogden_model_simple_invariants} is \textit{not} non-decreasing in $I_{C2}$. Differentiating again, we obtain the Hessian of the strain energy density function \eqref{eq:ogden_model_simple_invariants} with respect to these invariants:
\begin{equation}
    \label{eq:ogden_model_simple_invariants_hessian}
    \nabla^{2}_{\check{\Bio}} \check{\ps}_{O} = \begin{bmatrix}
    \mu J^{-2} I_{C1} & -\frac{\mu}{2}J^{-2} & -\mu J^{-3}(I_{C1}^{2} - I_{C2}) \\
    -\frac{\mu}{2}J^{-2} & 0 & \mu J^{-3}I_{C1}\\
    -\mu J^{-3}(I_{C1}^{2} - I_{C2}) & \mu J^{-3}I_{C1} & \mu J^{-4}(I_{C1}^{3}- 3I_{C1}I_{C2}) + \frac{\ka}{2}(1 + J^{-2})
    \end{bmatrix}
\end{equation}

The principal minors of the Hessian \eqref{eq:ogden_model_simple_invariants} are:
\begin{equation}
    \label{eq:ogden_model_simple_invariants_hessian_minors}
    \begin{split}
    M_{11} & =  \mu J^{-2} I_{C1} \ , \\
    M_{22} & =  -\frac{\mu}{4}J^{-4} <0 \ , \\
    M_{33} & = - \frac{\mu^{2}}{8}J^{-8}(2\mu(I_{C1}^{3}+ I_{C1}I_{C2}) + \ka J^{2}(1+J^{2})) \\
    & = - \Big [ \frac{\mu^{3}}{4}J^{-8}(I_{C1}^{3}+ I_{C1}I_{C2}) + \frac{\mu^{2}\ka}{8}J^{-6} (1 + J^{2}) \Big ]< 0 \ , 
    \end{split}
\end{equation}
two of which are negative for any $\mu, \ka >0$. This implies that the Hessian is \textit{not} positive semi-definite and, consequently, that that the strain energy density function \eqref{eq:ogden_model_simple_invariants} is \textit{not} convex in $I_{C1}, I_{C2}$. 

Hence we argue that an ICNN that is convex and non-decreasing in the invariants $I_{C1}, I_{C2}, J$ will not be able to represent the strain energy density function \eqref{eq:ogden_model_simple}.
On the contrary, for isotropic hyperelasticity, if we built an ICNN that satisfied \eqref{eq:polyconvexity_condition_isotropic} directly, then it would be able to represent any arbitrary isotropic hyperelastic strain energy density function, given a sufficient number of layers and neurons.
This can be achieved using an ICNN \eqref{eq:icnn_chen} with  the input $\hat{\by} = \BLA = \{ \la_{1}, \la_{2}, \la_{3}, \la_{1}\la_{2}, \la_{2}\la_{3}, \la_{3}\la_{1}, J, -J \}$ with symmetric enforcing $\bW_{i}^{(\hat{y})}$ through weight sharing for all the direct connections from the input (i.e. first layer and all the passthrough layers). Thus, $\bW_{i}^{(\la_{1})} = \bW_{i}^{(\la_{2})} = \bW_{i}^{(\la_{3})}$ and $\bW_{i}^{(\la_{1}\la_{2})} = \bW_{i}^{(\la_{2}\la_{3})} = \bW_{i}^{(\la_{3}\la_{1})}$ for $i = 0, \ldots, l-1$ in \eqref{eq:icnn_chen}.

\section{Derivation of the stress correction term for the isotropic hyperelastic neural network model}
\label{section:appendix_D}

The starting point for the derivation is due to the observation made in \citet{linden2023neur} that a stress correction term $\CN^{\text{\ stress}}_{0}$ of the form:

\begin{equation}
    \label{eq:stress_correction_term_non_polyconvex_D}
    \CN_{0}^{\ \text{stress}}(\bE) = - \partial_{\bE}\CN(\tilde{\BLA}(\bE))\Big|_{\bE = \Bzero}\colon\bE
\end{equation}

violates the polyconvexity and material symmetry conditions. As an alternative, they propose a stress correction term based on a weighted sum of derivatives method, where the wights are obtained via exploiting the chain rule and the behavior of derivative of strain invariants at undeformed state. Here we adapt this approach to the principal stretch based isotropic hyperelastic neural network model we have developed.

Recall the input to the model $\tilde{\BLA} = \{\la_{1}, \la_{2}, \la_{3}, \la_{1}\la_{2}, \la_{2}\la_{3}, \la_{3}\la_{1}, J, -J, -\ln J \}$ as proposed in Section
\ref{subsec:neural_network_for_material_modeling}. Due to the symmetry enforcement we have made, the derivative of the neural network output $\CN$ with respect to the input $\tilde{\BLA}$ respects the enforced symmetry. As, a result, we can consider a stress correction term for the strain energy function of the form:

\begin{equation}
    \label{eq:stress_correction_term_non_polyconvex_projection_D}
    \CN_{0}^{\text{\ stress}} = - \sum_{i=1}^{9} \partial_{\tilde{\LA}_{i}}\CN(\tilde{\BLA(\bE)})\Big|_{\bE= \Bzero}(\tilde{\LA}_{i} -\tilde{\LA}^{0}_{i})
\end{equation}

The corresponding stress contribution could be obtained by the chain rule as:
\begin{equation}
    \label{eq:stress_correction_term_non_polyconvex_projection_stress_D}
    \bS_{0}^{\text{\ stress}} = - \sum_{i=1}^{9} \partial_{\tilde{\LA}_{i}}\CN(\tilde{\BLA}(\bE))\Big|_{\bE= \Bzero} \partial_{\bE}\tilde{\LA}_{i}
\end{equation}

Due to symmetry enforcement, the above expression simplifies to:
\begin{equation}
    \label{eq:stress_correction_term_non_polyconvex_projection_stress_simplified_D}
    \bS_{0}^{\text{\ stress}} = - (\sum_{i=1}^{3} h_{1} \partial_{\bE}\tilde{\LA}_{i} + \sum_{j=4}^{6} h_{2} \partial_{\bE}\tilde{\LA}_{j} + h_{3} \partial_{\bE}\tilde{\LA}_{7} + h_{4} \partial_{\bE}\tilde{\LA}_{8} + h_{5} \partial_{\bE}\tilde{\LA}_{9})
\end{equation}
which when evaluated at $\bE = \Bzero$ gives:
\begin{equation}
    \label{eq:stress_correction_term_non_polyconvex_projection_stress_zero_D}
    \begin{split}
    \bS_{0}^{\text{\ stress}}\Big |_{\bE = \Bzero} &= - ( h_{1} \Bone + 2h_{2} \Bone + h_{3} \Bone - h_{4} \Bone - h_{5} \Bone) \\
    &= - ( h_{1}  + 2h_{2} + h_{3} - h_{4}  - h_{5} )\Bone  \quad \text{where} \\
    h_{1} &= \partial_{\tilde{\LA}_{1}} \CN \Big|_{\bE= \Bzero} =\partial_{\tilde{\LA}_{2}} \CN \Big|_{\bE= \Bzero} =\partial_{\tilde{\LA}_{3}} \CN \Big|_{\bE= \Bzero};\\
    h_{2} &= \partial_{\tilde{\LA}_{4}} \CN \Big|_{\bE= \Bzero} =\partial_{\tilde{\LA}_{5}} \CN \Big|_{\bE= \Bzero} =\partial_{\tilde{\LA}_{6}} \CN \Big|_{\bE= \Bzero};\\
    h_{3} &= \partial_{\tilde{\LA}_{7}}\CN \Big|_{\bE= \Bzero} ; \quad h_{4} = \partial_{\tilde{\LA}_{8}}\CN \Big|_{\bE= \Bzero} ; \quad h_{5} = \partial_{\tilde{\LA}_{9}}\CN \Big|_{\bE= \Bzero} 
    \end{split}
\end{equation}

Note that the stress contribution $\bS_{0}^{\text{\ stress}}$ coming from the correction term $\CN_{0}^{\text{\ stress}}$ is not polyconvex. However, we can exploit the fact that an affine transformation of $J$ preserves polyconvexity, and the fact that the derivative of $\partial_\bE J = J \bC^{-1}$, which when evaluated at $\bE = \Bzero$ becomes identity $\Bone$, to arrive at a polyconvex stress correction term. The polyconvex alternative that result in the requisite stress contribution at $\bE = \Bzero$ as given in \eqref{eq:stress_correction_term_non_polyconvex_projection_stress_zero_D} may be obtained by defining a stress correction term:

\begin{equation}
    \label{eq:stress_correction_term_polyconvex_projection_D}
    \CN_{0}^{\text{\ stress}} = - \sum_{i=1}^{9} \xi_{i}\partial_{\tilde{\LA}_{i}}\CN(\tilde{\BLA}(\bE))\Big|_{\bE= \Bzero} (J-1)
\end{equation}

where $\Bxi = \{\frac{1}{3}, \frac{1}{3}, \frac{1}{3}, \frac{2}{3}, \frac{2}{3}, \frac{2}{3}, 1, -1, 1\}$ are the weights that result in the requisite stress contribution as in \eqref{eq:stress_correction_term_non_polyconvex_projection_stress_zero_D}. The corresponding stress contribution can be obtained as:

\begin{equation}
    \label{eq:stress_correction_term_polyconvex_projection_stress_D}
    \bS_{0}^{\text{\ stress}} = - \sum_{i=1}^{9} \xi_{i} \partial_{\tilde{\LA}_{i}}\CN(\tilde{\BLA}(\bE))\Big|_{\bE= \Bzero} J \bC^{-1}
\end{equation}

It can be clearly seen that evaluation of \eqref{eq:stress_correction_term_polyconvex_projection_stress_D} at $\bE = \Bzero$ gives the desired stress correction term as in \eqref{eq:stress_correction_term_non_polyconvex_projection_stress_zero_D}:

\begin{equation}
    \begin{split}
    \label{eq:stress_correction_term_polyconvex_projection_stress_zero_D}
    \bS_{0}^{\text{\ stress}}\Big |_{\bE = \Bzero} &= - \sum_{i=1}^{9} \xi_{i} \partial_{\tilde{\LA}_{i}}\CN(\tilde{\BLA}(\bE))\Big|_{\bE= \Bzero} \Bone \\
    &= - \sum_{i=1}^{9} \xi_{i} h_{i} \Bone \\
    &= - ( h_{1}  + 2h_{2} + h_{3} - h_{4}  - h_{5} )\Bone
    \end{split}
\end{equation}

\section{Additional topology optimization examples assessing ML material model}
\label{app:additional_example_singleTO}

\subsection {Single scale topology optimization: Portal frame example}
\label{appsubsec:portal_frame_example_benchmark}

\begin{figure}[t!]
    \begin{minipage}{0.5\linewidth}
    \centering        
    \def\svgwidth{0.8\linewidth}
    {\footnotesize
\begingroup%
  \makeatletter%
  \providecommand\color[2][]{%
    \errmessage{(Inkscape) Color is used for the text in Inkscape, but the package 'color.sty' is not loaded}%
    \renewcommand\color[2][]{}%
  }%
  \providecommand\transparent[1]{%
    \errmessage{(Inkscape) Transparency is used (non-zero) for the text in Inkscape, but the package 'transparent.sty' is not loaded}%
    \renewcommand\transparent[1]{}%
  }%
  \providecommand\rotatebox[2]{#2}%
  \newcommand*\fsize{\dimexpr\f@size pt\relax}%
  \newcommand*\lineheight[1]{\fontsize{\fsize}{#1\fsize}\selectfont}%
  \ifx\svgwidth\undefined%
    \setlength{\unitlength}{905.6422148bp}%
    \ifx\svgscale\undefined%
      \relax%
    \else%
      \setlength{\unitlength}{\unitlength * \real{\svgscale}}%
    \fi%
  \else%
    \setlength{\unitlength}{\svgwidth}%
  \fi%
  \global\let\svgwidth\undefined%
  \global\let\svgscale\undefined%
  \makeatother%
  \begin{picture}(1,1.17986786)%
    \lineheight{1}%
    \setlength\tabcolsep{0pt}%
    \put(0,0){\includegraphics[width=\unitlength,page=1]{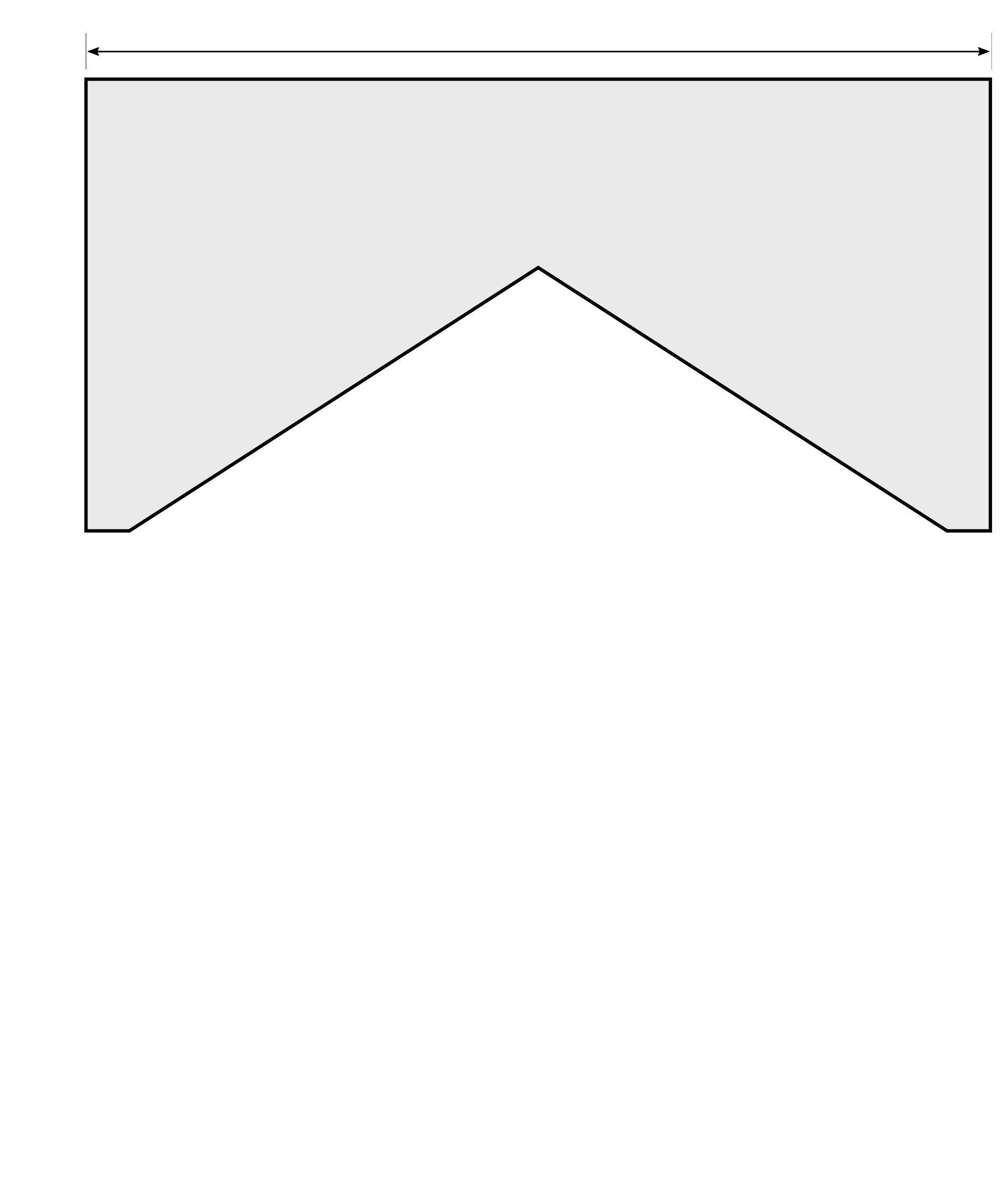}}%
    \put(0.52540057,1.14105735){\color[rgb]{0,0,0}\makebox(0,0)[lt]{\lineheight{1.25}\smash{\begin{tabular}[t]{l}$2L$\end{tabular}}}}%
    \put(0,0){\includegraphics[width=\unitlength,page=2]{portal_frame_bvp.pdf}}%
    \put(0.01253182,0.86573572){\color[rgb]{0,0,0}\makebox(0,0)[lt]{\lineheight{1.25}\smash{\begin{tabular}[t]{l}$L$\end{tabular}}}}%
    \put(0,0){\includegraphics[width=\unitlength,page=3]{portal_frame_bvp.pdf}}%
    \put(0.48,0.77324162){\color[rgb]{0,0,0}\makebox(0,0)[lt]{\lineheight{1.25}\smash{\begin{tabular}[t]{l}$\frac{7L}{12}$\end{tabular}}}}%
    \put(0.15000296,0.58){\color[rgb]{0,0,0}\makebox(0,0)[lt]{\lineheight{1.25}\smash{\begin{tabular}[t]{l}$\frac{11L}{121}$\end{tabular}}}}%
    \put(0.59101025,1.02){\color[rgb]{0,0,0}\makebox(0,0)[lt]{\lineheight{1.25}\smash{\begin{tabular}[t]{l}$\frac{L}{6}$\end{tabular}}}}%
    \put(0,0){\includegraphics[width=\unitlength,page=4]{portal_frame_bvp.pdf}}%
  \end{picture}%
\endgroup%

    \subcaption*{(a)}
    }
    \end{minipage}
    \hspace{-0.04\linewidth}
    \begin{minipage}{0.5\linewidth}
        \centering        
        \def\svgwidth{0.952\linewidth}
        \vspace{-112pt}
        {\scriptsize
        \vspace*{-11pt}
\begingroup%
  \makeatletter%
  \providecommand\color[2][]{%
    \errmessage{(Inkscape) Color is used for the text in Inkscape, but the package 'color.sty' is not loaded}%
    \renewcommand\color[2][]{}%
  }%
  \providecommand\transparent[1]{%
    \errmessage{(Inkscape) Transparency is used (non-zero) for the text in Inkscape, but the package 'transparent.sty' is not loaded}%
    \renewcommand\transparent[1]{}%
  }%
  \providecommand\rotatebox[2]{#2}%
  \newcommand*\fsize{\dimexpr\f@size pt\relax}%
  \newcommand*\lineheight[1]{\fontsize{\fsize}{#1\fsize}\selectfont}%
  \ifx\svgwidth\undefined%
    \setlength{\unitlength}{1080.66381836bp}%
    \ifx\svgscale\undefined%
      \relax%
    \else%
      \setlength{\unitlength}{\unitlength * \real{\svgscale}}%
    \fi%
  \else%
    \setlength{\unitlength}{\svgwidth}%
  \fi%
  \global\let\svgwidth\undefined%
  \global\let\svgscale\undefined%
  \makeatother%
  \begin{picture}(1,0.4800122)%
    \lineheight{1}%
    \setlength\tabcolsep{0pt}%
    \put(0,0){\includegraphics[width=\unitlength,page=1]{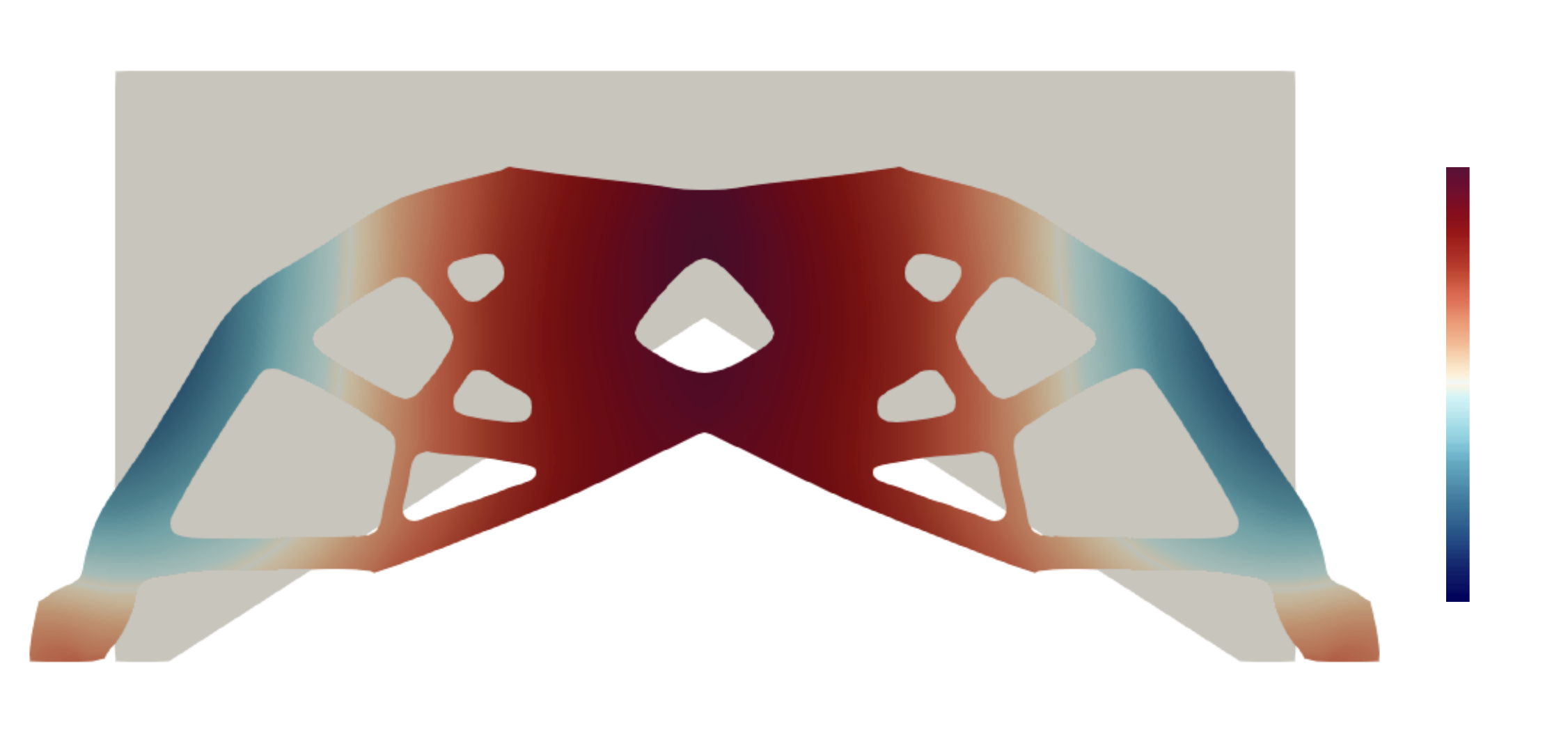}}%
    \put(0.95040486,0.09943219){\color[rgb]{0,0,0}\makebox(0,0)[lt]{\lineheight{1.25}\smash{\begin{tabular}[t]{l}3.\end{tabular}}}}%
    \put(0.95040486,0.37611398){\color[rgb]{0,0,0}\makebox(0,0)[lt]{\lineheight{1.25}\smash{\begin{tabular}[t]{l}20.\end{tabular}}}}%
    \put(0.94,0.034){\color[rgb]{0,0,0}\makebox(0,0)[t]{\lineheight{1.25}\smash{\begin{tabular}[t]{c}$||\bu||$\end{tabular}}}}%
    \put(0,0){\includegraphics[width=\unitlength,page=2]{portal_frame_final_design_ML.pdf}}%
  \end{picture}%
\endgroup%

        \vspace{-19pt}
        \subcaption*{(b)}
        }
    \end{minipage}

    \hspace{0.465\linewidth}
    \begin{minipage}{0.5\linewidth}
        \centering        
        \def\svgwidth{0.952\linewidth}
        \vspace{-11pt}
        {\scriptsize
        \vspace*{-112pt}
\begingroup%
  \makeatletter%
  \providecommand\color[2][]{%
    \errmessage{(Inkscape) Color is used for the text in Inkscape, but the package 'color.sty' is not loaded}%
    \renewcommand\color[2][]{}%
  }%
  \providecommand\transparent[1]{%
    \errmessage{(Inkscape) Transparency is used (non-zero) for the text in Inkscape, but the package 'transparent.sty' is not loaded}%
    \renewcommand\transparent[1]{}%
  }%
  \providecommand\rotatebox[2]{#2}%
  \newcommand*\fsize{\dimexpr\f@size pt\relax}%
  \newcommand*\lineheight[1]{\fontsize{\fsize}{#1\fsize}\selectfont}%
  \ifx\svgwidth\undefined%
    \setlength{\unitlength}{1080.66381836bp}%
    \ifx\svgscale\undefined%
      \relax%
    \else%
      \setlength{\unitlength}{\unitlength * \real{\svgscale}}%
    \fi%
  \else%
    \setlength{\unitlength}{\svgwidth}%
  \fi%
  \global\let\svgwidth\undefined%
  \global\let\svgscale\undefined%
  \makeatother%
  \begin{picture}(1,0.4800122)%
    \lineheight{1}%
    \setlength\tabcolsep{0pt}%
    \put(0,0){\includegraphics[width=\unitlength,page=1]{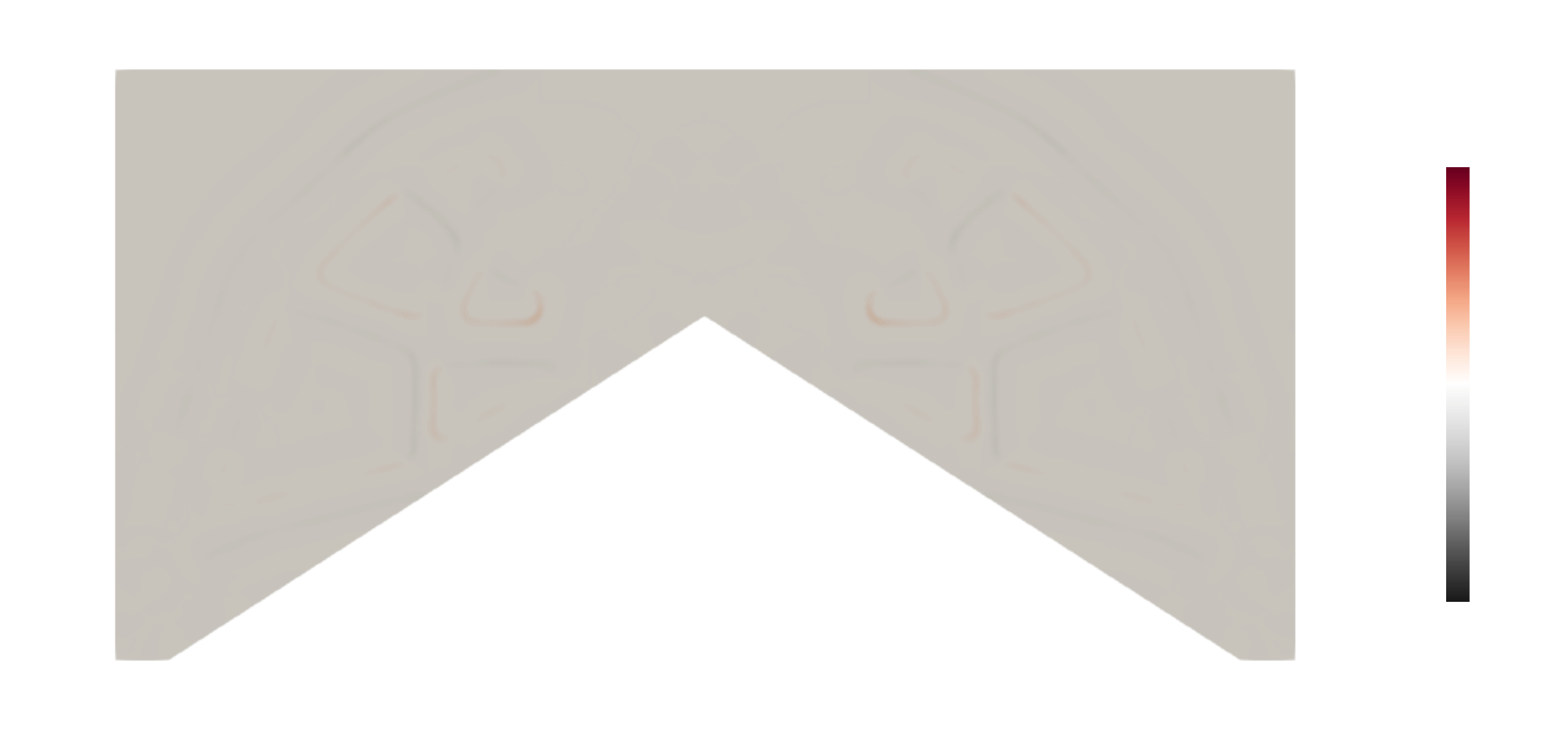}}%
    \put(0.95040486,0.09943226){\color[rgb]{0,0,0}\makebox(0,0)[lt]{\lineheight{1.25}\smash{\begin{tabular}[t]{l}-1.0\end{tabular}}}}%
    \put(0.95040486,0.37611405){\color[rgb]{0,0,0}\makebox(0,0)[lt]{\lineheight{1.25}\smash{\begin{tabular}[t]{l}1.0\end{tabular}}}}%
    \put(0.95040486,0.23731048){\color[rgb]{0,0,0}\makebox(0,0)[lt]{\lineheight{1.25}\smash{\begin{tabular}[t]{l}0.0\end{tabular}}}}%
    \put(0,0){\includegraphics[width=\unitlength,page=2]{portal_frame_final_design_ml_analytical_mse.pdf}}%
    \put(0.94,0.034){\color[rgb]{0,0,0}\makebox(0,0)[t]{\lineheight{1.25}\smash{\begin{tabular}[t]{c}$\de_{\rh}$\end{tabular}}}}%
  \end{picture}%
\endgroup%

        \vspace{-10pt}
        \subcaption*{(c)}
        }
    \end{minipage}
    \caption{(a) Portal frame BVP (b) Deformation of the design obtained using the ML model (c) Difference plot of the design obtained using the ML model to the design obtained using the ground truth phenomenological model.}
    \label{fig:portal_frame_bvp}
\end{figure}

\begin{figure}[t!]
    \begin{minipage}{0.4\linewidth}
        \centering
        \vspace{-230pt}
        \def\svgwidth{0.9\linewidth}
        {\footnotesize
        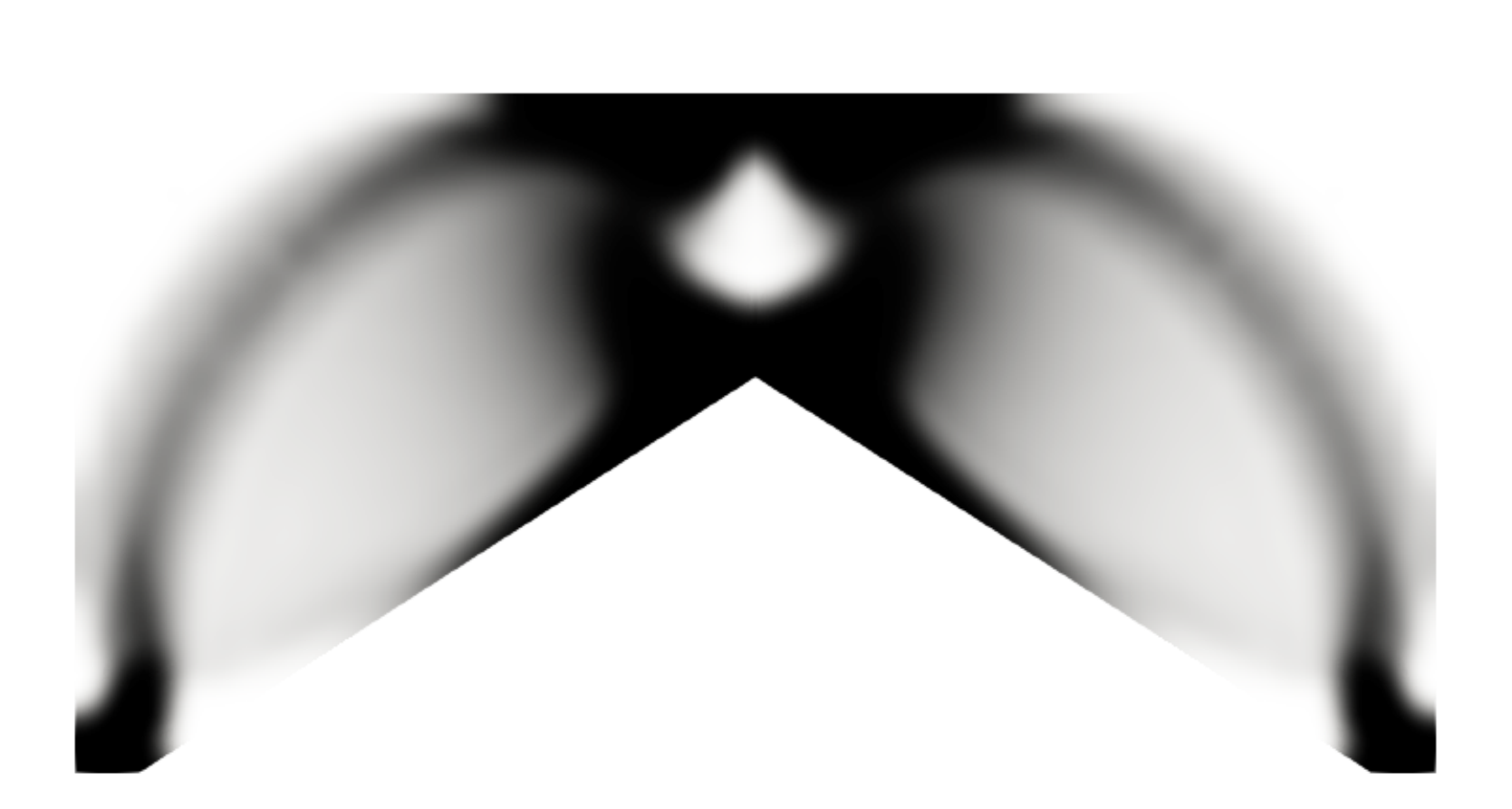
        }
        \vspace{-20pt}
        \subcaption*{$i = 19$}
    \end{minipage}
    \begin{minipage}{0.3\linewidth}
        \centering
        \def\svgwidth{\linewidth}
        \setlength{\figW}{8cm}
        \setlength{\figH}{6cm}
        {\scriptsize
        \input{figures/portal_frame_refined_history.tex}
        }
    \end{minipage}
    
    \begin{minipage}{0.4\linewidth}
        \centering
        \vspace{-325pt}
        \def\svgwidth{0.9\linewidth}
        {\footnotesize
        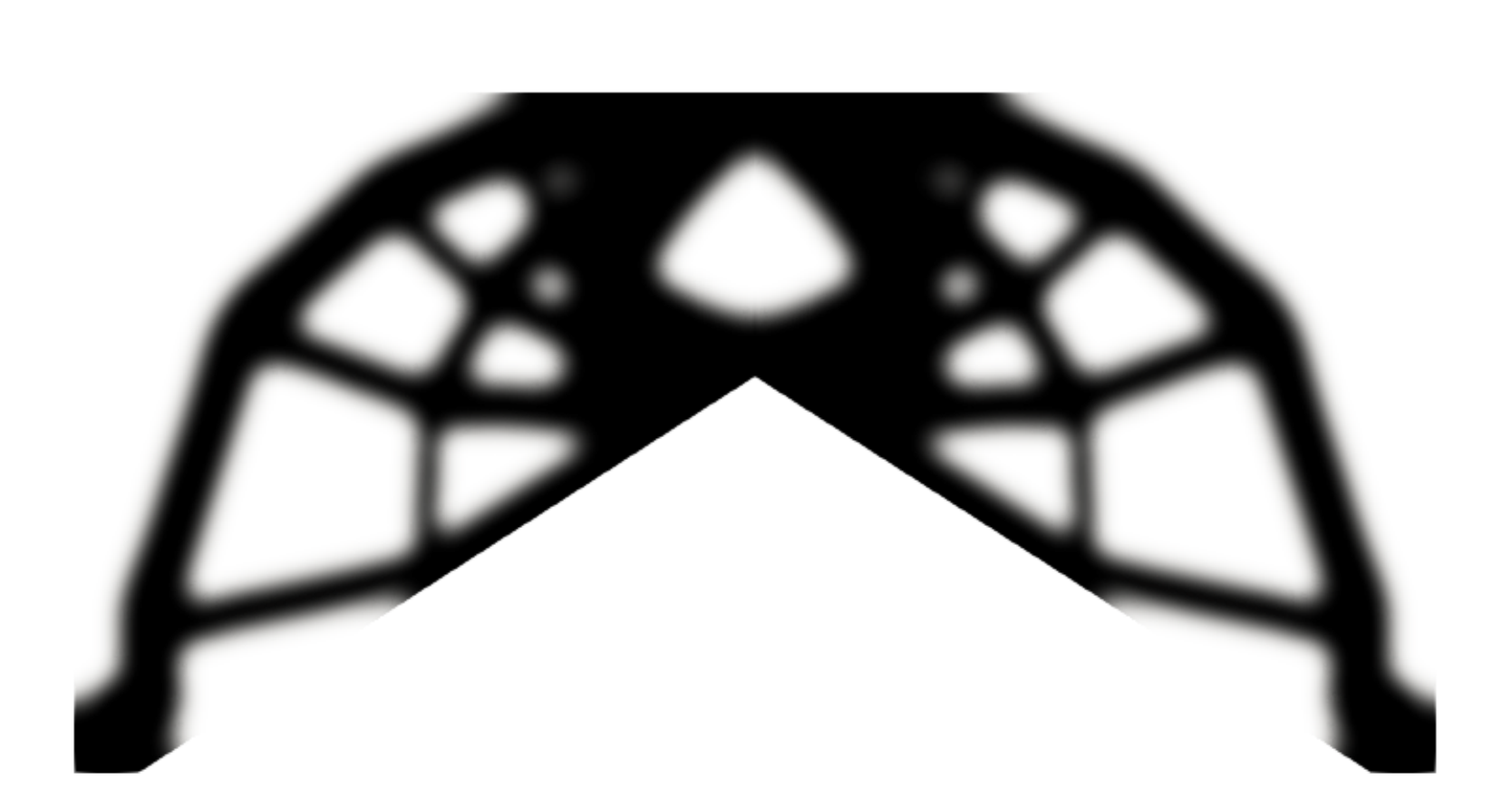
        }
        \vspace{-20pt}
        \subcaption*{$i = 97$}
    \end{minipage}

    \begin{minipage}{0.4\linewidth}
        \centering
        \def\svgwidth{0.9\linewidth}
        \vspace{-145pt}
        {\footnotesize
        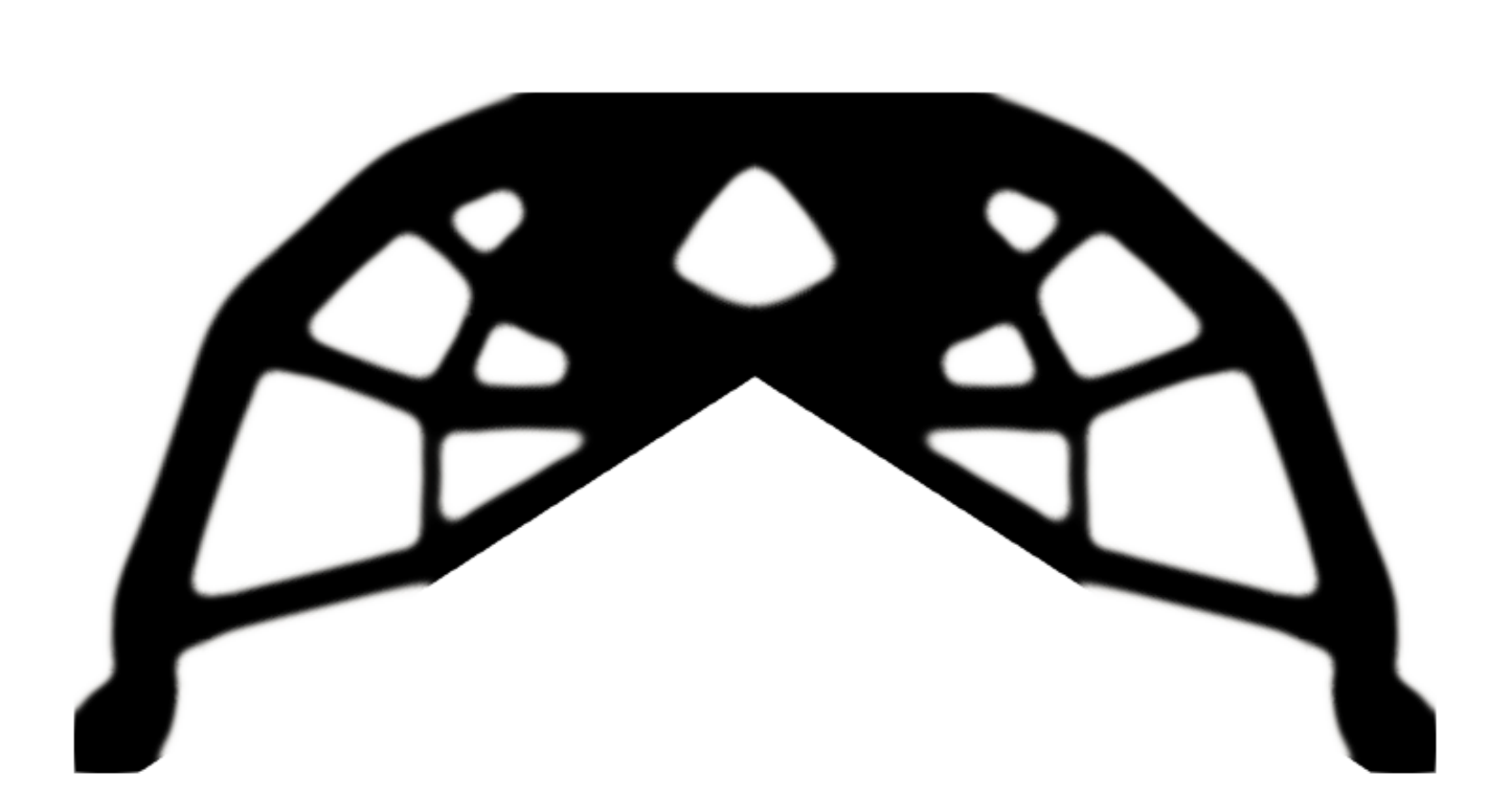
        \vspace{-20pt}
        \subcaption*{$i = 177$}
        }
    \end{minipage}
    \vspace{-10pt}
    \caption{Convergence history for portal frame TO problem.Visualized designs correspond to those corresponding to TO using ML model at iterations where critical continuation parameter updates occur.}
    \label{fig:portal_frame_topopt_benchmark}
\end{figure}

Next we consider a 2D portal frame with the design domain as shown in Figure \ref{fig:portal_frame_bvp}(a).
Due to the symmetry present in the geometry and loading conditions, only half of the domain is considered for the TO problem.
The support leg of the portal frame is fixed in the vertical direction and the horizontal displacement is constrained on the edge subject to the symmetry condition.
A vertically downward \textit{average applied displacement} $c_{N} = 0.2 L$ is applied at center of the portal frame.
The domain is discretized using 4 noded quadrilateral elements with a unstructured mesh of nominal edge length $0.5 \Umm$ adding up to a total of $28902$ elements.
The TO problem is formulated with the objective of maximizing the external work done by the applied displacement, subject to a material volume fraction constraint set by $g_{\max} = 0.5$.

The convergence history of the objective function for the portal frame in Figure \ref{fig:portal_frame_topopt_benchmark} shows that the optimization process using the ML model and the ground truth phenomenological model progresses in a similar manner.
The objective curves as well as the curves representing the continuation scheme overlap, which results in the intermediate as well as the final designs (see Figure \ref{fig:portal_frame_topopt_benchmark}) obtained using the two material models being almost identical.
The difference plot in Figure \ref{fig:portal_frame_bvp}(c) reinforces this observation.
Figure \ref{fig:portal_frame_bvp}(b) shows the magnitude of deformation of the final design obtained using the ML model in its deformed state.

\subsection {Multi-scale topology optimization: Portal frame example}
\label{appsubsec:multiscale_portal_frame_example_benchmark}

\begin{figure}[t!]
    \begin{minipage}{0.4\linewidth}
    \centering        
    \def\svgwidth{1.19\linewidth}
    {\scriptsize
    \vspace*{-70pt}
\begingroup%
  \makeatletter%
  \providecommand\color[2][]{%
    \errmessage{(Inkscape) Color is used for the text in Inkscape, but the package 'color.sty' is not loaded}%
    \renewcommand\color[2][]{}%
  }%
  \providecommand\transparent[1]{%
    \errmessage{(Inkscape) Transparency is used (non-zero) for the text in Inkscape, but the package 'transparent.sty' is not loaded}%
    \renewcommand\transparent[1]{}%
  }%
  \providecommand\rotatebox[2]{#2}%
  \newcommand*\fsize{\dimexpr\f@size pt\relax}%
  \newcommand*\lineheight[1]{\fontsize{\fsize}{#1\fsize}\selectfont}%
  \ifx\svgwidth\undefined%
    \setlength{\unitlength}{1080.66381836bp}%
    \ifx\svgscale\undefined%
      \relax%
    \else%
      \setlength{\unitlength}{\unitlength * \real{\svgscale}}%
    \fi%
  \else%
    \setlength{\unitlength}{\svgwidth}%
  \fi%
  \global\let\svgwidth\undefined%
  \global\let\svgscale\undefined%
  \makeatother%
  \begin{picture}(1,0.4800122)%
    \lineheight{1}%
    \setlength\tabcolsep{0pt}%
    \put(0,0){\includegraphics[width=\unitlength,page=1]{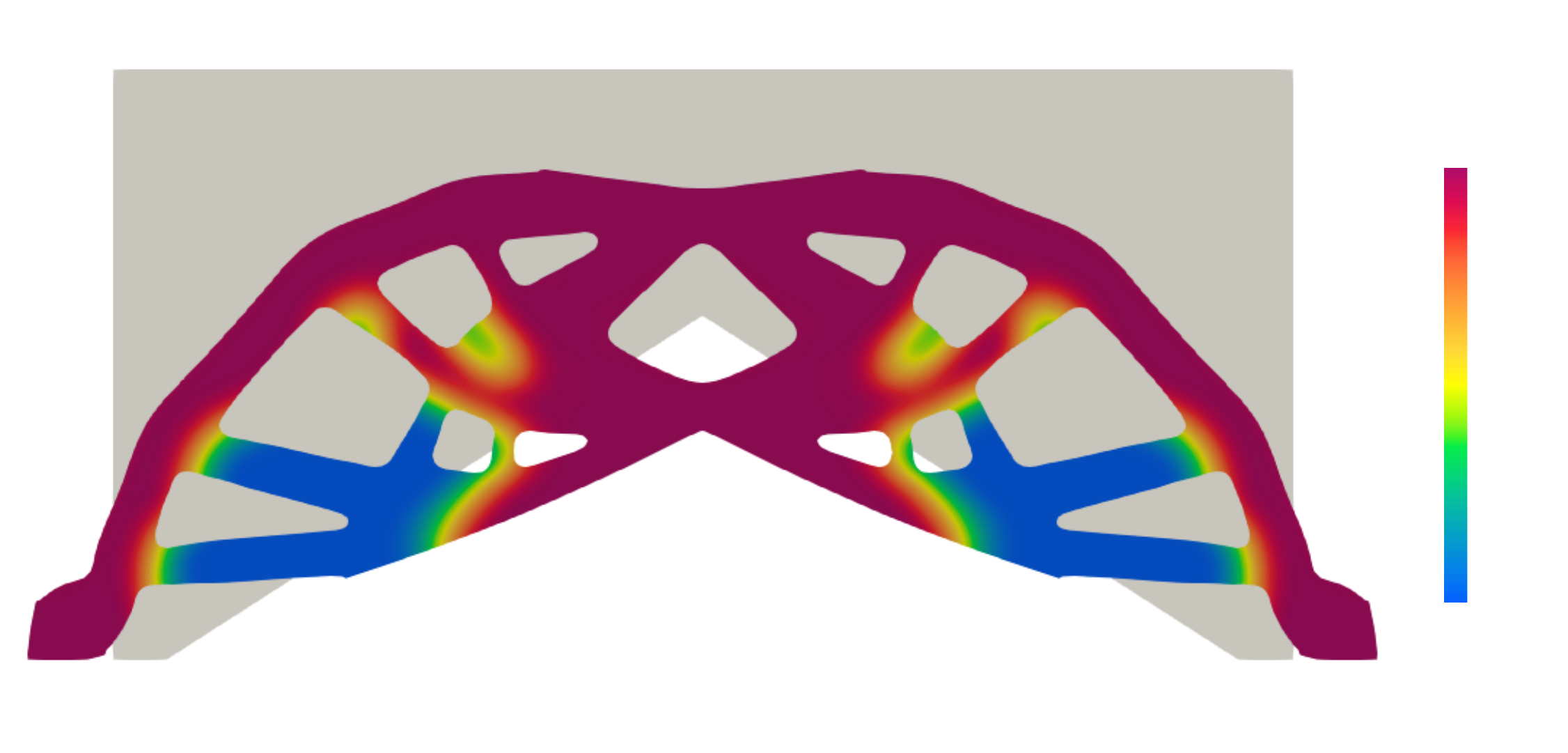}}%
    \put(0.94902358,0.09865776){\color[rgb]{0,0,0}\makebox(0,0)[lt]{\lineheight{1.25}\smash{\begin{tabular}[t]{l}0.1\end{tabular}}}}%
    \put(0.94902358,0.37533955){\color[rgb]{0,0,0}\makebox(0,0)[lt]{\lineheight{1.25}\smash{\begin{tabular}[t]{l}0.5\end{tabular}}}}%
    \put(0.94902358,0.16713419){\color[rgb]{0,0,0}\makebox(0,0)[lt]{\lineheight{1.25}\smash{\begin{tabular}[t]{l}0.2\end{tabular}}}}%
    \put(0.94902358,0.23653598){\color[rgb]{0,0,0}\makebox(0,0)[lt]{\lineheight{1.25}\smash{\begin{tabular}[t]{l}0.3\end{tabular}}}}%
    \put(0.94902358,0.30593777){\color[rgb]{0,0,0}\makebox(0,0)[lt]{\lineheight{1.25}\smash{\begin{tabular}[t]{l}0.4\end{tabular}}}}%
    \put(0.935,0.04){\color[rgb]{0,0,0}\makebox(0,0)[t]{\lineheight{1.25}\smash{\begin{tabular}[t]{c}$\al$\end{tabular}}}}%
    \put(0,0){\includegraphics[width=\unitlength,page=2]{portal_frame_final_design_micro.pdf}}%
  \end{picture}%
\endgroup%

    \vspace*{-10pt}
    \subcaption*{(a)}
    }
    \end{minipage}
    \hspace{0.08\linewidth}
    \begin{minipage}{0.30\linewidth}
        \centering
        \def\svgwidth{\linewidth}
        \setlength{\figW}{8cm}
        \setlength{\figH}{6cm}
        {\scriptsize
\begin{tikzpicture}

\definecolor{gray}{RGB}{128,128,128}
\definecolor{lightgray204}{RGB}{204,204,204}
\definecolor{mediumblue}{RGB}{0,0,205}
\pgfplotsset{grid style={dashed,lightgray204}}

\begin{axis}[
height=\figH,
legend cell align={left},
legend style={fill opacity=0.8, draw opacity=1, text opacity=1, draw=lightgray204},
tick align=outside,
tick pos=left,
width=\figW,
xlabel={Optimization Iteration Number},
xmajorgrids,
xmin=-9.25, xmax=194.25,
xtick = {0, 174},
ytick = {-6.62, -11.32, -9.14, -10.52},
xtick style={color=black},
ylabel={Objective},
ymajorgrids,
ymin=-11.551655, ymax=-6.2,
ytick style={color=black}
]
\addplot [very thick, magenta]
table {%
0 -6.62
1 -8.3886
2 -9.8705
3 -10.509
4 -10.801
5 -10.905
6 -10.976
7 -11.036
8 -11.093
9 -11.146
10 -11.191
11 -11.227
12 -11.255
13 -11.274
14 -11.289
15 -11.298
16 -11.306
17 -11.299
18 -11.307
19 -11.316
20 -6.6029
21 -6.9142
22 -7.2568
23 -7.4629
24 -7.6804
25 -7.8811
26 -8.092
27 -8.3133
28 -8.5435
29 -8.6933
30 -8.7891
31 -8.8765
32 -8.9559
33 -9.0261
34 -9.0878
35 -9.1462
36 -9.2031
37 -9.2596
38 -9.3153
39 -9.3704
40 -9.4287
41 -9.5019
42 -9.5736
43 -9.6369
44 -9.6863
45 -9.7305
46 -9.7651
47 -9.7929
48 -9.8126
49 -9.8276
50 -9.839
51 -9.8482
52 -9.8564
53 -9.8634
54 -9.8691
55 -9.0296
56 -9.0685
57 -9.093
58 -9.1101
59 -9.1235
60 -9.1342
61 -9.1426
62 -9.1492
63 -9.1547
64 -9.1596
65 -9.1638
66 -9.1673
67 -9.1703
68 -9.1729
69 -9.1753
70 -9.1775
71 -9.1794
72 -9.1811
73 -9.1827
74 -9.1843
75 -8.6682
76 -8.6786
77 -8.6835
78 -8.6878
79 -8.6918
80 -8.6955
81 -8.699
82 -8.7023
83 -8.7052
84 -8.7077
85 -8.71
86 -8.7121
87 -8.7141
88 -8.7161
89 -8.718
90 -8.72
91 -8.7218
92 -8.7237
93 -8.7255
94 -8.7272
95 -9.0348
96 -8.9733
97 -8.9754
98 -8.9768
99 -8.9781
100 -8.9792
101 -8.9804
102 -8.9817
103 -8.9829
104 -8.984
105 -8.9852
106 -8.9863
107 -8.9874
108 -8.9884
109 -8.9894
110 -8.9904
111 -8.9914
112 -8.9923
113 -8.9932
114 -8.9942
115 -9.6579
116 -9.5515
117 -9.5583
118 -9.5622
119 -9.5653
120 -9.5681
121 -9.5707
122 -9.573
123 -9.575
124 -9.5765
125 -9.5779
126 -9.579
127 -9.58
128 -9.5808
129 -9.5816
130 -9.5823
131 -9.583
132 -9.5837
133 -9.5844
134 -9.585
135 -10.283
136 -10.213
137 -10.223
138 -10.228
139 -10.232
140 -10.238
141 -10.242
142 -10.245
143 -10.247
144 -10.248
145 -10.25
146 -10.251
147 -10.252
148 -10.253
149 -10.254
150 -10.254
151 -10.255
152 -10.255
153 -10.256
154 -10.257
155 -10.516
156 -10.5
157 -10.503
158 -10.505
159 -10.507
160 -10.508
161 -10.51
162 -10.51
163 -10.511
164 -10.512
165 -10.513
166 -10.514
167 -10.515
168 -10.516
169 -10.516
170 -10.517
171 -10.518
172 -10.519
173 -10.52
174 -10.521
};
\addlegendentry{ML: varying microstructure}
\addplot [very thick, mediumblue, dash pattern=on 4pt off 2pt]
table {%
0 -6.62
1 -7.7338
2 -8.7694
3 -9.4678
4 -9.7943
5 -9.8665
6 -9.8975
7 -9.9163
8 -9.9288
9 -9.9379
10 -9.9447
11 -9.9503
12 -9.9551
13 -9.959
14 -9.9621
15 -9.9643
16 -9.9661
17 -9.9667
18 -9.9678
19 -9.968
20 -6.7915
21 -7.0798
22 -7.4088
23 -7.5545
24 -7.6415
25 -7.7085
26 -7.7636
27 -7.8089
28 -7.8451
29 -7.8754
30 -7.9015
31 -7.924
32 -7.9433
33 -7.9604
34 -7.9768
35 -7.993
36 -8.0095
37 -8.0261
38 -8.0418
39 -8.0479
40 -8.038
41 -8.0439
42 -8.0794
43 -8.1187
44 -8.1451
45 -8.165
46 -8.1848
47 -8.2069
48 -8.231
49 -8.2557
50 -8.284
51 -8.3142
52 -8.3441
53 -8.3724
54 -8.4004
55 -8.4318
56 -8.4684
57 -8.4993
58 -8.5207
59 -8.5376
60 -8.5503
61 -8.5612
62 -8.5706
63 -8.5783
64 -8.5851
65 -8.5909
66 -7.8283
67 -7.8572
68 -7.8787
69 -7.8914
70 -7.9002
71 -7.9074
72 -7.9135
73 -7.9182
74 -7.9215
75 -7.9241
76 -7.9263
77 -7.9283
78 -7.9302
79 -7.9319
80 -7.9335
81 -7.9349
82 -7.9364
83 -7.9378
84 -7.9391
85 -7.9404
86 -7.4286
87 -7.4385
88 -7.4463
89 -7.4524
90 -7.4578
91 -7.4628
92 -7.4674
93 -7.4714
94 -7.474
95 -7.4762
96 -7.4783
97 -7.4803
98 -7.4823
99 -7.4842
100 -7.486
101 -7.4879
102 -7.4897
103 -7.4914
104 -7.4929
105 -7.4944
106 -7.7812
107 -7.7296
108 -7.7326
109 -7.7343
110 -7.7358
111 -7.7372
112 -7.7385
113 -7.7397
114 -7.7409
115 -7.742
116 -7.743
117 -7.744
118 -7.7449
119 -7.7458
120 -7.7467
121 -7.7477
122 -7.7486
123 -7.7495
124 -7.7503
125 -7.7512
126 -8.3537
127 -8.2738
128 -8.2818
129 -8.2867
130 -8.2904
131 -8.2933
132 -8.2957
133 -8.2977
134 -8.2991
135 -8.3
136 -8.3007
137 -8.3013
138 -8.3019
139 -8.3024
140 -8.3028
141 -8.3033
142 -8.3037
143 -8.3041
144 -8.3045
145 -8.3049
146 -8.9263
147 -8.8666
148 -8.8714
149 -8.8799
150 -8.8816
151 -8.893
152 -8.8993
153 -8.9047
154 -8.9074
155 -8.9088
156 -8.9097
157 -8.9104
158 -8.911
159 -8.9116
160 -8.9122
161 -8.9127
162 -8.9132
163 -8.9136
164 -8.9139
165 -8.9143
166 -9.1432
167 -9.1253
168 -9.1246
169 -9.129
170 -9.1317
171 -9.1358
172 -9.1377
173 -9.1384
174 -9.1389
175 -9.1393
176 -9.1398
177 -9.1402
178 -9.1407
179 -9.1411
180 -9.1415
181 -9.1418
182 -9.1421
183 -9.1425
184 -9.1428
185 -9.1431
};
\addlegendentry{ML: fixed microstructure}
\end{axis}

\end{tikzpicture}
        \vspace{-10pt}
        \subcaption*{(c)}
        }
    \end{minipage}

    \begin{minipage}{0.4\linewidth}
        \def\svgwidth{1.19\linewidth}
        {\scriptsize
        \vspace{-60pt}
\begingroup%
  \makeatletter%
  \providecommand\color[2][]{%
    \errmessage{(Inkscape) Color is used for the text in Inkscape, but the package 'color.sty' is not loaded}%
    \renewcommand\color[2][]{}%
  }%
  \providecommand\transparent[1]{%
    \errmessage{(Inkscape) Transparency is used (non-zero) for the text in Inkscape, but the package 'transparent.sty' is not loaded}%
    \renewcommand\transparent[1]{}%
  }%
  \providecommand\rotatebox[2]{#2}%
  \newcommand*\fsize{\dimexpr\f@size pt\relax}%
  \newcommand*\lineheight[1]{\fontsize{\fsize}{#1\fsize}\selectfont}%
  \ifx\svgwidth\undefined%
    \setlength{\unitlength}{1080.66381836bp}%
    \ifx\svgscale\undefined%
      \relax%
    \else%
      \setlength{\unitlength}{\unitlength * \real{\svgscale}}%
    \fi%
  \else%
    \setlength{\unitlength}{\svgwidth}%
  \fi%
  \global\let\svgwidth\undefined%
  \global\let\svgscale\undefined%
  \makeatother%
  \begin{picture}(1,0.4800122)%
    \lineheight{1}%
    \setlength\tabcolsep{0pt}%
    \put(0,0){\includegraphics[width=\unitlength,page=1]{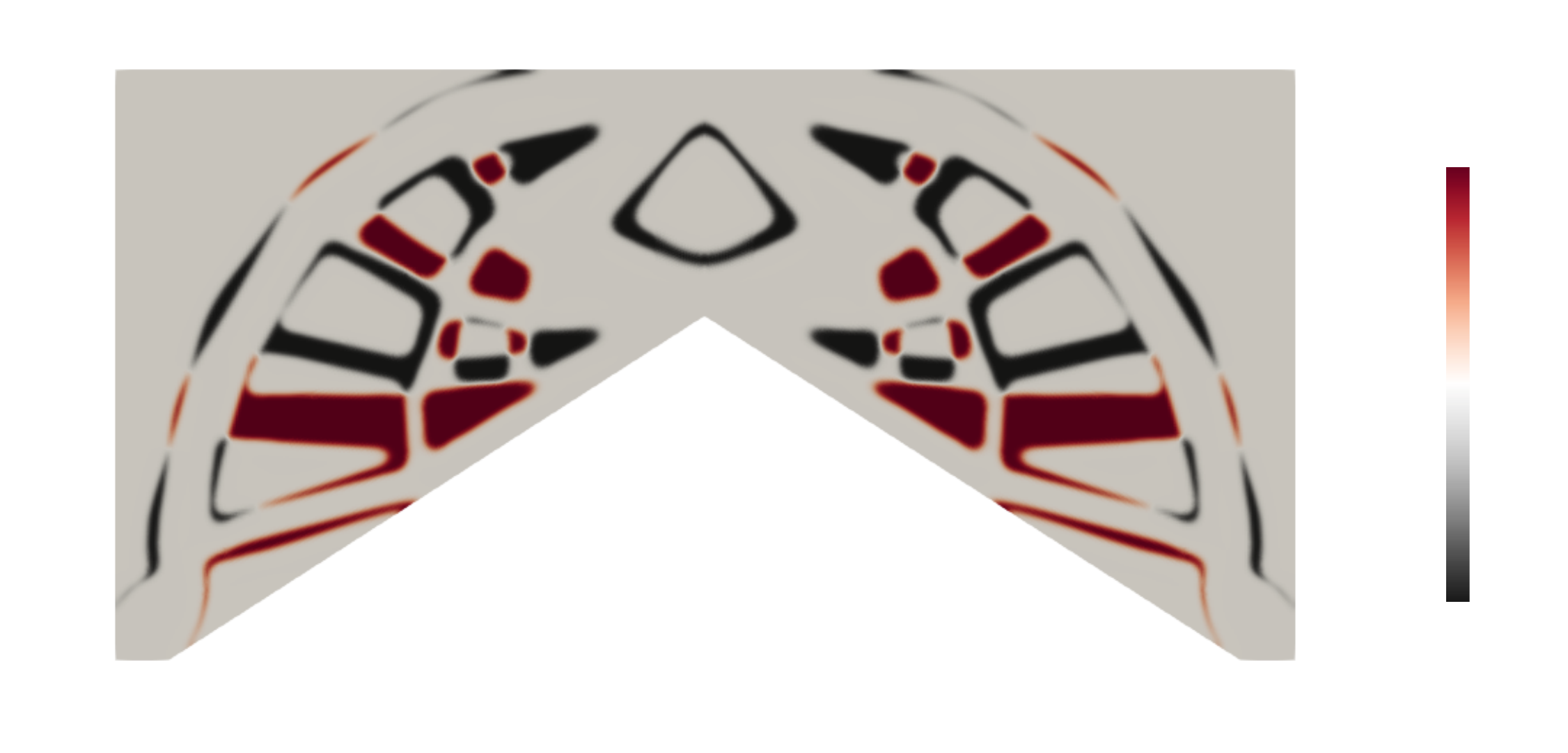}}%
    \put(0.95040486,0.09943231){\color[rgb]{0,0,0}\makebox(0,0)[lt]{\lineheight{1.25}\smash{\begin{tabular}[t]{l}-1.0\end{tabular}}}}%
    \put(0.95040486,0.3761141){\color[rgb]{0,0,0}\makebox(0,0)[lt]{\lineheight{1.25}\smash{\begin{tabular}[t]{l}1.0\end{tabular}}}}%
    \put(0.95040486,0.23731052){\color[rgb]{0,0,0}\makebox(0,0)[lt]{\lineheight{1.25}\smash{\begin{tabular}[t]{l}0.0\end{tabular}}}}%
    \put(0,0){\includegraphics[width=\unitlength,page=2]{portal_frame_final_design_uniform_micro_mse.pdf}}%
    \put(0,0){\includegraphics[width=\unitlength,page=3]{portal_frame_final_design_uniform_micro_mse.pdf}}%
    \put(0.935,0.04){\color[rgb]{0,0,0}\makebox(0,0)[t]{\lineheight{1.25}\smash{\begin{tabular}[t]{c}$\de_{\rh}$\end{tabular}}}}%
    \put(0,0){\includegraphics[width=\unitlength,page=4]{portal_frame_final_design_uniform_micro_mse.pdf}}%
  \end{picture}%
\endgroup%

        \vspace*{-15pt}
        \subcaption*{(b)}
        }
    \end{minipage}
    \caption{(a) Deformed configuration of the final design obtained using the ML model with varying microstructure (b) Difference plot between the design obtained with varying microstructure and the design obtained with fixed microstructure. (c) Convergence history for the portal frame TO problem with varying microstructure vs. fixed microstructure}
    \label{fig:portal_frame_micro_bvp}
\end{figure}

The portal frame example with varying microstructure and the baseline fixed microstructure are set up with the same side constraint bounds as the T bracket example discussed above.
Visible differences are observed in the final topologies between the varying microstructure and fixed microstructure cases, with the former containing $13$ void regions and the latter containing $15$.
The difference plot in Figure \ref{fig:portal_frame_micro_bvp}b highlights that although the external outline of the designs are similar, there is considerable rearrangement of the material in the internal regions of the designs. Similar to the T bracket example, allowing the microstructure to vary results in a lower objective value.
The convergence history for the portal frame example in Figure \ref{fig:portal_frame_micro_bvp}c shows that the baseline case with fixed microstructure undergoes a delayed update in the continuation scheme corresponding to the update in the penalty parameter $p=3$.
This induces a rightward shift in the convergence history when compared with the case with varying microstructure.
Consistent with the observations in the T bracket example, the final design with varying microstructure (Figure \ref{fig:portal_frame_micro_bvp}a) has material addition predominantly occurring close to regions with lower inclusion volume fraction $\al$ and material removal happening close to regions with higher inclusion volume fractions.

\subsection{Single scale topology optimization: 3D Benchmark example with fixed-fixed beam}
\label{appsubsec:fixed_fixed_beam_3D_example_benchmark}
\begin{table}[b!]
    \caption{Model and training hyperparameters}
    \label{tab:hyperparameters2}
    \begin{minipage}{0.3\linewidth}
    \centering
    \footnotesize
    \begin{tblr}{l r}
        \hline
        Hyperparameter & Value \\
        \hline[0.15em]
        No. hidden layers & $2$ \\
        No. neurons per layer & $8$ \\
        Initial learning rate & $1 \times 10^{-2}$ \\
        \hline[0.1em]
    \end{tblr} \end{minipage}
    \begin{minipage}{0.4\linewidth}
        \centering
        \footnotesize
        \begin{tblr}{l r}
        \hline
            Hyperparameter & Value \\
            \hline[0.15em]
            Exponential decay rate & $0.5$ \\
            Decay transition epoch interval & $1000$ \\
            Decay end value &  $5 \times 10^{-4}$ \\
            \hline[0.1em]
        \end{tblr}
    \end{minipage}
    \begin{minipage}{0.3\linewidth}
        \centering
        \footnotesize
        \begin{tblr}{l r}
        \hline
            Hyperparameter & Value \\
            \hline[0.15em]
            Max epochs & 15000 \\
            Batch size & 128 \\
            Early stopping patience & 5000 \\
            \hline[0.1em]
        \end{tblr}
    \end{minipage}
\end{table}

\begin{table}[b!]
    \begin{minipage}{\linewidth}
    \caption{Performance metrics of the trained model}
    \label{tab:model_performance2}
    \centering
    \footnotesize
    \begin{tblr}{c c r r r}
        \hline
        Model & Metric & Training Dataset & Validation Dataset & Test Dataset \\
        \hline[0.15em]
        \SetCell[r=3]{c} $\CM^{s}_{3D}$ & $1 - \mathrm{R}^{2}$ & $2.06 \times 10^{-5}$ & $2.03 \times 10^{-5}$ & $2.15 \times 10^{-5}$ \\
        & RMSE & $4.63 \times 10^{-3}$ & $4.60 \times 10^{-3}$ & $4.61 \times 10^{-3}$ \\
        & MAE & $3.05 \times 10^{-3}$ & $3.05 \times 10^{-3}$ & $3.05 \times 10^{-3}$ \\
        \hline[0.1em]
    \end{tblr}
    \end{minipage}
\end{table}

\begin{figure}[t!]
    \begin{minipage}{0.5\linewidth}
    \centering        
    \def\svgwidth{0.9\linewidth}
    {\footnotesize
    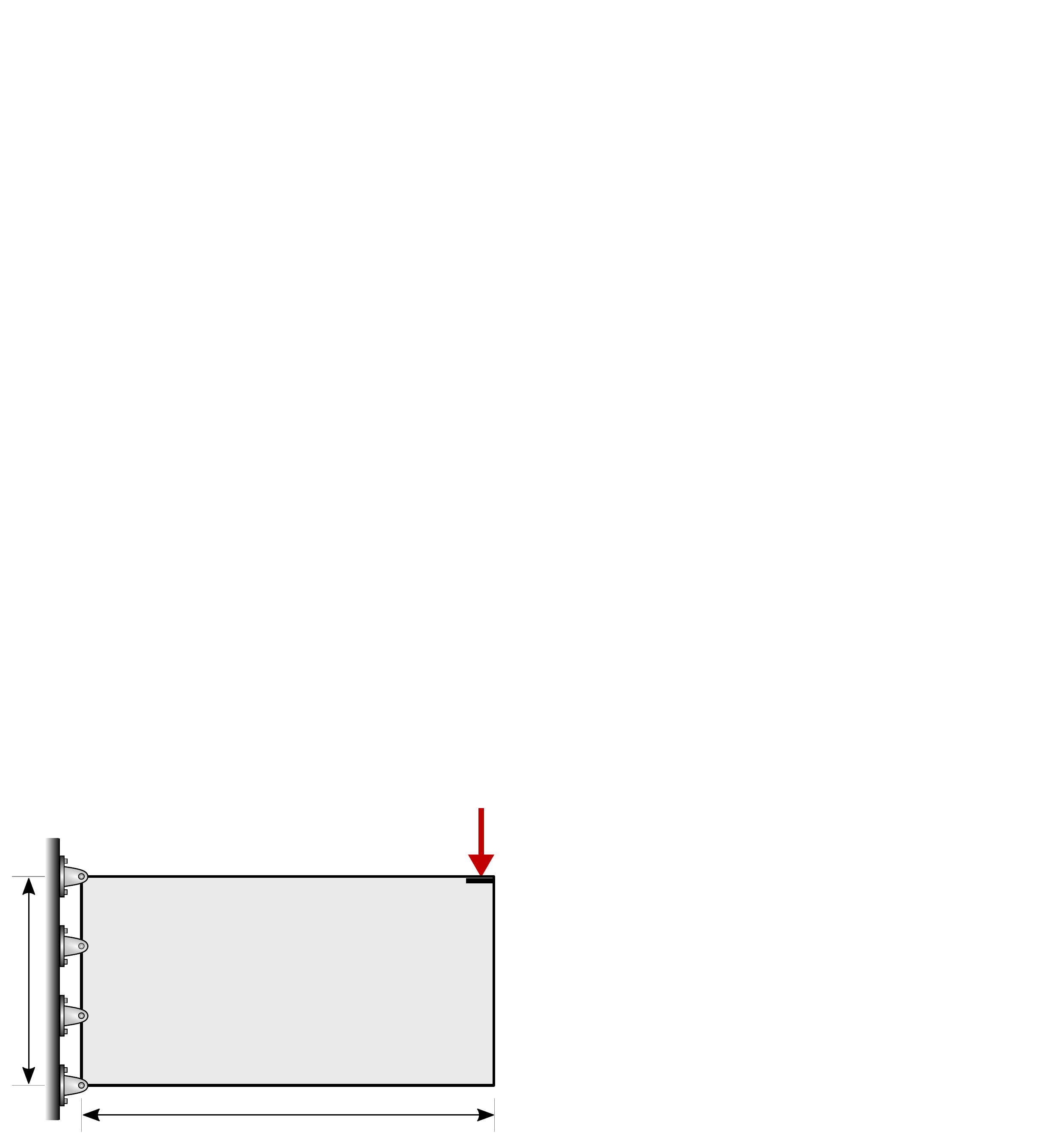
    \subcaption*{(a)}
    }
    \end{minipage}
    \begin{minipage}{0.5\linewidth}
        \centering        
        \def\svgwidth{0.9153\linewidth}
        {\footnotesize
        \vspace{-100pt}
\begingroup%
  \makeatletter%
  \providecommand\color[2][]{%
    \errmessage{(Inkscape) Color is used for the text in Inkscape, but the package 'color.sty' is not loaded}%
    \renewcommand\color[2][]{}%
  }%
  \providecommand\transparent[1]{%
    \errmessage{(Inkscape) Transparency is used (non-zero) for the text in Inkscape, but the package 'transparent.sty' is not loaded}%
    \renewcommand\transparent[1]{}%
  }%
  \providecommand\rotatebox[2]{#2}%
  \newcommand*\fsize{\dimexpr\f@size pt\relax}%
  \newcommand*\lineheight[1]{\fontsize{\fsize}{#1\fsize}\selectfont}%
  \ifx\svgwidth\undefined%
    \setlength{\unitlength}{1193.29203436bp}%
    \ifx\svgscale\undefined%
      \relax%
    \else%
      \setlength{\unitlength}{\unitlength * \real{\svgscale}}%
    \fi%
  \else%
    \setlength{\unitlength}{\svgwidth}%
  \fi%
  \global\let\svgwidth\undefined%
  \global\let\svgscale\undefined%
  \makeatother%
  \begin{picture}(1,0.70114332)%
    \lineheight{1}%
    \setlength\tabcolsep{0pt}%
    \put(0,0){\includegraphics[width=\unitlength,page=1]{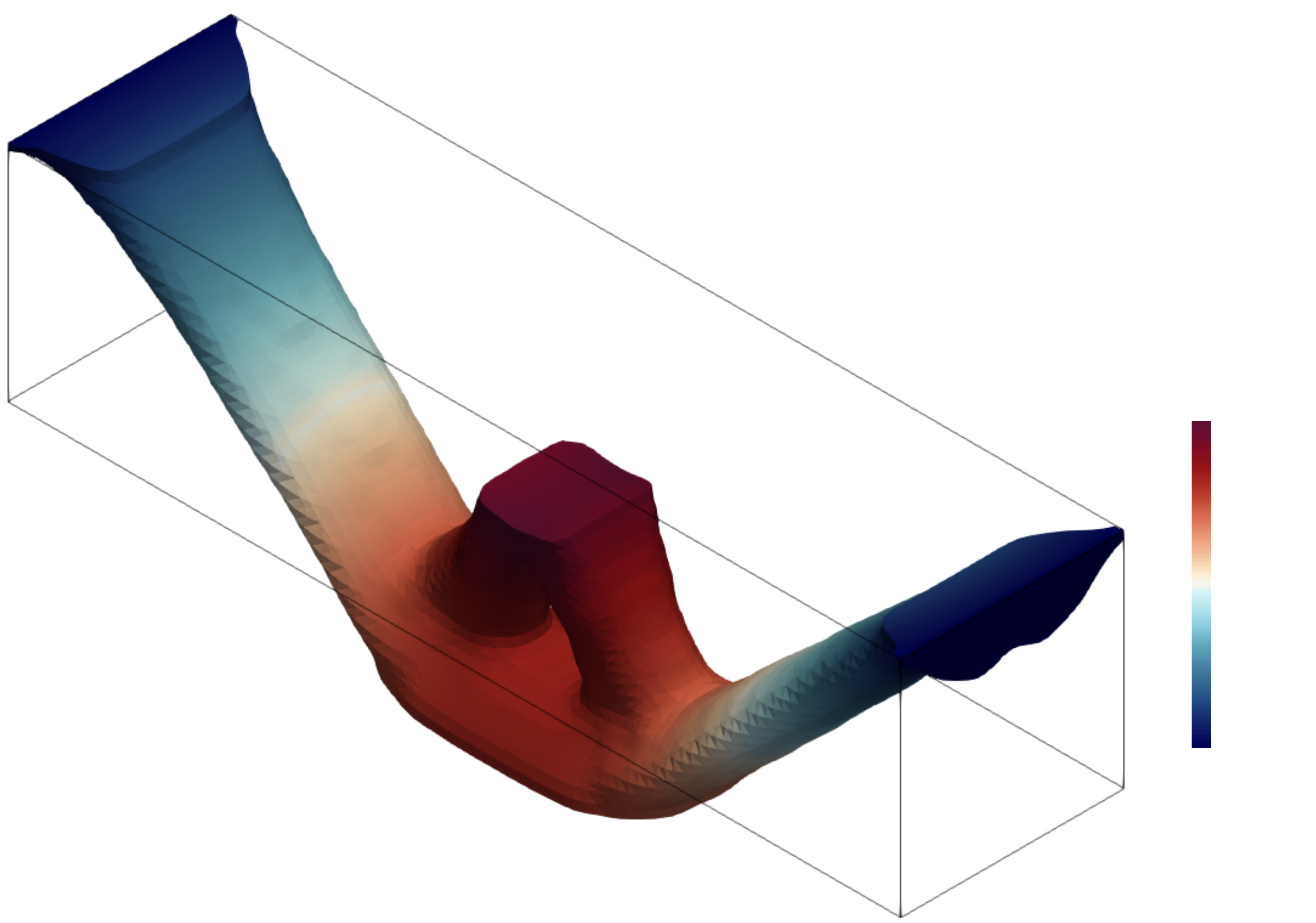}}%
    \put(0.93356982,0.13562585){\color[rgb]{0,0,0}\makebox(0,0)[lt]{\lineheight{1.25}\smash{\begin{tabular}[t]{l}0.\end{tabular}}}}%
    \put(0.93356982,0.38381194){\color[rgb]{0,0,0}\makebox(0,0)[lt]{\lineheight{1.25}\smash{\begin{tabular}[t]{l}31.\end{tabular}}}}%
    \put(0.915,0.05){\color[rgb]{0,0,0}\makebox(0,0)[t]{\lineheight{1.25}\smash{\begin{tabular}[t]{c}$||\bu||$\end{tabular}}}}%
    \put(0,0){\includegraphics[width=\unitlength,page=2]{fixed_fixed_beam_3D_final_design_ML.pdf}}%
  \end{picture}%
\endgroup%

        \vspace{-20pt}
        \subcaption*{\hspace{-100pt}(b)}
        }
    \end{minipage}

    \hspace{0.5\linewidth}
    \begin{minipage}{0.5\linewidth}
        \centering        
        \def\svgwidth{0.9153\linewidth}
        {\footnotesize
        \vspace{-80pt}
\begingroup%
  \makeatletter%
  \providecommand\color[2][]{%
    \errmessage{(Inkscape) Color is used for the text in Inkscape, but the package 'color.sty' is not loaded}%
    \renewcommand\color[2][]{}%
  }%
  \providecommand\transparent[1]{%
    \errmessage{(Inkscape) Transparency is used (non-zero) for the text in Inkscape, but the package 'transparent.sty' is not loaded}%
    \renewcommand\transparent[1]{}%
  }%
  \providecommand\rotatebox[2]{#2}%
  \newcommand*\fsize{\dimexpr\f@size pt\relax}%
  \newcommand*\lineheight[1]{\fontsize{\fsize}{#1\fsize}\selectfont}%
  \ifx\svgwidth\undefined%
    \setlength{\unitlength}{1198.14413228bp}%
    \ifx\svgscale\undefined%
      \relax%
    \else%
      \setlength{\unitlength}{\unitlength * \real{\svgscale}}%
    \fi%
  \else%
    \setlength{\unitlength}{\svgwidth}%
  \fi%
  \global\let\svgwidth\undefined%
  \global\let\svgscale\undefined%
  \makeatother%
  \begin{picture}(1,0.69963441)%
    \lineheight{1}%
    \setlength\tabcolsep{0pt}%
    \put(0,0){\includegraphics[width=\unitlength,page=1]{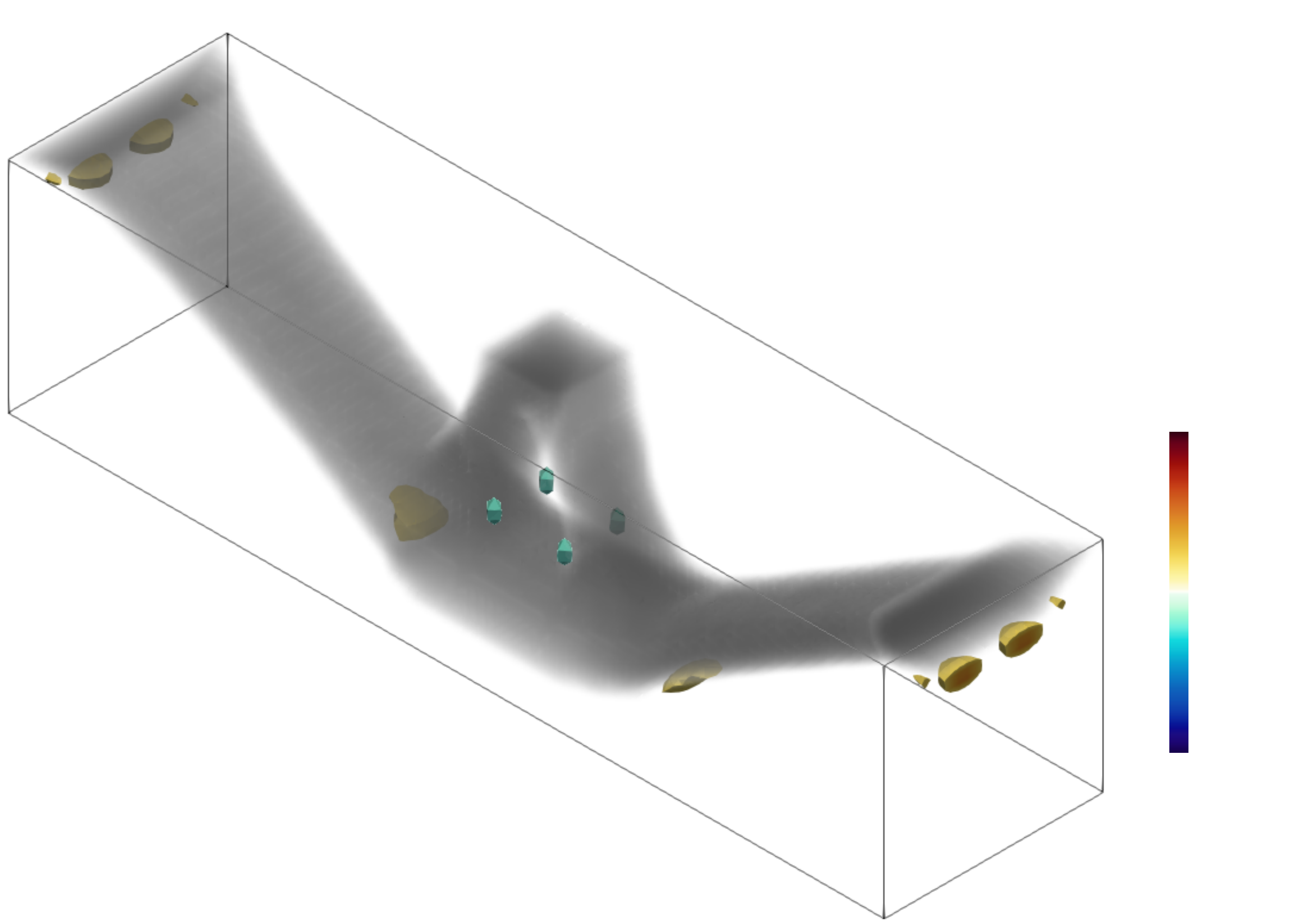}}%
    \put(0.91599053,0.13051945){\color[rgb]{0,0,0}\makebox(0,0)[lt]{\lineheight{1.25}\smash{\begin{tabular}[t]{l}-1.\end{tabular}}}}%
    \put(0.91599053,0.37394309){\color[rgb]{0,0,0}\makebox(0,0)[lt]{\lineheight{1.25}\smash{\begin{tabular}[t]{l}1.\end{tabular}}}}%
    \put(0.91599053,0.25177881){\color[rgb]{0,0,0}\makebox(0,0)[lt]{\lineheight{1.25}\smash{\begin{tabular}[t]{l}0.\end{tabular}}}}%
    \put(0.91,0.04){\color[rgb]{0,0,0}{\makebox(0,0)[t]{\lineheight{1.25}\smash{\begin{tabular}[t]{c}$\de_{\rh}$\end{tabular}}}}}%
    \put(0,0){\includegraphics[width=\unitlength,page=2]{fixed_fixed_beam_3D_final_design_analytical_ML_diff.pdf}}%
  \end{picture}%
\endgroup%

        \subcaption*{\hspace{-100pt}(c)}
        }
    \end{minipage}

    \caption{(a) Fixed-fixed beam BVP (b) Deformation of the design obtained using the ML model (c) Difference plot of the design obtained using the ML model to the design obtained using the ground truth phenomenological model. A transparent shadow of the design obtained using the phenomenological model is overlaid on the difference plot for reference.}
    \label{fig:fixed_fixed_beam_3D_bvp}
\end{figure}

Finally we consider a 3D fixed-fixed beam with the design domain as shown in Figure \ref{fig:fixed_fixed_beam_3D_bvp}(a).
The primary intention of considering this example is to assess the ``sensitivity" of the optimization process to the differences in the phenomenological model and the ML model, as well as to show that the framework is easily extended to 3D problems.
In order to train a 3D ML model, we employed the design of experiment approach described in Section \ref{subsec:doe_data_generation_for_ml} to generate the training dataset $\CD^{s}_{3D}$ consisting of $2^13 = 8192$ sample strain states by considering the spatial dimensionality $N_D =3$.

The beam is fixed along the left and the right faces and a downward \textit{average applied displacement} $c_{N} = 0.3 L$ is applied at the center of the beam's top surface.
Due to symmetry in the geometry and loading conditions, only a quarter of the domain is considered for this example.
The domain is discretized using 8-node hexahedral elements with a uniform mesh of edge length $2 \Umm$ adding up to a total of $16250$ elements.
A volume fraction constraint upper bound of $g_{\max} = 0.3$ is imposed in this example.
The optimization parameters are as shown in Table \ref{tab:TO_hyperparameters}, with the exception of the number of time steps, which is set to $15$ for this example.

We observe that the final objective values using the ML model and the ground truth phenomenological model are almost identical. 
The difference plot in Figure \ref{fig:fixed_fixed_beam_3D_bvp}(c) shows no significant material addition or deletion in the final design obtained using the ML model compared to the design obtained using the ground truth phenomenological model.
The convergence history for the ML model shows minor differences, however the lateral shift is caused by slight delays in convergence within each continuation segment.

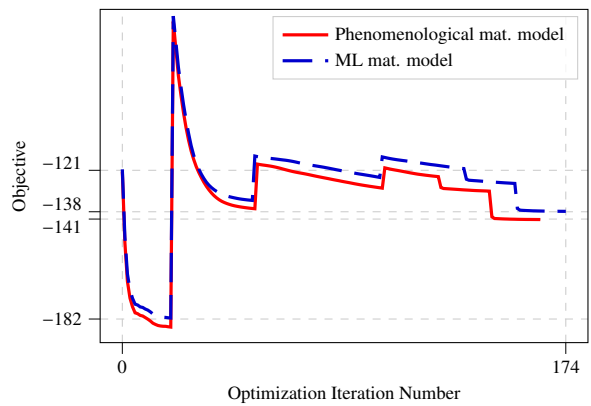
\begin{figure}[t!]
    \begin{minipage}{\linewidth}
    \centering        
    \def\svgwidth{0.4\linewidth}
    \setlength{\figW}{8cm}
    \setlength{\figH}{6cm}
    {\scriptsize
\begin{tikzpicture}

\definecolor{gray}{RGB}{128,128,128}
\definecolor{lightgray204}{RGB}{204,204,204}
\definecolor{mediumblue}{RGB}{0,0,205}
\pgfplotsset{grid style={dashed,lightgray204}}

\begin{axis}[
height=\figH,
legend cell align={left},
legend style={fill opacity=0.8, draw opacity=1, text opacity=1, draw=lightgray204},
tick align=outside,
tick pos=left,
width=\figW,
xlabel={Optimization Iteration Number},
xmajorgrids,
xmin=-8.7, xmax=182.7,
xtick = {0, 174},
ytick = {-121, -182, -138, -141},
xtick style={color=black},
ylabel={Objective},
ymajorgrids,
ymin=-191.67825, ymax=-55,
ytick style={color=black},
y tick label style={yshift={(\tick==-141)*-3pt}},
y tick label style={yshift={(\tick==-138)*3pt}},
y tick label style={yshift={(\tick==-121)*3pt}}
]
\addplot [very thick, red]
table {%
0 -121.55
1 -150.59
2 -166.54
3 -173.39
4 -176.99
5 -179.45
6 -179.81
7 -180.5
8 -180.76
9 -181.4
10 -182.05
11 -182.8
12 -183.71
13 -184.22
14 -184.65
15 -184.86
16 -184.97
17 -184.99
18 -185.17
19 -185.3
20 -59.944
21 -71.216
22 -82.23
23 -90.171
24 -97.362
25 -103.51
26 -108.69
27 -113.05
28 -116.54
29 -119.5
30 -121.88
31 -123.98
32 -125.91
33 -127.59
34 -129.04
35 -130.32
36 -131.41
37 -132.31
38 -133.08
39 -133.71
40 -134.2
41 -134.62
42 -134.98
43 -135.3
44 -135.56
45 -135.77
46 -135.96
47 -136.13
48 -136.28
49 -136.42
50 -136.55
51 -136.66
52 -136.76
53 -118.46
54 -118.65
55 -118.81
56 -118.95
57 -119.1
58 -119.26
59 -119.45
60 -119.67
61 -119.91
62 -120.18
63 -120.47
64 -120.75
65 -121.02
66 -121.28
67 -121.53
68 -121.75
69 -121.95
70 -122.14
71 -122.33
72 -122.53
73 -122.74
74 -122.97
75 -123.2
76 -123.42
77 -123.63
78 -123.84
79 -124.05
80 -124.25
81 -124.45
82 -124.67
83 -124.88
84 -125.08
85 -125.27
86 -125.49
87 -125.69
88 -125.88
89 -126.07
90 -126.28
91 -126.49
92 -126.68
93 -126.87
94 -127.05
95 -127.22
96 -127.38
97 -127.55
98 -127.7
99 -127.86
100 -128.01
101 -128.16
102 -128.32
103 -120
104 -120.24
105 -120.48
106 -120.69
107 -120.88
108 -121.06
109 -121.25
110 -121.45
111 -121.65
112 -121.85
113 -122.05
114 -122.23
115 -122.42
116 -122.59
117 -122.73
118 -122.87
119 -123.01
120 -123.13
121 -123.26
122 -123.38
123 -123.49
124 -123.6
125 -128.06
126 -128.46
127 -128.53
128 -128.6
129 -128.67
130 -128.74
131 -128.81
132 -128.88
133 -128.96
134 -129.03
135 -129.09
136 -129.15
137 -129.22
138 -129.28
139 -129.33
140 -129.37
141 -129.41
142 -129.45
143 -129.49
144 -129.55
145 -140.07
146 -140.77
147 -140.82
148 -140.87
149 -140.92
150 -140.96
151 -140.99
152 -141.03
153 -141.05
154 -141.07
155 -141.1
156 -141.12
157 -141.13
158 -141.15
159 -141.16
160 -141.16
161 -141.17
162 -141.18
163 -141.19
164 -141.2
};
\addlegendentry{Phenomenological mat. model}
\addplot [very thick, mediumblue, dash pattern=on 11.1pt off 4.8pt]
table {%
0 -120.6
1 -148.34
2 -163.55
3 -170.06
4 -173.51
5 -175.91
6 -176.24
7 -177.03
8 -177.25
9 -177.9
10 -178.34
11 -179.06
12 -179.93
13 -180.46
14 -180.85
15 -181.1
16 -181.26
17 -181.34
18 -181.49
19 -181.59
20 -57.735
21 -68.279
22 -78.756
23 -86.722
24 -93.982
25 -100.28
26 -105.93
27 -110.89
28 -114.84
29 -117.95
30 -120.5
31 -122.6
32 -124.38
33 -125.91
34 -127.13
35 -128.12
36 -128.95
37 -129.66
38 -130.29
39 -130.8
40 -131.21
41 -131.55
42 -131.83
43 -132.08
44 -132.31
45 -132.53
46 -132.72
47 -132.88
48 -133.01
49 -133.14
50 -133.25
51 -133.36
52 -115.15
53 -115.38
54 -115.56
55 -115.72
56 -115.86
57 -115.99
58 -116.11
59 -116.22
60 -116.34
61 -116.47
62 -116.61
63 -116.74
64 -116.86
65 -117.01
66 -117.19
67 -117.38
68 -117.6
69 -117.84
70 -118.08
71 -118.29
72 -118.5
73 -118.69
74 -118.88
75 -119.05
76 -119.22
77 -119.39
78 -119.56
79 -119.74
80 -119.94
81 -120.14
82 -120.35
83 -120.56
84 -120.77
85 -120.97
86 -121.16
87 -121.36
88 -121.55
89 -121.75
90 -121.94
91 -122.14
92 -122.33
93 -122.51
94 -122.67
95 -122.84
96 -123.01
97 -123.18
98 -123.36
99 -123.54
100 -123.73
101 -123.92
102 -115.51
103 -115.74
104 -115.97
105 -116.18
106 -116.37
107 -116.54
108 -116.7
109 -116.84
110 -116.98
111 -117.11
112 -117.26
113 -117.42
114 -117.57
115 -117.73
116 -117.88
117 -118.02
118 -118.17
119 -118.33
120 -118.48
121 -118.63
122 -118.75
123 -118.88
124 -119.02
125 -119.17
126 -119.31
127 -119.45
128 -119.59
129 -119.73
130 -119.86
131 -119.98
132 -120.1
133 -120.2
134 -120.29
135 -124.66
136 -125.07
137 -125.15
138 -125.23
139 -125.3
140 -125.38
141 -125.46
142 -125.55
143 -125.63
144 -125.7
145 -125.78
146 -125.86
147 -125.93
148 -126
149 -126.07
150 -126.14
151 -126.2
152 -126.26
153 -126.33
154 -126.39
155 -136.65
156 -137.35
157 -137.42
158 -137.48
159 -137.53
160 -137.57
161 -137.62
162 -137.64
163 -137.67
164 -137.69
165 -137.72
166 -137.74
167 -137.77
168 -137.78
169 -137.8
170 -137.81
171 -137.81
172 -137.82
173 -137.83
174 -137.83
};
\addlegendentry{ML mat. model}
\end{axis}

\end{tikzpicture}
    }
    \end{minipage}

    \caption{Convergence history for the fixed-fixed beam 3D TO benchmark.}
    
\end{figure}

\newpage

\section{Nomenclature}
\vspace{-20pt}
\printnomenclature




\end{document}